\documentclass[a4paper,onecolumn,11pt]{quantumarticle}
\pdfoutput=1
\usepackage[utf8]{inputenc}
\usepackage[english]{babel}
\usepackage[T1]{fontenc}
\usepackage{amsmath}
\usepackage{hyperref}
\usepackage{mathtools}
\usepackage{tikz}
\usepackage{lipsum}
\newcommand{\pin}{\par\noindent}
\begin{document}

\title{Holographic entanglement renormalisation of topological order in a quantum liquid}

\author{Anirban Mukherjee}
\affiliation{Indian Institute of Science Education and Research Kolkata,India}
\affiliation{Ames Laboratory, Ames, Iowa 50011, USA}
\email{anirbanm@ameslab.gov}
\orcid{0000-0001-5515-7054}
\author{Siddhartha Lal}
\orcid{0000-0002-5387-6044}
\affiliation{Indian Institute of Science Education and Research Kolkata,India}
\email{slal@iiserkol.ac.in}
\maketitle

\begin{abstract}
  We introduce a novel momentum space entanglement renormalization group (MERG) scheme for the topologically ordered (T.O.) ground state of the 2D Hubbard model on a square lattice (\cite{anirbanmotti,anirbanmott2}) using a unitary quantum circuit comprised of non-local unitary gates. At each MERG step, the unitary quantum circuit disentangles a set of electronic states, thereby transforming the tensor network representation of the many-particle state. By representing the non-local unitary gate as a product of two-qubit disentangler gates, we provide an entanglement holographic mapping (EHM) representation for MERG. Using entanglement based measures from quantum information theory and complex network theory, we study the emergence of topological order in the bulk of the EHM. We also demonstrate that the MERG is equivalent to a stabiliser quantum error correcting code. The MERG reveals distinct holographic entanglement features for the normal metallic, topologically ordered insulating quantum liquid and Ne\'{e}l antiferromagnetic symmetry-broken ground states of the 2D Hubbard model at half-filling found in Ref.\cite{anirbanmotti}, clarifying the essence of the entanglement phase transitions that separates the three phases. An MERG analysis of the quantum critical point of the hole-doped 2D Hubbard model found in Ref.\cite{anirbanmott2} reveals the evolution of the many-particle entanglement of the quantum liquid ground state with hole-doping, as well as how the collapse of Mottness is responsible for the emergence of d-wave superconductivity. We perform an information theoretic analysis of the EHM network, demonstrating that the information bottleneck principle is responsible for the distillation of entanglement features in the heirarchical structure of the EHM network. As a result, we construct a deep neural network (DNN) architecture based on our EHM network, and employ it for predicting the onset of topological order. We also demonstrate that the DNN is capable of distinguishing between the topologically ordered and gapless normal metallic phases.
\end{abstract}
\section{Introduction}
\label{sec:intro}
\pin
Classifying quantum many-body systems at criticality, as well as away from it,
based on entanglement features of the many-particle wavefunction remains a challenging problem. It bears implications for many-body theory, quantum information theory and quantum field theory~\cite{calabrese2004,ryu2006aspects,ryu2006,vidal2007,
amico2008entanglement,eisert2010,laflorencie2016quantum,swingle2010,
brown2016,nishioka2018,zengchenwen}. The renormalization group (RG) formalism~\cite{wilson1971a,wilson1971b,fisher1974} is central to meeting this challenge, as it involves characterizing various phases of matter and the phase transitions between them. Interpreting the RG length scale as an extra holographic dimension~\cite{balasubramanian1999,de2000,heemskerk2009,lee2010,casini2011,heemskerk2011,lee2014}  has allowed the realization of $d+1$ dimensional space in the bulk by starting from a $d$ dimensional quantum field theory at the boundary. Indeed, there has been a surge in the literature on the emergence of spatial geometry via the holographic renormalization of entanglement features~\cite{van2010building,evenbly2011,nozaki2012,swingle2012a,swingle2012b,qi2013,lee2016,hayden2016}   (e.g., mutual information, entanglement entropy) has been observed. Given that a diverse array of gapped and gapless quantum liquids can be studied via entanglement renormalization (ER)~\cite{aguado2008,giovannetti2008,pfeifer2009,bridgeman2015,swingle2016,wen2016,gu2016,gerster2017}, it is imperative to ask the question: can we classify various quantum liquids solely based on the attributes of entanglement features generated via ER?~   
\par\noindent
Various kinds of entanglement RG (ERG) have been implemented in the recent past, mostly in terms of renormalisation schemes implemented on real space lattices. We review the construction of some of them here. The multiscale entanglement renormalization group ansatz 
(MERA) is a tensor network defined as the stacking of $N$ transformation layers comprising unitaries and isometries acting on a system of $2^{N}$ qubits~\cite{vidal2007,vidal2008,evenbly2009}. The input array of qubits at the d-dimensional {\emph boundary} (i.e., at the ultraviolet (UV) energy scale of the network) are in general entangled and comprise the initial many-body state. These input qubits are gradually transformed by a series of duplet layers stacked along the holographic RG direction from UV to infrared (IR) energy scales. Each of the duplet layers is composed of the following: layer $1$ is a tensor product of (real space) local unitary disentangler gates with two input and two output legs each, while layer $2$ is a product of isometries with two input legs and one output leg each. Layer $2$ is, in essence, a projective transformation that removes the disentangled qubits generated at each RG step from the complete Hilbert space. As a result, the number of legs at the output from each layer is exactly half of the number of input legs. MERA and its other variants~\cite{haegeman2013,you2016,kim2017} allow the holographic generation of spatial geometry~ \cite{evenbly2011,swingle2012a,swingle2012b,qi2013,lee2016,hyatt2017} along the RG direction, starting from a tensor network~\cite{orus2019tensor,orus2014practical,biamonte2017} representation of the many-body state it acts on. There is a generalization of MERA, i.e., the entanglement holographic mapping (EHM)~\cite{qi2013,lee2016}, in which each transformation layer is comprised of only unitaries (i.e., layer $1$ discussed above). As a consequence, the EHM is also a tensor network formed from a collection of the unitary transformation layers. This should be contrasted with MERA, where only the many body states have a tensor network representation. In EHM, the disentangled qubits are present in the bulk, allowing a fuller description of the space-time geometry emergent via the RG flow. The time dynamics of the EHM tensor network is encoded in the inverse energy scales of the decoupled degrees of freedom included in its bulk, allowing for the computation of equal time, as well as unequal time, bulk correlators~\cite{lee2016}. This demonstrates the equivalence of the EHM to a quantum renormalization group (QRG, see Ref. \cite{lee2014quantum,lee2016} for recent attempts). As we will discuss below, in our implementation of the EHM via a unitary RG~\cite{anirbanmotti,anirbanmott2,pal2019}, the appearance of an energy scale for quantum fluctuations naturally encodes the quantum dynamics of the emergent space. 
\par\noindent
An outstanding challenge for ERG has been in its application to strongly correlated electronic systems~\cite{corboz2009,barthel2009}. Although there are several works on constructing tensor networks, e.g., MERA or EHM, for systems of noninteracting fermions~\cite{corboz2009,lee2016,haegeman2018rigorous}, there are only a few on systems of interacting fermions~\cite{evenbly2010,murg2010simulating}. Thus, by and large, the implementation of ERG on highly entangled systems resulting from strong electronic correlations remains unresolved. Our work thus aims for the construction a momentum-space entanglement renormalization group (MERG)~\cite{balasubramanian2012momentum} technique based on the unitary transformation based RG (URG) that has recently been applied to the 2D Hubbard model~\cite{anirbanmotti} and quantum XXZ kagome antiferromagnet~\cite{pal2019}. At every step of the URG, we perform an unitary operation $U$ that, similarly to QRG~\cite{lee2014quantum,lee2016}, MERA and EHM systematically disentangles the electronic states at the UV from those at IR. This leads to the iterative block diagonalization of the Hamiltonian along the RG flow, approaching a model at the IR fixed point which is simpler than the bare model one started with.
\pin
Recently, in Ref.\cite{anirbanmotti}, we have obtained the wavefunction for an insulating symmetry-preserved Mott liquid ground state of the 2D Hubbard model at half-filling analytically from the low-energy fixed point theory. The ground state manifold is found to be topologically degenerate~\cite{wen1990}, with a nonlocal Wilson loop that commutes with the elements of the stabilizer group for the ground state manifold~\cite{gottesman1997,nielsen2002quantum}. We have also validated quantitatively the effective Hamiltonian and ground state wavefunction for the Mott insulator obtained from the URG procedure by benchmarking with high accuracy the ground state energy per particle, as well as the doublon fraction, against other numerical methods~\cite{leblanc2015solutions}. The parent metallic normal state of the Mott liquid was found to be a gapless marginal Fermi liquid. The Mott liquid was also observed to be unstable towards a Ne\'{e}l antiferromagnetic ground state upon including the effects of a symmetry-breaking perturbation in the URG analysis. Here, we will analyse the entanglement properties of the Mott liquid, marginal Fermi liquid and Ne\'{e} antiferromagnet ground states of the 2D Hubbard model using the MERG method. Further, in Ref.\cite{anirbanmott2}, we found that d-wave superconductivity was emergent from a quantum critical point (QCP) associated with the collapse of Mottness upon doping holes into the Mott liquid ground state. We will, therefore, also investigate how the entanglement features of the Mott liquid evolves with hole-doping, such that its dominant symmetry-breaking instability at half-filling towards the Ne\'{e}l antiferromagnet is replaced by that towards d-wave superconductivity.
\par\noindent
In the present work, the MERG scheme is implemented as a {\emph reverse unitary RG}. Similar reverse RG approaches that involve the re-entangling of hitherto disentangled degrees of freedom has been discussed for some tensor network RG approaches~\cite{swingle2012b,wen2020}. For this, we start from the many-body state at the $IR$ fixed point, and approach the $UV$ (i.e., towards the eigenstate of the parent model) by iterative applications of the inverse of the unitary transformations of the URG. The reconstruction method is validated by showing a reduction in the uncertainty of the energy eigenvalue of the reconstructed state with respect to that of the parent model. By representing the many-body states at each RG step as quantum circuits/tensor network~\cite{biamonte2017}, and by representing each unitary transformation step as a product of two-local universal gates~\cite{divincenzo1995two}, we obtain an equivalent EHM description for the MERG. Further, such a decomposition guarantees that URG is a version of a Clifford stabiliser code~\cite{gottesman1998heisenberg} that obeys the Gottesman-Knill theorem. Therefore, the reverse RG can be visualized as a entanglement RG flow along the reverse holographic direction~\cite{swingle2012b,lee2016}. Much like the EHM, MERG is a version of MERA that preserves spectral weight during the RG flow~\cite{lee2016}; this is precisely why we are able to implement both forward and reverse RG constructions. Importantly, note that at each RG step, the $U$ transformations are not determined variationally as in MERA~\cite{evenbly2009} or EHM. Instead, they are determined directly from the form of the Hamiltonian obtained at the previous step~\cite{anirbanmotti}. The URG procedure also encodes naturally an energy scale for quantum fluctuations ($\omega$), accounting for the variation in entanglement content as well as the energy contributions of the decoupled states within the bulk of the tensor network. Thus, similar to EHM~\cite{lee2016}, the bulk correlation functions in the MERG are naturally time dynamical. It is also noteworthy that each transformation layer in MERG has a finite-depth that quantifies the circuit complexity of the corresponding unitary gate. This is in contrast with MERA, where every RG step has only unit depth. On the other hand, this aspect of MERG is similar to deep MERA (dMERA)~\cite{kim2017}, where every RG step has a finite depth $D$. 
\par\noindent
We will now discuss the entanglement features pertinent to the ERG program, and then provide a comparative study with the features observed in MERG. One important property relates to the entanglement entropy of a region $R$ in the input state at the UV scale $|\Psi\rangle$, i.e., the boundary of MERA or EHM tensor network. It was shown in Ref.\cite{swingle2012b} that the entanglement entropy is bounded from above by the number of legs that must be cut for the isolation of the region $R$ from its complementary part. Upon descending further into the tensor network, the number of legs that need to be cut are reduced exponentially. The depth of the MERA/EHM quantum circuit/tensor network quantifies the circuit complexity of the state $|\Psi\rangle$, i.e., the minimum number of two-local and one-local universal gates required to obtain the entangled state from a separable state~\cite{brown2018second}. As we are disentangling qubits in the MERG program, the circuit complexity is  greatly reduced in the bulk of the EHM. Therefore, the gain in efficiency in obtaining the entanglement features of a region $R$ in $|\Psi\rangle$ is tied to the reduction in circuit complexity of state $|\Psi\rangle$ and the unitary transformation $U$. This attribute of the EHM conforms to the Ryu-Takayanagi formula~\cite{ryu2006}, where the entanglement entropy of a region is determined by the minimal surface (whose length is the number of links cut) of the causal cone enveloping it~\cite{swingle2012b,beny2013causal}. This observation motivates the manifestation of a holographic duality within the ERG framework~\cite{lee2016}. In this work, we show that the EHM constructed from MERG has features similar to the Ryu-Takayanagi relation; indeed, we find that the entanglement entropy obtained by isolating a region $R$ after every layer of transformation is bounded from above by the length of the minimal surface of the causal cone that surrounds it. We show that both the Mott insulating and the normal metallic states obtained from MERG respect the Ryu-Takayanagi relation. Further, we find that in MERG, the structure of the causal cone deep in the bulk of the EHM is determined by degrees of freedom residing close to the Fermi surface. This is similar to the finding that Fermi surface wave vectors play a crucial role in determining the entanglement entropy for gapless metallic quantum critical systems~\cite{swingle2010,swingle2012conformal}.
\par\noindent
Another essential feature of tensor networks such as EHM and MERA is that they encode the entanglement content of the many-body state geometrically~\cite{lee2016,hyatt2017}. This can be quantified in terms of a distance measure such as the negative logarithm of the quantum mutual information (MI) between pairs of qubits $(i,j)$. It is important to recall that the quantum mutual information characterises the total (i.e., quantum as well as classical) correlations between a pair of qubits~\cite{groisman2005}. The lower the entanglement of the pair, the higher is the distance and vice-versa. Note that the information theoretic distance between a pair of qubits $(i,j)$ obtained from the bulk of EHM is, in general, different from that obtained at the boundary, implying that a curved space-time geometry can be encoded into the EHM. For instance, an information theoretic distance proportional to the logarithm of the physical distance indicates the generation of hyperbolic space-time geometry from entanglement along the RG holographic direction~\cite{hyatt2017}. We further note that while Ref.\cite{wolf2008area} established that the mutual information is an upper bound for various two-point correlation functions, it has been shown that that the {\emph entanglement geometry} generated from MI asymptotically (i.e. deep within the tensor network) of the EHM encodes the single-particle bulk correlation function~\cite{lee2016}. In the EHM obtained by us from the MERG, we compute the mutual information between pairs of pseudospins $i$ and $j$ (each of which is a composite of two single electron states in momentum-space with opposite spin).
\pin
Upon approaching the Fermi energy in the bulk of our EHM, the MI is thus related to the four fermion bulk correlators of a system of interacting electrons, rather than the two fermionic kind constructed from noninteracting degrees of freedom in the EHM of Ref.\cite{lee2016}. This indicates that our EHM encodes the spacetime geometry generated by the emergent degrees of freedom at low energies. We find that for the Mott insulator, the largest MI pertains to a singlet state formed by pseudospin pairs deep in the IR. This is also displayed by a finite correlation between the pseudospins in this particular case. On the other hand, we find that for a gapless metallic state, the correlation between momentum-space pseudospins vanishes in the IR. This is also reflected in our finding of a spacetime that collapses to a singularity in the IR for the insulating case, and a spacetime that expands for the metallic phase. While the former indicates the condensation of singlet pairs with real-space short-ranged entanglement in the Mott insulating state~\cite{anirbanmotti}, the latter reflects on the scaling of the ERG towards a gapless Fermi surface~\cite{swingle2012b,swingle2010,lee2016}. Further, our findings appear to indicate that the metal-insulator transition between the Mott insulator and the gapless normal state acts as a horizon between the two entanglement spacetimes corresponding to these states, consistent with the suggestion of Ref.\cite{lee2014,lee2016horizon}. We also note that, very recently, quantum information theoretic studies have been carried out within the cluster dynamical mean-field theory (CDMFT) framework of the Mott metal-insulator transition at half-filling~\cite{tremblay6,tremblay8} and the pseudogap-correlated metal transition upon hole-doping~\cite{tremblay7}. Interestingly, these works appear to find signatures of these transitions even within the local entropy and total mutual information between a single site and the rest of the lattice.
\par\noindent
We end by mentioning some applied aspects of our work. Both MERA and EHM have interpretations as quantum error correcting codes~\cite{gottesman1997,almheiri2015,pastawski2015,kim2017entanglement}: along the reverse holographic direction from IR to UV, the unitary transformation layer acts as an encoding map that re-entangles the stabiliser codeword (IR fixed point) with the (hitherto) disentangled degrees of freedom~\cite{ferris2014}. Conversely, the passage from UV to IR involves disentangelement of qubits at UV scale and acts a decoder of the stabiliser codeword in the IR. This has implication for fault tolerant quantum computing and topological quantum error correction codes~\cite{kitaev2003fault,ferris2014,bombin2006}. In our present work, the codework space~ is formed by the two-fold topologically degenerate ground states of the Mott insulator~, such that the forward ($U$) and reverse ($U^{\dagger}$) unitary transformations either distill the codeword space from the space of decoupled qubits, or spreads the entanglement content of the former across the latter. We demonstrate these error correction features of the MERG in this work. In this way, MERG provides a platform for designing noise resilient topological error correction~\cite{bombin2012strong,kim2017noise,kim2017}
\par\noindent
A second important perspective of the MERG is as a deep neural network (DNN) architecture. We recall that a DNN generalisation of the restricted Boltzmann machine (RBM) is comprised of a stack of several hidden layers of neurons that generate simpler representations of the input data while preserving its essential features \cite{salakhutdinov2009}. For example, an input data with $N$ features (i.e., with a $N$-dimensional feature vector) undergoes dimensional reduction such that irrelevant features are discarded, and relevant features are distilled from one layer to the next. This has been demonstrated as being equivalent to the variational RG framework of Ref.\cite{kadanoff2000statistical,kadanoff1976}, where UV degrees of freedom are coarse-grained in an iterative fashion, thereby distilling the IR theory~\cite{beny2013,mehta2014exact}. These works show that variational aspect of the Kadanoff RG is equivalent to the usage of training data in constructing the optimal weight matrices for the hidden layers of the DNN using the steepest descent algorithm. Tishby and Zaslavsky~\cite{tishby2015deep} have shown that such DNN architectures follow the Information Bottleneck (IB) principle: an optimally transformed representation of the input data is one in which the mutual information between the output and input data is reduced, while preserving the essential components of the target feature vector. Similarly, the disentangling and isometry layers of tensor networks such as MERA \cite{gao2017efficient,lu2019efficient,liu2019machine} lead to a lower-dimensional representation of the original many-particle wavefunction they act on: the qubits disentangled at every layer are projected out, while preserving the entanglement features pertaining to degrees of freedom in the IR~\cite{beny2013deep}.
\par\noindent
This suggests that the EHM of resulting from our MERG is equivalent to a unitary realisation of a DNN based purely on unitary transformations. As our MERG is based on analytic expressions for the unitary disentanglers of the URG formalism, the equivalent DNN does not need optimisation via a variational procedure. Indeed, we will demonstrate in this work that the MERG based DNN follows the optimal IB trajectory. Further, this allows for the holographic reconstruction of the many-body state of the parent Hamiltonian (i.e, input feature vector of the DNN/ UV boundary of the MERG tensor network) starting from the essential features obtained from the target vector (RG fixed point/ IR bulk of the MERG tensor network) by reversing the flow of information across the DNN. In this sense, the RG flow of the URG represents a supervised DNN~\cite{tishby2015deep}, while the inverse RG flow represents a generative DNN~\cite{you2018machine,hu2019machine}. Recent works have highlighted similar relations between DNNs and tensor networks \cite{cohen2016expressive,levine2019quantum,li2018neural} on the one hand, and the holographic duals of DNNs \cite{hashimoto2018deep}. Importantly, we construct a DNN based on the MERG that can classify Mott insulating and normal metallic phases by distinguishing their entanglement features. In this way, the DNN we construct is sensitive to the metal-insulator transition that lies between these two phases of quantum matter.
\par\noindent
The rest of the work is organised as follows. In Section \ref{RGSetupSection}, we present the 2D Hubbard model briefly, as well as the unitary RG formalism of Ref.\cite{anirbanmotti} that is pertinent to the present work. We then formulate the MERG scheme for the topologically ordered insulating ground state of the 2D Hubbard model in Section \ref{MERG}. This is followed by developing quantum circuit representation of the many-particle state as well as the renormalisation procedure, leading to a numerical validation of the MERG. We then present the entanglement holographic mapping (EHM) for the MERG, allowing for a study of the emergence of the topologically ordered ground state from the entanglement RG and its visulation as a quantum error correcting code. In Section \ref{ERGflow}, we compute the ERG flow using several measures of entanglement (e.g., mutual information), developing insight on the distinction between the entanglement signatures of a gapped insulating Mott liquid ground state, a gapless normal metallic state and the Ne\'{e}l antiferromagnetic state. Further, in Section \ref{pathtodwave}, we analyse the evolution of the many-particle entanglement features of the Mott liquid at half-filling and the QCP arrived through hole-doping. This will shed light on how d-wave superconductivity arises from the collapse of Mottness at the QCP. Finally, in Section \ref{EHMtoDNN}, we use the EHM to develop a deep neural network that can classify the insulating and metallic phases based on their entanglement features. We conclude in Section \ref{conclusions}. Finally, details of the URG method, properties of the normal metallic, Ne\'{e}l antiferromagnetic and d-wave superconducting phases, and the theory at the quantum critical point are presented in the appendices.
\section{Preliminaries}\label{RGSetupSection}
\subsection{The model}
\pin
The 2D Hubbard model on a square lattice at $1/2$-filling, with nearest neighbour hopping (strength $t$) and on-site Hubbard repulsion (strength $U_{0}$), is described by the Hamiltonian
\begin{eqnarray}\centering
\hat{H}&=&\sum_{\mathbf{k}, \sigma}\epsilon_{0\mathbf{k}}c^{\dagger}_{\mathbf{k}\sigma}c_{\mathbf{k}\sigma}+U_{0}\sum_{\mathbf{r}}\hat{\tau}_{\mathbf{r}\uparrow}\hat{\tau}_{\mathbf{r}\downarrow}\label{Hubbard Hamiltonian}~,
\end{eqnarray}
where $c^{\dagger}_{\mathbf{k}\sigma}/c_{\mathbf{k}\sigma}$ are the electron creation/annihilation operator with wave-vector $\mathbf{k}$ and spin $\sigma$. The operator $\hat{\tau}_{\mathbf{r}\sigma}=\hat{n}_{\mathbf{r}\sigma}-\frac{1}{2}$, where $\hat{n}_{\mathbf{r}\sigma}=c^{\dagger}_{\mathbf{r}\sigma}c_{\mathbf{r}\sigma}$ is the number operator at lattice site $\mathbf{r}=j_{1}\hat{x}+j_{2}\hat{y}$ and $\epsilon_{0\mathbf{k}}$ is the bare dispersion. The hopping term is number diagonal in momentum-space, with a dispersion  $\epsilon_{0\mathbf{k}}=-2t(\cos k_{x}+\cos k_{y})$. 
On the other hand, the Hubbard repulsion term is diagonal in position space, and is therefore off-diagonal in the momentum basis. This is responsible for quantum fluctuations in the one particle dispersion ($\Delta(\epsilon_{\mathbf{k}\sigma}\hat{n}_{\mathbf{k}\sigma})$). Below, we lay out the unitary renormalization group procedure introduced in Ref.\cite{anirbanmotti} that block diagonalizes the Hamiltonian by recursively resolving the quantum fluctuations.
\subsection{Unitary Renormalization Group} (URG)\label{URG-Formal}
\pin
In earlier works \cite{anirbanmotti,anirbanmott2,anirbanurg1}, we introduced a unitary operator based RG that block diagonalizes the Hamiltonian iteratively and performs a systematic disentanglement of electronic degrees of freedom in the rotated eigenspace. A typical workflow of this unitary RG procedure is given in Fig.\ref{flowchartRG} (see Appendix \ref{App-1} for derivation of the unitary operator),
\begin{figure}[tbp]
\centering 
\includegraphics[width=.5\textwidth]{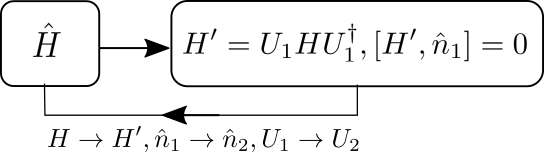}
\caption{\label{flowchartRG} Schematic diagram of the iterative unitary RG procedure. Each RG step generates a number occupation operator $\hat{n}_{1}$ that commutes with the Hamiltonian.}
\end{figure}
such that the $(j)$th step involves a unitary operation $U_{(j)}$ given by
\begin{eqnarray}\centering
U_{(j)}&=&\frac{1}{\sqrt{2}}(1+\eta_{(j)}-\eta^{\dagger}_{(j)})\label{UnitaryOpDef}\\
\eta^{\dagger}_{(j)}&=&\frac{1}{\hat{\omega}_{(j)}-Tr_{j}(H_{(j)}^{D}\hat{n}_{j})\hat{n}_{j}}c^{\dagger}_{j}Tr_{j}(H_{(j)}c_{j})~.\label{eta-operator}
\end{eqnarray}
The operators $\eta_{(j)}$, $\eta^{\dagger}_{(j)}$ fulfill the algebra $\lbrace\eta_{(j)},\eta^{\dagger}_{(j)}\rbrace=1$, $[\eta_{(j)},\eta^{\dagger}_{(j)}]=1-2\hat{n}_{(j)}$. The renormalized Hamiltonian \cite{anirbanmotti,anirbanmott2,anirbanurg1} is given by
\begin{eqnarray}
H_{(j-1)}&=&U_{(j)}H_{(j)}U^{\dagger}_{(j)}~,\nonumber\\
&=&Tr_{j}(H_{(j)}\hat{n}_{j})\hat{n}_{j}+\lbrace Tr_{j}(c^{\dagger}_{j}H_{(j)})c_{j},\eta^{\dagger}_{(j)}\rbrace +h.c.~.
\end{eqnarray}
From the Hamiltonian flow equation, we can extract the entire family of non-perturbative RG equations for various one-, two-, three-particle as well as higher-order vertices~ \cite{anirbanurg1}. The non-perturbative nature of the RG equations stems from the appearance of the renormalized single-particle self-energy and $n$-particle correlation energies in the denominator of the expression for $\eta_{(j)}$ (eq.\eqref{eta-operator}). Indeed, upon expanding the denominator in $\eta_{(j)}$ around the diagonal part of the parent Hamiltonian ($H^{D}_{0}$), 
\begin{eqnarray}
\eta_{(j)}\approx\left[\frac{1}{\hat{\omega}-H^{D}_{0}}+\frac{1}{\hat{\omega}-H^{D}_{0}}\Delta H^{D}_{(j)}\frac{1}{\hat{\omega}-H^{D}_{0}}+\ldots\right]\lbrace Tr_{j}(c^{\dagger}_{j}H_{(j)})c_{j},\eta^{\dagger}_{(j)}\rbrace~,
\end{eqnarray}
we can recover the one-loop BCS, ZS, ZS' diagrammatic contribution to the two-particle vertex RG equation~\cite{anirbanmotti}. The non-perturbative nature of the RG equations allow us to obtain stable fixed points for the Hamiltonian flow, from which we can extract the effective Hamiltonian and various renormalized couplings and parameters.
\pin
The operator $\hat{\omega}_{(n)}$ in the expression for $\eta^{\dagger}_{(n)}$ (eq.\eqref{eta-operator}) accounts for the residual quantum fluctuations due to remnant entanglement in the rotated eigenspace, and is defined as follows
\begin{eqnarray}\centering
\hat{\omega}_{(j)}=H^{D}_{(j)}+\Delta H_{(j)}~, \label{omegaOpdef}
\end{eqnarray}
where $H^{D}_{(j)}$ represents the diagonal component of $H_{(j)}$ and $\Delta H_{(j)}=H_{(j-1)}-H_{(j)}$ represents the renormalization of the Hamiltonian. Furthermore, the good occupation quantum numbers of the disentangled electronic states label the Hamiltonian's eigenstates in the rotated eigenspace. The eigenvalues ($\{\omega\}$) of the operator $\omega_{(n)}$ form a set of quantum energyscales. The URG framework thus generates effective Hamiltonian RG flows at various $\omega$ energyscales such that, at each $\omega$, a sub-part of the complete Hamiltonian spectrum is renormalized as well as block diagonalised.
\par\noindent
Note that eq.\eqref{omegaOpdef} is an equivalent way to write the Hamiltonian RG flow equation: $\Delta H_{(j)}=\hat{\omega}_{(j)}-H^{D}_{(j)}$. For performing further analytical calculations and numerical simulations, we choose $\hat{\omega}_{j}$ to be a diagonal matrix and ignore all higher order off-diagonal contributions accumulated in $\Delta H$. Importantly, the eigenvalues $\omega_{0,k}=\epsilon_{k}+\Sigma^{0}_{k}$ of $\hat{\omega}$ are in one-to-one correspondence with the energy eigenvalues $\epsilon_{k}$ of the single-particle part $H_{1}$ of the many-body Hamiltonian, say, $H=H_{1}+H_{2}$ (and $H_{2}$ comprises the two-particle and higher order interactions). Here, $\Sigma^{0}_{\mathbf{k}}$ represents the bare self-energy and the momentum $k$ labels the eigenstates $|k\rangle$ of $H_{1}$. In order to quantify the cost incurred upon ignoring the off-diagonal renormalization contributions within $\omega$, we now define a \textit{cost function} ($C_{j}$): 
\begin{eqnarray}
C_{j}=|\langle\Psi_{G,j}|(\hat{\omega}_{0}-H^{D}_{(j)}-\Delta H_{(j)})|\Psi_{G,j}\rangle|^{2}~,\label{cost_function}
\end{eqnarray}
where $|\Psi_{G,j}\rangle$ is the 
ground state wavefunction of the renormalized Hamiltonian $H_{(j)}$ for step $j$. After the first step of the renormalization procedure, the cost is $C_{1}=|\langle\Psi_{G,0}|\sum_{k}\Sigma_{k}\hat{n}_{k}-\Delta H_{0}|\Psi_{G,0}\rangle|^{2}$. At the RG fixed point ($j^{*}$)~\cite{anirbanmotti,anirbanurg1,glazek2004}, the cost function is easily seen to vanish
\begin{equation}
|\langle\Psi_{G,j^{*}}|\omega_{k}-H^{D}_{j^{*}}|\Psi_{G,j^{*}}\rangle=0=\Delta H_{j^{*}}~\Rightarrow~C_{j^{*}}=0~.
\end{equation}
In this manner, the URG procedure is a deterministic optimization approach where the cost function vanishes at the RG fixed point, and the exact ground state of the fixed point Hamiltonian is found.
\pin
We briefly discuss below certain important aspects of the URG formalism:
\par\noindent
\begin{enumerate}
\item[1.] \textbf{Spectrum preservation}\\
By definition, the unitary transformations of URG are spectrum preserving
\begin{eqnarray}
H|\Psi_{n}\rangle= E_{n}|\Psi_{n}\rangle\to U_{n}HU_{n}^{\dagger}U_{n}|\Psi_{n}\rangle=E_{n}U_{n}|\Psi_{n}\rangle~. 
\end{eqnarray}
As a result, the ground state energy of the fixed point Hamiltonian $H_{*}|\Psi_{G,*}\rangle =E_{g}|\Psi_{G,*}\rangle$ matches that of the parent Hamiltonian $H$.
\item[2.]\textbf{Preservation of symmetries}\\
The unitary operator $U_{(j)}$ (eqs.\eqref{UnitaryOpDef},\eqref{eta-operator}) is solely constructed from the terms of the Hamiltonian. It manifestly respects, therefore, the symmetries of the parent Hamiltonian $H$. For instance, if $H$ is particle number conserving, $[H,N]$, where $N=\sum_{k}\hat{n}_{k}$ is the total number operator, then $U_{(j)}\hat{N}U_{(j)}^{\dagger}=\hat{N}$.
\item[3.]\textbf{Nature of rotation generators}\\
The generator of the many-particle transformation, $i(\eta-\eta^{\dagger})$, corresponds to a non-local controlled-Y gate, where the control bit is the single electronic state to be disentangled and the collection of target electronic states is comprised of all the other electronic states that remain coupled to one another. Furthermore, a Jordan-Wigner transformation of the electronic states can lead to a representation of $i(\eta-\eta^{\dagger})$ involving a weighted sum of Pauli strings, where every term  will have an odd number of Y gates. The real valued representation of $U$ in URG ensures that there is no dynamics in the global phase dynamics of the many-particle wavefunction. Instead, the URG leads to changes in the relative phases between the different configurations that are in superposition. At a later point in this work, we present the $U_{(j)}$ for the 2d Hubbard model as a quantum circuit that can be written completely in terms of one- and two-qubit gates (eq.\eqref{decompU}). 
\end{enumerate}
\par\noindent
We adapt the RG procedure to the 2D Hubbard model by setting up a labelling scheme for the states in momentum space. The states are labelled by two indexes: distance $\Lambda$ from the noninteracting Fermi surface ($\epsilon_{\mathbf{k}}=0$, FS), and the direction normal to the FS ($\hat{s}=\frac{\nabla\epsilon_{\mathbf{k}}}{|\nabla\epsilon_{\mathbf{k}}|}|_{\epsilon_{\mathbf{k}}=E_{F}}$) such that $\mathbf{k}_{\Lambda\hat{s}}=\mathbf{k}_{F\hat{s}}+\Lambda\hat{s}$. This unveils a natural scheme for labelling the states in terms of distances $\Lambda_{N}>..>\Lambda_{j}>\Lambda_{j-1}>..>0$. The RG transformations then disentangle electronic states farthest from the FS, gradually scaling towards Fermi energy $E_{F}$. At step $j$, all the states on the curve $\Lambda_{j}$ are completely disentangled via a unitary rotation $U_{(j)}$. The resulting  Hamiltonian $H_{(j-1)}=U_{(j)}H_{(j)}U_{(j)}^{\dagger}$ is off-diagonal only for states residing within a window $\Lambda<\Lambda_{j}$ around the erstwhile FS.
The disentanglement of an entire curve at distance $\Lambda_{j}$ is represented via a product of unitary rotations $U_{(j)}=\prod_{l\in(\hat{s},\sigma=\uparrow/\downarrow)}U_{(j,l)}$, where $U_{(j,l)}$ disentangles one state $|\mathbf{k}_{\Lambda_{j}\hat{s}}\sigma\rangle$ on the curve $\Lambda_{j}$. The form of $U_{(j,l)}=\frac{1}{\sqrt{2}}[1+\eta_{(n)}-\eta^{\dagger}_{(n)}]$ where $n=(j,l)$ and $\eta_{(n)}$, $\eta^{\dagger}_{(n)}$ have the same definitions as presented in eq.\ref{UnitaryOpDef}.
\subsection{Normal and topologically ordered (T.O.) insulating phases of the 2D Hubbard model at $1/2$-filling}
\begin{figure}
\centering
\includegraphics[width=.5\textwidth,origin=c,clip]{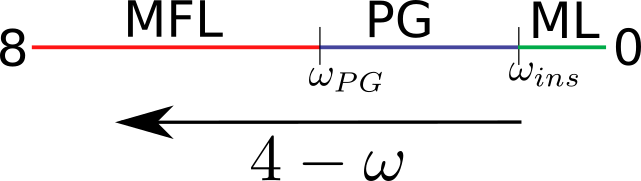} 
\caption{\label{phDiag} Schematic representation of the $T=0$ RG phase diagram of the 2D Hubbard model at $1/2$-filling as a function of quantum fluctuation energyscale ($8<4-\omega<0$, for a tight-binding bandwidth of $8t,~t=1$). Red, blue and green regions represent the marginal Fermi liquid metal (MFL), pseudogap (PG) and the Mott insulating liquid (ML) phases respectively. $\omega_{PG}$ and $\omega_{ins}$ are energy scales for transitions that initiate and end the PG phase.}
\end{figure}
\pin
As shown schematically in Fig.\ref{phDiag}, the renormalization procedure reveals a $T=0K$ phase diagram with the $x$-axis given by the one-particle quantum fluctuation energyscale $\omega$ within $\hat{\omega}_{(n)}$. This tracks the single-particle spectral function, i.e., for high energies $4-\omega>\omega_{PG}$ (where the tight-binding bandwidth is $8t$ with $t=1$), the spectrum is gapless and corresponds to the marginal Fermi liquid phase~\cite{anirbanmotti,anirbanmott2} (red region in Fig.\ref{phDiag}). For $\omega_{ins}<4-\omega<\omega_{PG}$, the spectrum is pseudogapped, while the spectrum is completely gapped for $4-\omega<\omega_{ins}$. It is in this final regime that the insulating Mott liquid state resides, with both spin and charge excitations being gapped~\cite{anirbanmotti,anirbanmott2}. The Hamiltonian for the MFL is found to be
\begin{eqnarray}\centering
H^{*}_{1}&=&\sum_{j,l}\epsilon_{j, l}\hat{n}_{j, l}+\frac{1}{8}\sum_{j,j',l}R^{*}_{ll'}\hat{n}_{j, l}\hat{n}_{j, l'}(1-\hat{n}_{j', l})~,\label{normalState} 
\end{eqnarray}
where $j,l=\Lambda_{j},\hat{s},\sigma$ and $j,l'=-\Lambda+\delta, T\hat{s}$. The first and second terms in Hamiltonian eq.\ref{normalState} describes the 1 particle and 2 particle-1 hole dispersions respectively. The 2 particle-1 hole dispersion $R^{*}_{ll'}=\omega -1/2(\epsilon_{j*,l}+\epsilon_{j*,l'})$ arises from the interplay between forward scattering processes involving electronic pairs with different net momentum. About the fluctuation scale $\frac{W}{2}-\omega_{ins}$, the system undergoes a transition into a Mott insulating phase (green region in the phase diagram Fig.\ref{phDiag}) described by the fixed point Hamiltonian
\begin{eqnarray}\centering
\hat{H}^{*}_{2}&=&\sum_{\hat{s}}K^{*}_{\hat{s}}\bigg[\mathbf{A}_{*, \hat{s}}\cdot\mathbf{A}_{*, -\hat{s}} - \mathbf{S}_{*, \hat{s}}\cdot\mathbf{S}_{*, -\hat{s}}\bigg]~,\label{fixed_point_Ham_mott}
\end{eqnarray}
where the charge and spin type psuedospins within the emergent fixed point window $\Lambda^{*}_{\hat{s},\omega}$, $\mathbf{A}_{*, \hat{s}}$ and $\mathbf{S}_{*, \hat{s}}$ respectively, are defined as
\begin{eqnarray}
\centering
\hspace*{-2cm}
\mathbf{A}_{*,\hat{s}} &=& \sum_{\Lambda<\Lambda^{*}_{\hat{s},\omega}}\mathbf{A}_{\Lambda,\hat{s}}~~,\mathbf{S}_{*,\hat{s}}=\sum_{\substack{\Lambda<\Lambda^{*}_{\hat{s},\omega}}}\mathbf{S}_{\Lambda,\hat{s}}~,\mathbf{A}_{\Lambda,\hat{s}}=f^{c;\dagger}_{\Lambda,\hat{s}}\frac{\boldsymbol{\sigma}}{2}f^{c}_{\Lambda,\hat{s}}~~,\mathbf{S}_{\Lambda,\hat{s}}=f^{s;\dagger}_{\Lambda,\hat{s}}\frac{\boldsymbol{\sigma}}{2}f^{s}_{\Lambda,\hat{s}}~,
\label{pseudospins}
\end{eqnarray}
where $f^{c;\dagger}_{\Lambda,\hat{s}} =\left[c^{\dagger}_{\Lambda,\hat{s},\sigma} ~c_{-\Lambda,T\hat{s},-\sigma}\right]$~and~$f^{s;\dagger}_{\Lambda,\hat{s}} = \left[c^{\dagger}_{\Lambda,\hat{s},\sigma}~ c^{\dagger}_{\Lambda - 2\Lambda^{*}_{\hat{s}},T\hat{s},-\sigma}\right]$ are the spinorial representation for a pair of electrons. Note that the pairing is between two electronic states with net momentum $\mathbf{Q}$; these pairs carry the highest spectral weight among all finite-momentum pairs, and condense into pseudospins at low-energies~\cite{anirbanmotti}.
\pin
$K^{*}_{\hat{s}}$ is the final magnitude of the backscattering coupling. The degenerate ground state configurations of the Mott insulating phase are labelled by \textit{either} the eigenvalues of the three operators $\mathbf{A}_{*}=\mathbf{A^{*}}_{\hat{s}}+\mathbf{A^{*}}_{-\hat{s}}$ , $\mathbf{A^{*}}_{\hat{s}}$ and $\mathbf{A^{*}}_{-\hat{s}}$, \textit{or} their $\mathbf{S}$ counterparts 
\begin{eqnarray}
\hspace*{3cm}&&|\Gamma_{+}\rangle = |A_{*}=0,A^{*}_{\hat{s}}=A^{*}_{-\hat{s}}=N^{*}_{\hat{s}},S_{\Lambda\hat{s}}=0\rangle~,~\nonumber\\
\hspace*{3cm}&&|\Gamma_{-}\rangle = |S_{*}=2N^{*}_{\hat{s}},S^{z}_{*}=0,S^{*}_{\hat{s}}=S^{*}_{-\hat{s}}=N^{*}_{\hat{s}},A_{\Lambda\hat{s}}=0\rangle~.~~~~~~
\end{eqnarray}
The two configurations $|\Gamma_{+}\rangle$ and $|\Gamma_{-}\rangle$ are constructed such that $|\Gamma_{+}\rangle$ is an eigenstate of the first term in $H^{*}_{2}$ (and has null contribution from the second term), while $|\Gamma_{-}\rangle$ is the eigenstate of the second part (and has no contribution from the first). Importantly, we note that the true degeneracy between these ground states is only achieved in the thermodynamic limit. Further, the RG procedure has been quantitatively validated in Refs.\cite{anirbanmotti,anirbanmott2} by benchmarking the ground state energy per site of the Hubbard model at various dopings against other numerical methods across values of the Hubbard repulsion ($U$) with magnitudes ranging from weak to strong~\cite{leblanc2015solutions}. 
\par\noindent
To characterize the topological features of the gapped two degenerate subspaces, we construct a nonlocal operators $W$
\begin{eqnarray}\centering
W&=&\exp\left[i\frac{\pi}{2}(|\Gamma_{+}\rangle\langle\Gamma_{+}|-|\Gamma_{-}\rangle\langle\Gamma_{-}|-1)\right]~.
\label{wilsonloop}
\end{eqnarray}
$W$ commutes with the $SU(2)\times SU(2)$ pseudospin rotational invariant Hamiltonian $H^{*}_{2}$ in the projected subspace of the states $|\Gamma_{+}\rangle$ and $|\Gamma_{-}\rangle$. The two degenerate ground states are adiabatically connected via a twist operator/ nonlocal gauge transformation $\hat{O}|\Gamma_{\pm}\rangle=|\Gamma_{\mp}\rangle$. As these two states are protected by a many body gap, adiabatic passage between these degenerate ground states involve the creation of charge-$1/2$ excitations~\cite{wen1990,oshikawa2006,chen2010local}; this is seen from the anticommutation relation $\lbrace O,W \rbrace=1$. Similar twist-translation relation operator relations have also been  found recently by some of us for quantum liquid ground states in frustrated quantum antiferromagnets~\cite{pal2020topological,pal2019magnetization}. In the next section, we introduce the entanglement renormalization scheme using the unitary transformations eq.\ref{UnitaryOpDef}. This will reveal nontrivial entanglement features of gapped topological order, as well as distinguishing it from a gapless state of matter.
\section{MERG construction for the topologically ordered insulating ground state of the 2D Hubbard model}\label{MERG}
\pin
The ground state wavefunction of a parent model such as the 2D Hubbard model is in general difficult to obtain. However, once available, it allows for the computation of various correlation functions for, e.g., characterizing the low energy features of the system as being either gapless or gapped. We will now demonstrate how this task can be carried out for the 2D Hubbard model using the unitary  renormalization group (URG) method. We have already seen above that the URG approach can be carried out in a non-perturbative manner, helping obtain stable fixed point theories at which the RG flows terminate. Further, the stable fixed point Hamiltonians (eq.\ref{fixed_point_Ham_mott}) are comparatively easier to solve either analytically or numerically than their parent model counterparts~\cite{anirbanmott2}.
\par
\begin{figure}
\centering
\includegraphics[width=.75\textwidth]{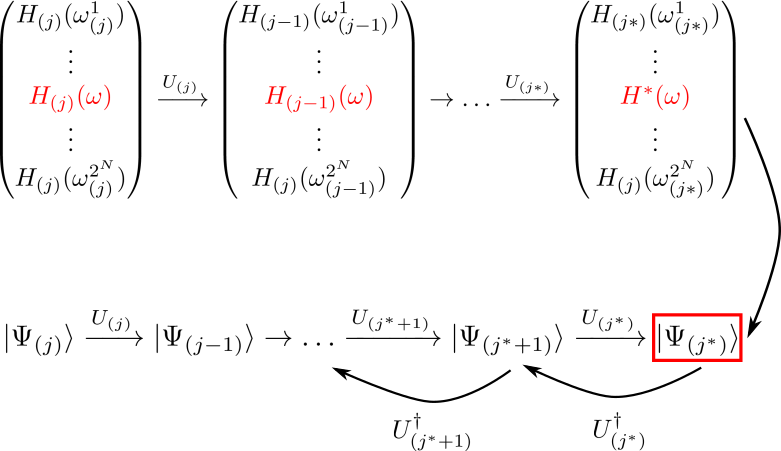}
\caption{\label{URGflowScheme} Upper row: URG flow scheme for Hamiltonians, terminating at fixed point Hamiltonians, e.g. $H^{*}(\omega)$, characterised by fluctuation energy scale $\omega$. Lower line: entanglement renormalization RG flow constructed via the inverse unitary transformations $U^{\dagger}_{(j^{*}+n)}s$ on the eigenstate $|\Psi^{*}\rangle$ (red bordered box) of $H^{*}(\omega)$.} 
\end{figure}

\noindent
As shown in Fig.\ref{URGflowScheme}, the precise program we follow is as follows. We begin by implementing the URG \cite{anirbanmotti} by applying a sequence of unitary transformations to the parent Hamiltonian $H$. As demonstrated above, successive unitary transformations lead to a fixed point Hamiltonian $H^{*}(\omega)$ and its ground eigenstate $|\Psi^{*}\rangle$ at a given quantum fluctuation scale $\omega$. Now, by reversing the unitary transformation, we can reconstruct the (a priori unknown) ground eigenstate $|\Psi\rangle$ of the parent model.  
\par
\begin{figure}
\centering
\includegraphics[width=0.5\textwidth]{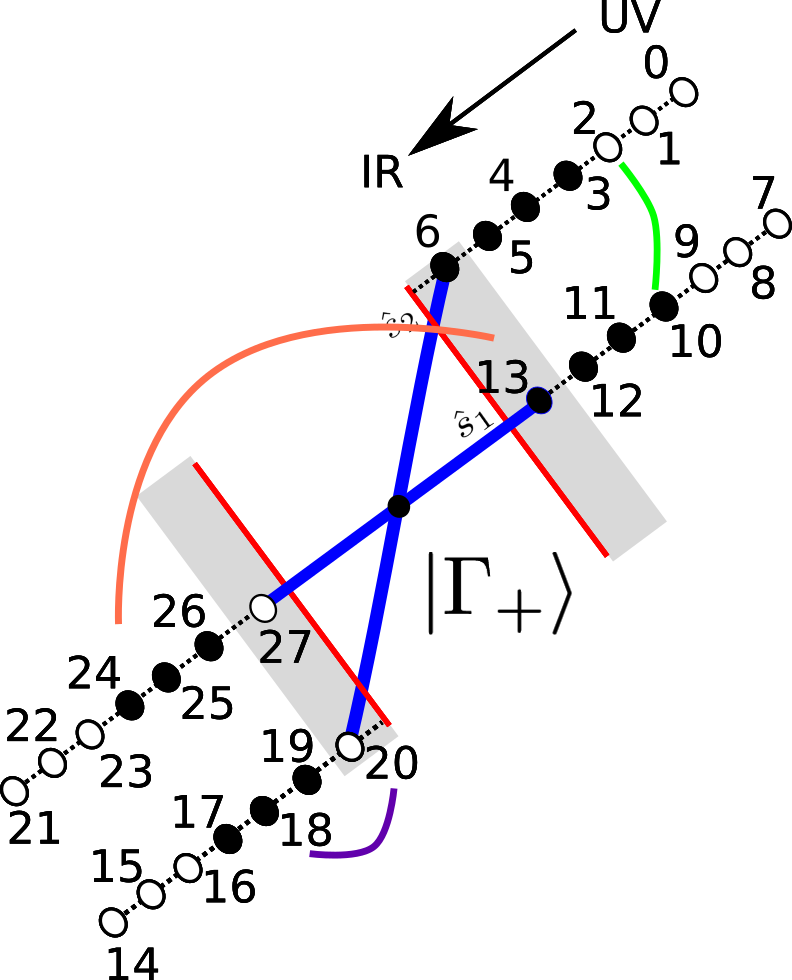}
\caption{\label{GSrep} Representation of the ground state in the MERG construction. The blue lines indicate the singlets formed out of pairs of pseudospins $(6,20)$ and $(13,27)$. Dark/white circles represent up/down pseudospins. The arrow labelled UV to IR depicts the high to low variation of the pseudospin dispersion. The orange, green and purple arrows represents the backscattering, tangential scattering, forward scattering processes respectively.}
\end{figure}

\noindent
In this way, we realize URG as a momentum space entanglement renormalization group (MERG) procedure carried out for the 2D Hubbard model in Refs.\cite{anirbanmotti,anirbanmott2}. In those works, we computed various vertex RG equations for the 2D Hubbard model from a URG analysis. By numerically solving the URG flow equations on a momentum-space grid of dimension $2048\times 2048$, we obtained the effective Hamiltonian and the ground state at the Mott liquid stable fixed point. In order to perform a quantitative benchmarking of the URG method, we also carried out a finite size scaling of the ground state energies computed for momentum- space grid sizes ranging from $1024\times 1024$ to $32768\times 32768$ and achieved excellent agreement with the answers obtained from various numerical methods~\cite{leblanc2015solutions}. URG computations involving eigenfunctions (rather than Hamiltonians) face, however, a considerable challenge: the number of configurations of a many-body system of interacting qubits grows exponentially with an increase in system size (i.e., the number of qubits). Therefore, in the present work, we consider a simplified construction of the effective theory obtained in Ref.\cite{anirbanmotti,anirbanmott2}. 
\pin
We begin by solving the Hamiltonian RG equations numerically on a $2048 \times 2048$ momentum-space grid. As shown in Fig.\ref{GSrep}, we then consider an effective problem with a simplified $k$-space of only four outward directions normal to the FS, $\hat{s}_{1}=(\pi/4,3\pi/4)$, $\hat{s}_{2}=(\pi/2,\pi/2)$, $\hat{s}_{3}=(-\pi/4,-3\pi/4)$, $\hat{s}_{4}=(-\pi/2,-\pi/2)$. $14$ electrons reside at low-energies along each normal $\hat{s}$, and are paired up to form $7$ pseudospins ($\mathbf{A}$ in eq.\ref{pseudospins}). Altogether, this leads to a system of 28 pseudospins (i.e., 56 electrons). The pseudospin states are labelled as follows (see Fig.\ref{GSrep}): the states labelled $0,...,6$ are along $\hat{s}_{1}$, $7,...,13$ are along $\hat{s}_{2}$, $14,...,20$ are along $\hat{s}_{3}$ and $21,...,28$ are along $\hat{s}_{4}$. Next, we prepare the system in the eigenstate $|\Gamma_{+}\rangle$ of the Hamiltonian $H^{*}_{2}$. We consider here a simple construction of $|\Gamma_{+}\rangle$ (see Fig.\ref{GSrep}):~two pair of singlets formed out of pair of backscattering pseudospins $(6,20)$ (involving $\hat{s}_{1}$ and $\hat{s}_{3}$) and $(13,27)$ (involving $\hat{s}_{2}$ and $\hat{s}_{4}$). The black/white circles in Fig.\ref{GSrep} represent up/down configurations of the disentangled pseudospins. Altogether, $|\Gamma_{+}\rangle$ is represented as ($A^{z}_{i}|1_{i}\rangle =1/2|1_{i}\rangle$, $A^{z}_{i}|0_{i}\rangle =-1/2|0_{i}\rangle$), 
\begin{eqnarray}
|\Gamma_{+}\rangle = \frac{1}{2}\prod_{n=[0,3]}\prod_{i=[7n,7n+2],j=[7n+3,7n+5]}|0_{i}1_{j}\rangle(|1_{6}0_{20}\rangle-|0_{6}1_{20}\rangle)(|1_{13}0_{27}\rangle-|0_{13}1_{27}\rangle)~,\label{many-body-state}
\end{eqnarray}
and where we have $N=6$ pseudospins along each of the four $\hat{s}_{i}$ directions normal to the FS.The two singlets that comprise the initially entangled subspace can be seen from the second and third terms above, while the product of $|0_{i}1_{j}\rangle$ states correspond to the disentangled pseudospin states. Similarly in terms of the 28 spin type pseudospin qubits $\mathbf{S}_{\Lambda,\hat{s}}$ the ground state is given by ($S^{z}_{i}|\Uparrow_{i}\rangle =1/2|\Uparrow_{i}\rangle$, $S^{z}_{i}|\Downarrow_{i}\rangle =-1/2|\Downarrow_{i}\rangle$),
\begin{eqnarray}
|\Gamma_{-}\rangle = \frac{1}{2}\prod_{n=[0,3]}\prod_{\substack{i=[7n,7n+2], j=[7n+3,7n+5] }}|\Downarrow_{i}\Uparrow_{j}\rangle(|\Uparrow_{6}\Downarrow_{20}\rangle +|\Downarrow_{6}\Uparrow_{20}\rangle)(|\Uparrow_{13}\Downarrow_{27}\rangle+|\Downarrow_{13}\Uparrow_{27}\rangle)~,~~~~\label{many-body-state2}
\end{eqnarray}
Note that $\mathbf{A}$ pseudospin operators annhilate the eigenstates of $\mathbf{S}$, and vice versa~\cite{anirbanmotti}.
\par\noindent
We now reintroduce the dominant quantum fluctuations in the state $|\Gamma_{+}\rangle$ in the form of tangential scattering, forward scattering and backscattering processes via the successive application of the reverse unitary maps $U^{\dagger}_{(j*+n)}$ (see Fig.\ref{URGflowScheme}). Importantly, we note that the fluctuations are being added only within the projected subspace of charge pseudospins. We note that for this specific case, the unitary operation $U_{(j)}$ disentangles four pseudospins labelled $n(N+1)+N-j$ (where $n=0,1,2,3$ is index for the four normal directions, and there are $N=6$ pseudospins along each normal direction)
\begin{eqnarray}\centering
 U_{(j)}&=&U_{4N-j+3}U_{3N-j+2}U_{2N-j+1}U_{N-j}~.~~~\label{unitary_op}
 \end{eqnarray}
In the above, $U_{n(N+1)+N-j}$ disentangles the pseudospin along the normal $\hat{s}_{n+1}$ at distance $\Lambda_{n(N+1)+N-j}$ from the reference non interacting FS, and is defined as follows
\begin{eqnarray}\centering
U_{n(N+1)+N-j}=\frac{1}{\sqrt{2}}\left[1+A^{-}_{n(N+1)+N-j}\frac{1}{\hat{\omega}-H^{D}_{1(j)}}C_{j}B^{+}_{j}-A^{+}_{n(N+1)+N-j}\frac{1}{\hat{\omega}-H^{D}_{0(j)}}C_{j}B^{-}_{j}\right]~,~~~\label{U_ren}
\end{eqnarray}
where $\mathbf{B}_{j}=\left[ \mathbf{A}_{0j} ~ \mathbf{A}_{1j} ~ \mathbf{A}_{2j} ~ \mathbf{A}_{3j}\right]^{T}$, $\mathbf{A}_{nj}=\sum_{l=1}^{N+1}\mathbf{A}_{nl+N-j}$ and the coupling matrix
\begin{eqnarray}\centering
C_{j}=\left(\begin{array}{cccc}
V^{(j)}_{1} & L^{(j)} & K^{(j)}_{1} & 0\\
L^{(j)}  & V^{(j)}_{2} & 0 & K^{(j)}_{2}\\
K^{(j)}_{1} & 0 & V^{(j)}_{1} & L^{(j)}\\
0 & K^{(j)}_{2} & L^{(j)} & V^{(j)}_{2}
 \end{array}\right)~.
\end{eqnarray}
$H^{D}_{1(j)}$ and $H^{D}_{0(j)}$ are the diagonal parts of the Hamiltonian in the projected subspaces of $\uparrow$/$\downarrow$ configurations of pseudospin $j$, e.g., 
\begin{eqnarray}\centering
H^{D}_{1(j)}=\sum_{l=1,i=[1,4]}^{j}\epsilon_{\Lambda_{l}\hat{s}_{i}}A^{z}_{\Lambda_{l}\hat{s}_{i}}+B_{j}^{z}CB_{j}^{z}
\end{eqnarray} 
In the coupling matrix, $V_{(j)}$, $L_{(j)}$ and $K_{(j)}$ are forward (purple line), backward (orange line) and tangential (green line) scattering couplings respectively displayed in Fig.\ref{GSrep}. Finally, the MERG flow equation is given by
\begin{eqnarray}\centering
H_{(j)}=U^{\dagger}_{(j)}H_{(j-1)}U_{(j)}~,\label{ren_H}
\end{eqnarray}
where the renormalized Hamiltonian 
\begin{eqnarray}\centering
H_{(j)}=\mathbf{B}_{j}C_{j}\mathbf{B}_{j}^{T}~.\label{pseudospin_rep_H}
\end{eqnarray}
From the MERG equation (eq.\ref{ren_H}) and $U_{(j)}$ (eq.\ref{U_ren})(l=1,2), we obtain the coupling RG equations accounting for the quantum fluctuations arising from the scattering of a pseudospin pair
\begin{eqnarray}\centering
V_{l}^{(j-1)}&=&V_{l}^{(j)}+\frac{(V_{l}^{(j)})^{2}}{\frac{1}{2}(\epsilon_{\Lambda_{j}\hat{s}}+\epsilon_{\Lambda_{j} -\hat{s}})-\omega-\frac{1}{4}V_{l}^{(j)}}~,\nonumber\\
K_{l}^{(j-1)}&=&K_{l}^{(j)}+\frac{(K_{l}^{(j)})^{2}}{\omega - \frac{1}{2}(\epsilon_{\Lambda_{j}\hat{s}}+\epsilon_{\Lambda_{j} -\hat{s}})-\frac{1}{4}K_{l}^{(j)}}~,\nonumber\\
L^{(j-1)}&=&L^{(j)}+\frac{N_{F}^{2}(L^{(j)})^{2}}{ \frac{1}{2N'}\sum_{s}(\epsilon_{\Lambda_{j-1}\hat{s}}+\epsilon_{\Lambda_{j} -\hat{s}})-\omega-\frac{1}{4}L^{(j)}}~.\label{RGeqns}
\end{eqnarray}
In the above, $l=1,2$ represents the two normal directions $\hat{s}_{1}$ and $\hat{s}_{2}$.
Additionally, note that we choose $W/2-\omega=0$ in implementing the MERG procedure, corresponding to the lowest quantum fluctuation (QF) scale lying well within the Mott insulating phase shown in the phase diagram Fig.\ref{phDiag}. For the QF energy $W/2-\omega=0$, the $V$ and $L$ vertices are RG irrelevant (eq.\ref{RGeqns}), while the $K$ backscattering vertices are RG relevant leading to the effective Hamiltonian $H_{2}^{*}$ (eq.\ref{fixed_point_Ham_mott}).
\par\noindent
In the next subsection, we will represent the fixed point ground state $\Gamma_{+}$, as well as the sequence of states $|\Psi_{(j)}\rangle's$ connected to it via $U^{\dagger}_{(j)}$'s, as many-body quantum circuits. As quantum circuits are very generally known to be examples of tensor networks \cite{biamonte2017,gao2017efficient}, this will help us realize a momentum space tensor network renormalization scheme for the MERG procedure.  
\subsection{Quantum circuit network representation of the T.O. ground state and its renormalization}
\begin{figure}
\centering
\includegraphics[width=0.6\textwidth]{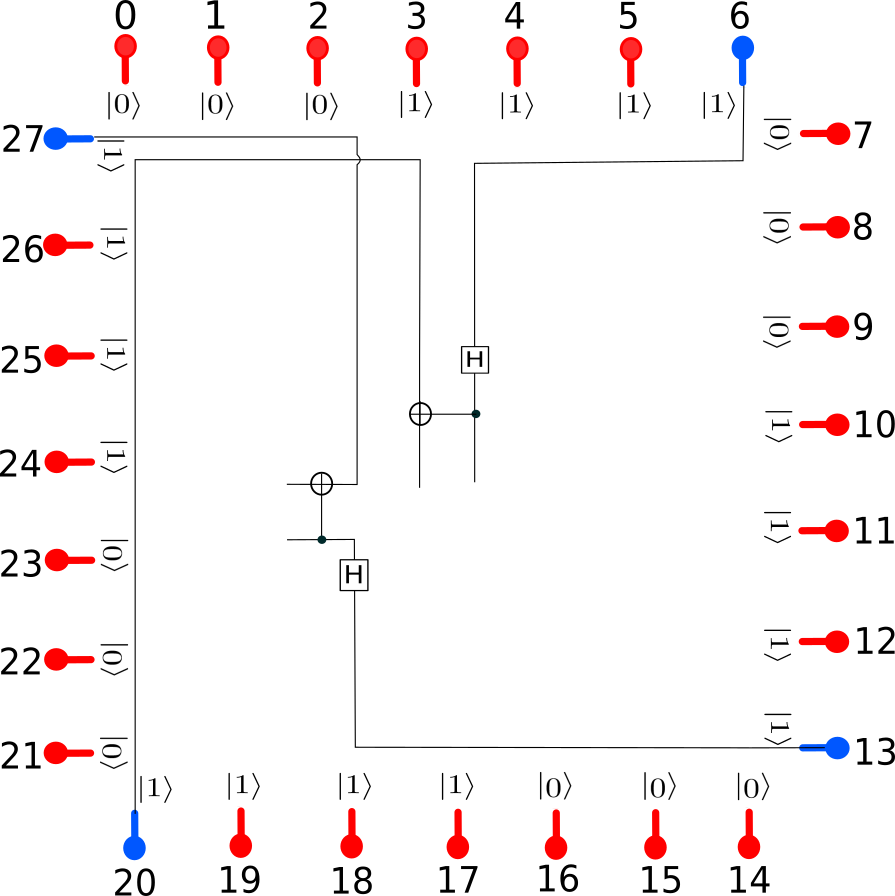}
\caption{\label{FPGSTN}Quantum circuit representation of $|\Gamma_{+}\rangle$ in terms of one- and two-pseudospin gates.}
\end{figure}

\pin
As seen above, the many body ground state $|\Gamma_{+}\rangle$ at the RG fixed point has a simple entanglement structure involving only two maximally entangled pairs (Fig.\ref{URGflowScheme}). As can be seen in Fig.\ref{FPGSTN}, this allows a easily computable quantum circuit description involving two controlled not (C-NOT) gates acting between pseudospins (6,20) and (13,27), such that the first pseudospin in each bracket (6 and 13) is a target bit and the second (20 and 27) a control bit. The Hadamard gate $H$ acts on the control bit, rotating its state $|1\rangle$ to $\sqrt{2^{-1}}(|0\rangle-|1\rangle)$. This is followed by a C-NOT gate acting on 20 and 27, leading to the singlet states (blue qubits in Fig.\ref{URGflowScheme}) in eq.\ref{many-body-state}. The disentangled states (red qubits in Fig.\ref{URGflowScheme}) have either up or down spin configurations, as represented in eq.\ref{many-body-state}.  
\begin{figure}
\centering
\includegraphics[width=0.6\textwidth]{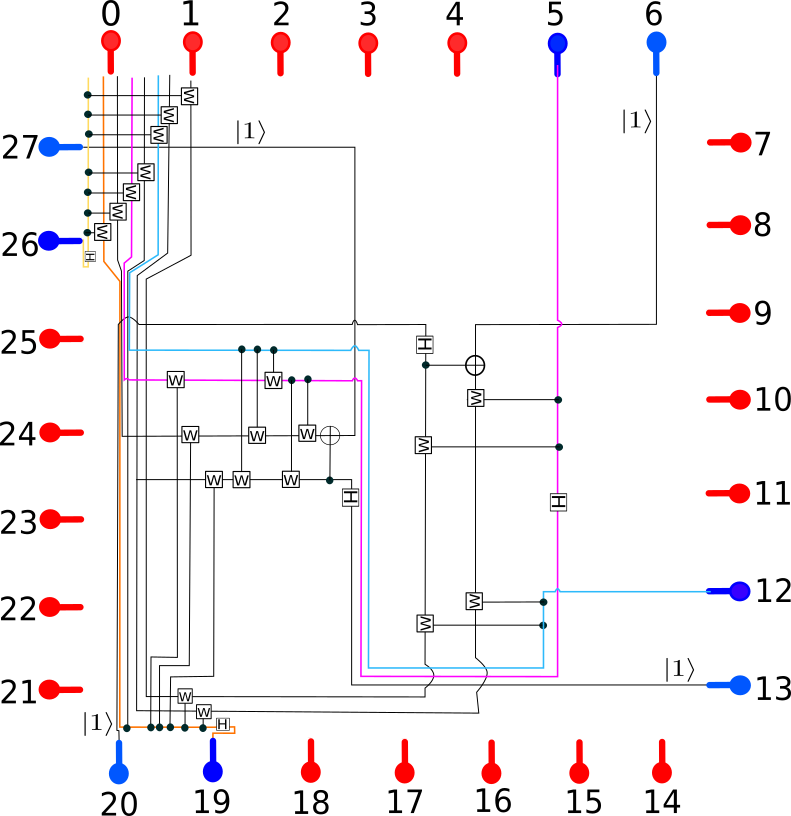}
\caption{\label{1RGTN}Quantum circuit representation of the many-body state after 1 reverse RG step in terms of one- and two-pseudospin gates. The purple, cyan, orange and yellow lines represent the entangling pathways of the qubits/pseudospins indexed 5, 12, 19 and 26.}
\end{figure} 
\par\noindent
The quantum circuit description for the states generated via reverse RG is (Fig.\ref{URGflowScheme})
\begin{eqnarray}\centering
U^{\dagger}_{(j)}|\Psi_{(j-1)}\rangle=|\Psi_{(j)}\rangle~,\label{reverse-URG-State}
\end{eqnarray}
and can be obtained via a decomposition of $U_{(j)}$ eq.\ref{unitary_op} as a product of one/two qubit gates\cite{divincenzo1995two}
\begin{eqnarray}\centering
U_{(j)}&=&U_{4N-j+3}U_{3N-j+2}U_{2N-j+1}U_{N-j}\nonumber\\
U_{\alpha}&=&U_{\alpha,N}\otimes U_{\alpha,N-1}\otimes\ldots\otimes U_{\alpha,N-j-1} \otimes U_{\alpha,2N+1}\otimes U_{\alpha,2N}\otimes\ldots\otimes U_{\alpha,2N-j}\otimes U_{\alpha,3N+2}\otimes\nonumber\\
&&\otimes U_{\alpha,3N+1}\otimes\ldots\otimes U_{\alpha,3N-j+1}\otimes U_{\alpha,4N+3}\otimes U_{\alpha,4N+2}\otimes\ldots\otimes U_{\alpha,4N+2-j}H_{\alpha}~,~~~~~~~~~\label{decompU}
\end{eqnarray}
where $\alpha=n(N+1)+N-j$, n=$0,1,2,3$ and $N=6$. The above decomposition of the unitary circuit into two local CNOT gates, Hadamard gates and phase gates implies that URG is a version of a Clifford stabiliser code~\cite{gottesman1998heisenberg} that obeys the Gottesman-Knill theorem, i.e., such a quantum circuit can be simulated on a classical computer in polynomial time.
\par
\begin{figure}
\centering
\includegraphics[width=0.75\textwidth]{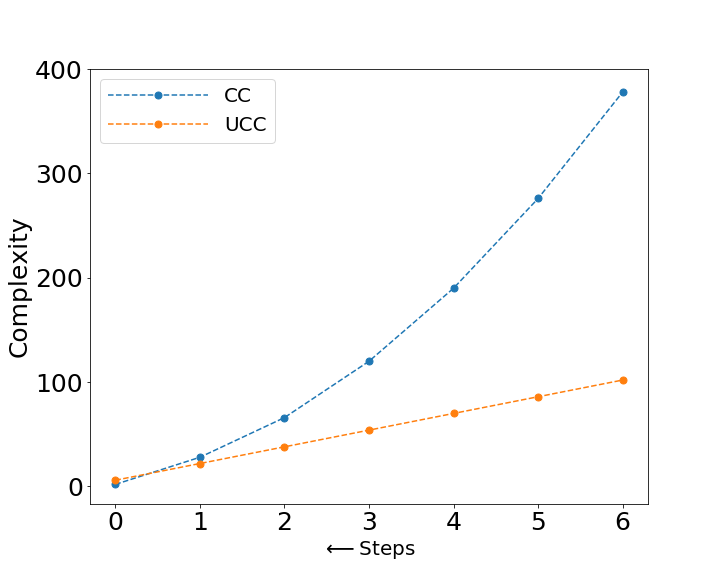} 
\caption{\label{fig:10}Orange curve $CC$ represents circuit complexity of the many-body state, blue curve $UCC$ represents complexity of the unitary transformation $U$ (or equivalently $D$ depth of $U$) along the RG flow trajectory.}
\end{figure}

\noindent
The unitary transformation $U_{N-j}$ disentangles qubit labelled $N-j$ (see Fig.\ref{GSrep} for the state labels), with similar definitions for the other unitary transformations. $H_{\alpha}$ represents the Hadamard gate. The individual two qubit gates $U_{n(N+1)+N-j,m(N+1)+N-l}$ can be represented via a controlled-$U$ rotation with $\alpha=n(N+1)+N-j$ as the control bit and $\beta=m(N+1)+N-l$ as the target bit
\begin{eqnarray}\centering
U_{\alpha,\beta} &=& \frac{1}{\sqrt{2}}\left[|1_{\alpha}\rangle\langle 1_{\alpha}|\otimes W_{\beta}+|0_{\alpha}\rangle\langle 0_{\alpha}|\otimes I_{2}\right]~.
\end{eqnarray}
The form of the single qubit rotation operations, $W_{\beta}=\exp\left(\frac{i}{2}\boldsymbol{\sigma_{\beta}}\cdot\hat{\mathbf{n}}\right)$, can be obtained by solving the reverse RG eq.\ref{reverse-URG-State} using the decomposition of the unitary operator eq.\ref{decompU} and the following representation of the $|\Psi_{(j)}\rangle$
\begin{eqnarray}\centering
|\Psi_{(j)}\rangle = a|1_{n(N+1)+N-j}\rangle|\Phi\rangle +b|0_{n(N+1)+N-j}\rangle|\chi\rangle~,
\end{eqnarray} 
where 1 and 0 represent the configurations of the pseudospin labelled $n(N+1)+N-j$. The states $|\Phi\rangle$ and $|\chi\rangle$ represent the configurations of the rest of the pseudospins. The quantum circuit representation for the state $|\Psi_{(j^{*}+1)}\rangle$ is shown in Fig.\ref{1RGTN}. The number of one/two-local unitary gates needed to obtain the quantum circuit description of the state $|\Psi_{(j^{*}+n)}\rangle$ quantifies the circuit complexity (CC)~\cite{brown2018second}. The CC for the circuit designs of $|\Psi_{(j)}\rangle$ at every RG step is found to be
\begin{eqnarray}\centering
CC(j)=8j^{2}-2j-4~.
\end{eqnarray}
In Figures \ref{FPGSTN} and \ref{1RGTN}, the quantum circuit description for the states $|\Psi_{(1)}\rangle$ and $|\Psi_{(2)}\rangle$ are thus found to possess the circuit complexity $CC(1)=2$ and $CC(2)=22$ respectively. By using the many body states generated via the reverse RG steps discussed earlier, the variation of the circuit complexity along the RG flow is numerically verified by the blue curve in Fig.\ref{fig:10}. The orange curve in Fig.\ref{fig:10} represents the decrement in complexity of the unitary disentangling operation at every RG step.
With this construction, we have established MERG as a quantum circuit/tensor network renormalization group scheme. In the next subsection, we validate the RG formalism by displaying its capability in reconstructing the ground state of the parent model from the effective model.  
\subsection{Eigenstate reconstruction within the MERG scheme}
\begin{figure}
\centering
\includegraphics[width=0.7\textwidth]{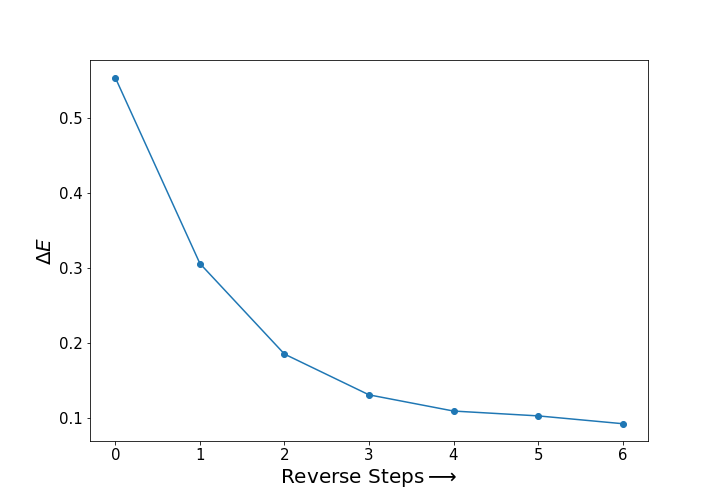}
\caption{\label{stateCC_RG}Plot displaying the reduction in energy uncertainty $\Delta E$ (with onsite repulsion $U_{0}=8$) upon reversing the RG steps.} 
\end{figure}

\pin
MERG, being a unitary map involving inverse unitary transformations on the eigenstates of the effective model, allows us to reconstruct the eigenstates of the parent Hubbard model. In this work, we account for only the dominant fluctuations in the charge pseudospin subspace of the operators. Deviation of the various pseudospin scattering couplings $K,V,$ and $L$ from a uniform magnitude of $U_{0}$ of the parent model can be clearly seen in eq.\ref{pseudospin_rep_H}. Upon reversing the RG steps, this deviation should decrease. We quantify this decrease through the reduction in the pseudospin interaction energy fluctuations in the entangled subspace $\Delta E_{(j)} = \langle\Psi_{(j)}|(\Delta H_{j})^{2}|\Psi_{(j)}\rangle$ (see Fig.\ref{stateCC_RG}). After six reverse RG steps, the fluctuations have reduced to $10\%$ of its initial value (see Fig.\ref{stateCC_RG}), and we expect a further fall in the value of the fluctuations with increasing system size. This validates the eigenstate reconstruction within the MERG scheme. 
\subsection{Entanglement holographic mapping (EHM) representation for the topologically ordered phase}
\begin{figure}
\centering
\includegraphics[width=0.6\textwidth]{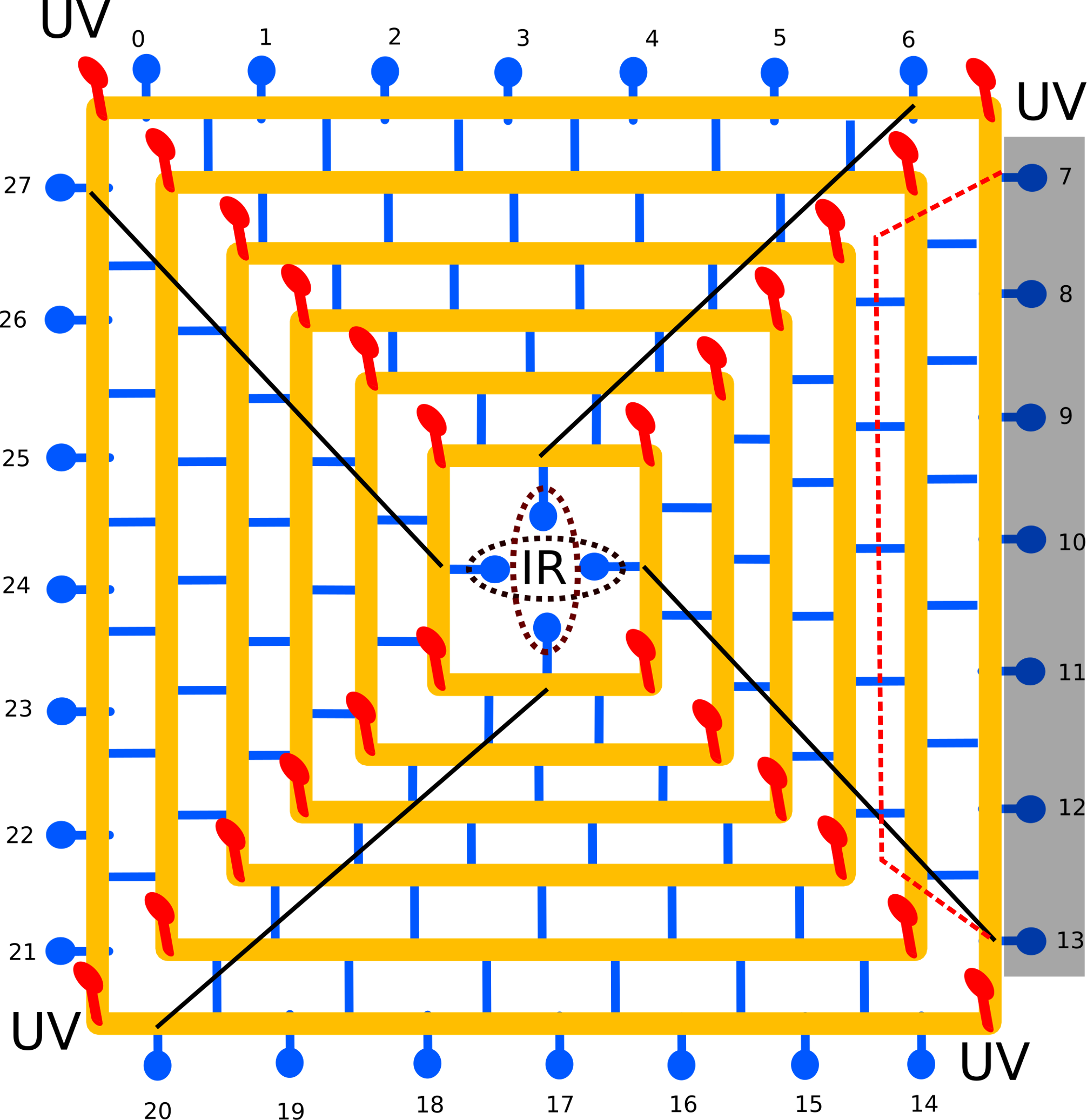} 
\caption{\label{EHM}Entanglement holographic mapping (EHM) representation of MERG. The blue legs represent the holographic boundary made of the physical pseudospin qubits. The yellow block represents a nonlocal unitary disentangler which iteratively maps the boundary (blue) qubits to the bulk (red) qubits. The black lines represent the passage of the qubits $6,13,20,27$ from UV to IR. The grey box represents the boundary region $7-13$ in the UV that is isolated within the bulk of the EHM by the the minimal surface/causal cone (dashed red line) after the second RG step.}
\end{figure}

\pin
In the earlier section, we have established the URG procedure as a tensor network RG by providing quantum circuit description of many-body states \cite{biamonte2017}. In Fig.\ref{EHM}, we describe the architecture of the tensor network RG through an equivalent entanglement holographic mapping (EHM) representation~\cite{qi2013,lee2016}. Similar to the case of spectrum bifurcation renormalization group (SBRG)~\cite{you2016entanglement}, unitary maps preserves the canonical fermion anticommutation relations. 
\par\noindent
At each RG step shown in Fig.\ref{EHM}, the nonlocal unitary operator (yellow block in Fig.\ref{EHM}) disentangles four pseudospins in the {\it holographic boundary} at high energies (UV), mapping them onto the red {\it emergent bulk physical qubits} at lower energies (i.e., towards IR). The first layer of unitary transformation disentangles the pseudospins located farthest from the FS (labelled $0, 7, 14, 21$), the next step pseudospins distentangled the pseudospins $1, 8, 15, 22$ and so on, eventually scaling towards the FS. Alongside these disentanglement RG steps, the pseudospins belonging to the entangled space undergo entanglement resharing. Deep in the IR regime, the system enters into an Mott liquid phase characterized by pseudospins pairing up as singlets (the dotted ovals pairing pseudospins (6,20) and (13,27) in Fig.\ref{EHM}).
\par\noindent
Similar to our earlier discussions, an important feature of the nonlocal unitary operation $U_{(j)}$ (yellow block in Fig.\ref{EHM}) is that it can be decomposed as a product of local 2-qubit disentanglers. Below we present the decomposition for the operator $U_{N-j}$
\begin{eqnarray}\centering
U_{(j)}&=&U_{4N-j+3}U_{3N-j+2}U_{2N-j+1}U_{N-j}\nonumber\\
U_{N-j}&=&U_{N-j,N}\ldots U_{N-j,N-j+1}U_{N-j,2N+1}\ldots U_{N-j,2N-j+1}U_{N-j,3N+2}\ldots U_{N-j,3N-j+2}\nonumber\\
&\times &U_{N-j,4N+3}\ldots U_{N-j,4N-j+3}~.
\label{U_decomp_disentangle}
\end{eqnarray}
As an outcome of the complete transformation $U_{(j)}$, the four pseudospins labelled $N-j,2N-j+1,3N-j+2,4N-j+3$ are disentangled, and the \emph{circuit complexity} of the unitary transformation is given by $UCC_{j}=16j-10$. It is interesting to note that the circuit complexity for the tensor network representation of the MERG quantum circuit ($UCC_{j}\sim O(j)$) is substantially less than that of the tensor network representation of the quantum state itself ($CC\sim O(j^{2})$).
\begin{figure}
\centering
\includegraphics[width=\textwidth]{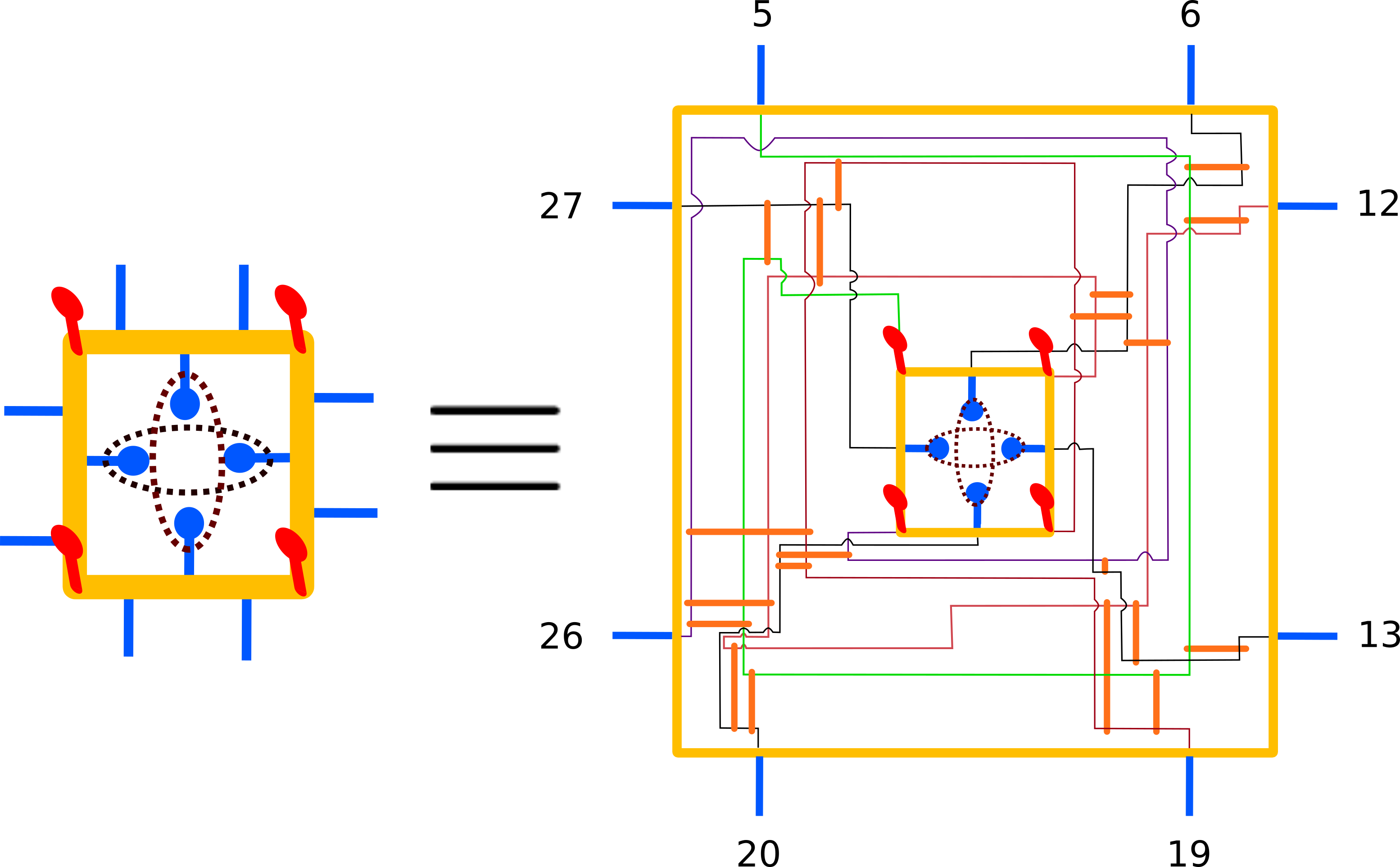} 
\caption{\label{fig:9}Decomposition of nonlocal unitary disentangler (yellow block) into arrangement of local two-qubit/pseudospin disentanglers (orange blocks). The green, pink, brown and violet lines represents the Hilbert space for pseudospins labelled 5, 12, 19 and 26 respectively. The orange block operates on the lines crossing its edges.}
\end{figure}
\par\noindent
We now present an algorithm for obtaining the form of the individual disentangler $U_{\alpha,\beta}$ (where $\alpha=N-j$,$\beta=N$).
For the RG step $j$, $|\Psi_{(j)}\rangle$ is the input state and $U_{(j)}|\Psi_{(j)}\rangle=|\Psi_{(j-1)}\rangle$ is the rotated output state. We first perform a decomposition of both the input and output states in terms of the up (1)/down (0) spin basis states labelled $\alpha$, $\beta$ and the subsystem configuration of the rest of the pseudospins
 \begin{eqnarray}\centering
 |\Psi_{(j)}\rangle &=& \sum_{m,n=[1,n_{max}]}D^{m,n,(j)}_{\alpha,\beta}|m\rangle_{\alpha,\beta}|\Phi_{(j)}^{n}\rangle~,\nonumber\\
 |\Psi_{(j-1)}\rangle &=& \sum_{m',n'=[1,n'_{max}]}D^{m',n',(j-1)}_{\alpha,\beta}|m'\rangle_{\alpha,\beta}|\Phi_{(j-1)}^{n'}\rangle~.\label{Psidecomp}
 \end{eqnarray}
Here, the states $|\Phi_{n}\rangle$ and $|m\rangle$ are orthonormalized, with $m$ belonging to the configuration set of two pseudospins $m\in \lbrace 00,01,10,11\rbrace$. Additionally, we note that in the state $|\Psi_{(j-1)}\rangle$, the configuration of pseudospin $\alpha$ is restricted to $0$ or $1$; this restricts $m'$ to one of the two subsets $m'\in\lbrace 00,01\rbrace$ and $m'\in\lbrace 10,11\rbrace$. 
\par\noindent 
We now construct an auxillary state $|\rho\rangle$ by extracting the two pseudospin configuration of $\alpha,\beta$ and the state $|\Psi_{(j-1)}\rangle$, and embedding it into a higher dimensional Hilbert space that constitutes all the subsystem states $|\Phi_{(j)}^{n}\rangle$. If $n'_{max}>n_{max}$, then the states $|\Phi_{(j)}^{n}\rangle$ are augmented by the states $|\chi_{n_{max}+1}\rangle$,$\ldots$,$|\chi_{n'_{max}}\rangle$ orthogonal to them. 
This leads to the following $|\rho\rangle$ for $n'_{max}\leq n_{max}$
 \begin{eqnarray}\centering
 |\rho\rangle &=& \sum_{m',n'}D^{m',n',(j-1)}_{\alpha,\beta}|m'\rangle_{\alpha,\beta}|\Phi_{(j)}^{n'}\rangle~.\label{aux_state_1}
 \end{eqnarray}
On the other hand, if $n'_{max}\geq n_{max}$,
 \begin{eqnarray}\centering
 |\rho\rangle &=& \sum_{m',n'=[1,n_{max}]}D^{m',n',(j-1)}_{\alpha,\beta}|m'\rangle_{\alpha,\beta}|\Phi_{(j)}^{n'}\rangle +\sum_{m',p=[n_{max}+1,n'_{max}]}D^{m',p,(j-1)}_{\alpha,\beta}|m'\rangle_{\alpha,\beta}|\chi_{p}\rangle~.~~~~~~~~~\label{aux_state_2}
 \end{eqnarray}
We now need to find an unitary transformation $U_{\alpha,\beta}$ that maps $|\Psi_{(j)}\rangle$ to $|\rho\rangle$. This can be constructed by first performing Gram-Schmidt orthonormalization to find a state $|\rho_{\perp}\rangle$ that is orthogonal to $|\Psi_{(j)}\rangle$
\begin{eqnarray}\centering
|\rho^{\perp}\rangle =\mathcal{N}(|\rho\rangle -|\Psi_{(j)}\rangle\langle\Psi_{(j)}|\rho\rangle)~,
\end{eqnarray}
where $\mathcal{N}$ is the normalisation factor. The inner product $\langle\Psi_{(j)}|\rho\rangle$ can then be determined from eq.\ref{aux_state_1} and eq.\ref{aux_state_2}, allowing us to determine the state $|\rho^{\perp}\rangle$ exactly. From here, we can construct the unitary operation
\begin{eqnarray}\centering
U_{\alpha,\beta}=\exp\left(\tau\left(|\rho_{\perp}\rangle\langle\Psi_{(j)}|-|\Psi_{(j)}\rangle\langle\rho_{\perp}|\right)\right)~,
\end{eqnarray}
where the many-body rotation angle is $\tau=\cos^{-1}(\langle\Psi_{(j)}|\rho\rangle)$. 
\par\noindent
In order to find the next unitary operator in the decomposition, we recast $|\rho\rangle$ in the pseudospin basis of another set of states $\alpha', \beta'$ (as in eq.\ref{Psidecomp}). A similar representation for $|\Psi_{(j-1)}\rangle$ is found in the pseudospin basis of $\alpha',\beta'$, following which the steps outlined above are followed once again to obtain the disentangler $U_{\alpha',\beta'}$. In this way, the entire decomposition given by eq.\ref{U_decomp_disentangle} is obtained. A quantum circuit description for a particular nonlocal unitary operator $U_{(j)}$ (yellow block in Fig.\ref{EHM}) is given in Fig.\ref{fig:9}. The orange blocks on the right panel of Fig.\ref{fig:9} disentangles a pair of pseudospins. The existence of a quantum circuit design of the nonlocal disentangler $U_{(j)}$ as a combination of two local disentanglers justifies a holographic description for MERG~\cite{lee2016}.
\par\noindent
We close this section by providing a comparison between the MERG architecture Figs.\ref{EHM} and \ref{fig:9} and other tensor network methods, e.g., multiscale entanglement renormalization group ansatz (MERA)~\cite{evenbly2011} and deep MERA~\cite{kim2017noise}. The first important difference worth noting is that while each transformation layer in MERG is composed only of unitaries, in MERA and DMERA, every transformation layer is composed of a layer of two-local unitaries and a layer of isometries. Next, both MERA and DMERA involve variational parameters in the transformation layers. On the other hand, MERG does not have variational parameters: the unitary transformations are solely determined via the form of the Hamiltonian that is block diagonalized and the choice of quantum fluctuation energy scale $\omega$ (eq.\ref{UnitaryOpDef}). Finally, in MERG and DMERA, each unitary transformation layer 
is composed of many sublayers of two local unitary disentanglers gates; this is defined as the depth of the unitary gate $D$. For instance, in DMERA, the depth $D$ is equal to the number of variational parameters employed in the transformation layer. However, in MERG, the depth at each unitary RG step ($D_{j}$) is equal to the circuit complexity, i.e., $D_{j}=UCC_{j}=16j-10$ (Fig.\ref{fig:10}).    
\subsection{Emergence of T.O. ground state and quantum error correcting code}
\begin{figure}
\centering
\includegraphics[scale=0.4]{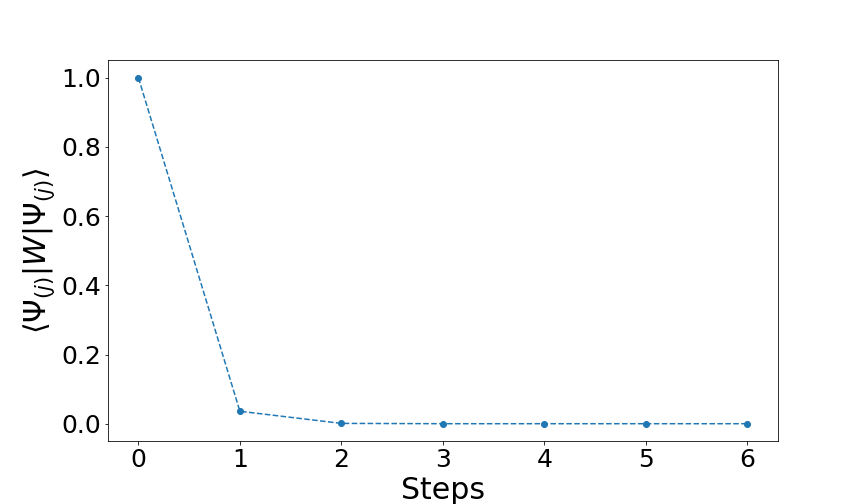}
\caption{\label{fig:11} Evolution of the expectation value of the Wilson loop along the RG trajectory.}
\end{figure}

\pin
We will now demonstrate that the topologically ordered ground state manifold of the Mott liquid states $\lbrace |\Gamma_{+}\rangle, |\Gamma_{-}\rangle\rbrace$ (eq.\ref{many-body-state}, eq.\ref{many-body-state2}) can be associated with a stabilizer group $G$ in the space of the four qubits $|6\rangle$, $|13\rangle$, $|20\rangle$, $|27\rangle$ that belong to the emergent space shown in Fig.\ref{GSrep}. That is, every element $g\in G$ will satisfy the property $g|\Gamma_{\pm}\rangle =|\Gamma_{\pm}\rangle$~ \cite{gottesman1997,nielsen2002quantum}. In the ground state $|\Gamma_{+}\rangle$, the qubit pairs ($|6\rangle$, $|20\rangle$) and ($|13\rangle$, $|27\rangle$) form a pair of $\mathbf{A}$ pseudospin singlets (eq.\ref{pseudospins}), and at the IR fixed point of the URG, these are in tensor product with the rest of the disentangled qubits. Similarly, in $|\Gamma_{-}\rangle$, the qubit pairs ($|6\rangle$, $|20\rangle$) and  ($|13\rangle$, $|27\rangle$) form a pair of $\mathbf{S}$ pseudospin triplets at the IR fixed point (which the decoupled qubits are in tensor product with). The elements of the stabilizer group $G$ can then be constructed out of the stabilizer groups $G_{1}$ and $G_{2}$ in the subspace of qubits ($|6\rangle$, $|20\rangle$) and ($|13\rangle$, $|27\rangle$) respectively
\begin{eqnarray}\centering
G_{1}&=&\left\lbrace -4A^{z}_{6}A^{z}_{20}-4S^{z}_{6}S^{z}_{20},-4A^{x}_{6}A^{x}_{20}+4S^{x}_{6}S^{x}_{20}\right\rbrace ,\nonumber\\
G_{2}&=&\left\lbrace -4A^{z}_{13}A^{z}_{27}-4S^{z}_{13}S^{z}_{27},-4A^{x}_{13}A^{x}_{27}+4S^{x}_{13}S^{x}_{27}\right\rbrace ,\nonumber\\
G&=&\left\lbrace  g_{1}\otimes g_{2}, g_{1}\in G_{1}, g_{2}\in G_{2}\right\rbrace ~.\label{stabilizer}
\end{eqnarray}
It can be easily seen that the eigenstates $|\Gamma_{\pm}\rangle$ (eqs.\ref{many-body-state} and \ref{many-body-state2}) are left invariant by the action of group elements of $G$. Furthermore, the elements $g\in G$ commute with the Wilson loop operator $W$ (eq.\ref{wilsonloop}): $[W,g]=0$. Thus, the eigenvalues of $W$ $(\pm 1)$ are good quantum numbers that label the eigenstates $| \Gamma_{\pm}\rangle$: 
$W| \Gamma_{\pm}\rangle = \pm |\Gamma_{\pm}\rangle$. In this way, we observe that the MERG program yields the topological Wilson loop operator $W$ as well as the stabilizer group $G$ (eqs.\ref{stabilizer}) associated with the topologically ordered manifold. Below, this will allow us to construct strategies for topological quantum error correction.
\pin
Fig.\ref{fig:11} shows that the Wilson loop expectation value $\langle\Psi_{(j)}|W|\Psi_{(j)}\rangle$ has minimal growth in the first few RG steps, followed by a sharp rise between the second last and final RG steps, ending at $\langle\Psi_{(j^{*})}|W|\Psi_{(j^{*})}\rangle=1$. This shows that in the state $|\Gamma_{+}\rangle$, the pseudospins form a pair of maximally entangled singlets that are distilled out from a (bare) state where the entanglement is initially spread among many members. For example, in the second last RG step, there are $CC(2)=22$ entangled bonds compared to just the two singlet bonds in $|\Gamma_{+}\rangle$. This explains the sharp distillation of topological order via the URG procedure.
\par\noindent 
In the language of topological quantum error correcting codes, the 
the topologically ordered states form the \emph{codewords}, and the
unitaries $U_{j}$ of the URG \emph{decode} them. On the other hand, the reverse unitaries $U^{\dagger}_{j}$'s of the MERG sequentially \emph{encode} the codeword $|\Gamma_{+}\rangle$ in a higher dimensional Hilbert space by re-entangling them with the decoupled degrees of freedom (Fig. \ref{URGflowScheme}). The identity $U_{j}U^{\dagger}_{j}=1$ describes the encoding-decoding program. An example state obtained by the encoding step: $U^{\dagger}_{j^{*}+1}|\Psi_{j^{*}}\rangle=|\Psi_{j^{*}+1}\rangle$ is represented in Fig.\ref{1RGTN}. 
The decoding strategy $U_{(j)}|\Psi_{(j)}\rangle=|\Psi_{(j-1)}\rangle$ (forward RG) leads to the \emph{stabilizer} qubits (disentangled pseudospins) which can maintain a non-trivial error syndrome in the presence of external noise in the encoded state $|\Psi_{(j)}\rangle$. In this way, MERG can generate an encoding-decoding program using the form for $U$ (eq.\ref{UnitaryOpDef}) for the case of a parametric noise that can be modelled into a Hamiltonian $H_{1}$ different from $H$ (eq.\ref{Hubbard Hamiltonian}). Thus, MERG is realized as a quantum error correction code that resolves the internal quantum fluctuations, leading to a topological codeword manifold~\cite{ferris2014}. Further, this platform paves the way for optimal error correction strategies in cases of more general environmental noise. In the next section, we will track the many-body entanglement features along the RG flow that leads to the topological ordered state. 
\section{Entanglement RG flow towards the topologically ordered ground state}\label{ERGflow}
\pin
The entanglement features~\cite{you2018machine} of a many-body state are formed by a collection of various quantifiers, e.g., mutual information, entanglement entropy, Renyi entropy, purity etc. for the Schmidt (entanglement) spectra obtained for all possible different biparititions. In the present context, the topologically ordered state~\cite{wen1990topological} is characterized by emergence of (real space) short ranged strongly entangled singlet pairs (similar to the short ranged resonating valence bonds (RVB)~\cite{kivelson1987topology}) in the low energy subspace. For example, in the Mott liquid ground state wavefunction eq.\ref{many-body-state}, the members of the pairs (6,20) and (13,27) (Fig.\ref{GSrep}) are maximally entangled and belong to different sides of the  $E_{F}=0$ energy boundary (FS); the large $k$-space separation ensures that these are short ranged pairs in real space. In this section, we investigate the RG crossover into a system of strongly entangled pairs $|\Gamma_{+}\rangle$ amidst scattering between all pseudospin pairs by tracking a subset of their \emph{entanglement features}.
\subsection{Definitions and Computation method}
\pin
 We first define the entanglement features that has been used for the analysis.
\par\noindent
1.~\emph{Mutual information}($MI$) between a pair of pseudospins  characterizes the \emph{strength of entanglement}~\cite{hyatt2017,lee2016} among the members $i$ and $j$
 \begin{eqnarray}\centering
 I(i:j)=-Tr(\rho_{i}\ln\rho_{i})-Tr(\rho_{j}\ln\rho_{j})+Tr(\rho_{ij}\ln\rho_{ij})~,\label{MI}
 \end{eqnarray}
where $\rho_{i}$,$\rho_{j}$ are the $1$-pseudospin reduced density matrices (RDM) and $\rho_{ij}$ is the $2$-pseudospin RDM. More precisely, if pseudospins $(i,j)$ are individually strongly entangled with the rest of the pseudospins, then $\rho_{ij}$ has a huge spread in its probability eigenvalues. This lowers the MI content among $(i,j)$ and characterizes a weakly entangled pair. On the other hand, if the pseudospins are strongly entangled as a pair, then $\rho_{ij}$ has a smaller spread and characterizes a strongly entangled pair. For example, the singlet pair $(i,j)=(6,20)$ in eq.\ref{many-body-state} has maximum possible mutual information $I(i:j)=2\ln 2$, and clearly $\rho_{ij}=1$ is a pure state with zero spread.
\par\noindent
2.~\emph{Information distance} is a distance measure between pseudospins computed as the negative logarithm of the $MI$
 \begin{eqnarray}\centering
 d(i,j)=-\ln\frac{I(i:j)}{2\ln 2}~.
 \end{eqnarray}
$d(i,j)$ is defined such that strongly entangled pairs have smaller information distance. The maximally entangled pairs $(6,20)$ in $|\Gamma_{+}\rangle$ has zero information distance $d(6,20)=0$, while the pair of disentangled pseudospins $(1,2)$ has $d(1,2)\to\infty$. In a many-body state, $|\Psi\rangle$ is the collection of all possible information distances $d(i,j)$, and describes an effective \emph{spacetime geometry}~\cite{lee2016}. This provides a link between the quantum circuit network descriptions of state $|\Psi_{(j)}\rangle$ and the entanglement spacetime geometries $d(i,j)'s$ obtained at each RG step. In this way, the strongest entangled members in the state $|\Psi_{(j)}\rangle$ characterizes the \emph{geodesic}-~$d_{g}=\min_{i,j}d(i,j)$ of the spacetime geometry.
\par\noindent
3.~ \emph{Purity} is defined as $Tr(\rho^{2})$, and characterizes the spread in the probability eigenvalues of the RDM. Purity is useful for classifying the amount of entanglement information lost in obtaining the RDM, i.e., the partial tracing over the rest of the entangled degrees of freedom. 
\par\noindent
4.~The \emph{Schmidt spectrum}/entanglement spectrum~\cite{laflorencie2016quantum} is a set of numbers $\{\lambda_{i}\}$ obtained from a Schmidt decomposition of $|\Psi\rangle$ across a bipartition, e.g., between 2 and $N-2$ pseudospins
\begin{eqnarray}\centering
|\Psi\rangle =\sum_{i=1}^{k}\lambda_{i}|\phi_{i}\rangle_{2}|\Phi_{i}\rangle_{N-2}~, \label{SchmidtSpec}
\end{eqnarray}  
where $|\phi_{i}\rangle$ are the mutually orthogonal two-pseudospin states and $|\Phi_{i}\rangle$ are the orthogonal many-body configurations of the $N-2$ pseudospins. For such a bipartition of $2$ and $N-2$ pseudospins, the quantity $k$ is given by $k<min(d_{A},d_{B})=4$, where $d_{A},d_{B}$ are the subsystem Hilbert space dimensions.
\par\noindent
5.~The \emph{Physical distance} $d_{p}(i,j)$ between pseudospin pairs in momentum space is obtained by first computing the $k$-space distance using the Euclidean metric
 \begin{equation}\centering\centering\centering
 d(\mathbf{k}_{i},\mathbf{k}_{j})=\sqrt{(k_{ix}-k_{jx})^{2}+(k_{iy}-k_{jy})}^{2}~,
 \end{equation}
and then inverting $d(\mathbf{k}_{i},\mathbf{k}_{j})$ to obtain the relation $d(i,j)=1/d(\mathbf{k}_{i},\mathbf{k}_{j})$.
\par\noindent
The computation method we apply is as follows. We first apply the MERG technique on the state $|\Psi^{*}\rangle$ to obtain all the states $|\Psi_{(j)}\rangle$ in the steps outlined in Fig.\ref{URGflowScheme}. This is followed by a computation of the 
one- particle and two-particle Schmidt spectra (as in eq.\ref{SchmidtSpec}) for all possible pairs of pseudospins. The various entanglement features listed above can then be computed numerically from the Schmidt spectra. 
\subsection{Results on emergence of strongly entangled short-distance pairs}\label{EmStrongShort}
\begin{figure}
\centering
\includegraphics[width=\textwidth]{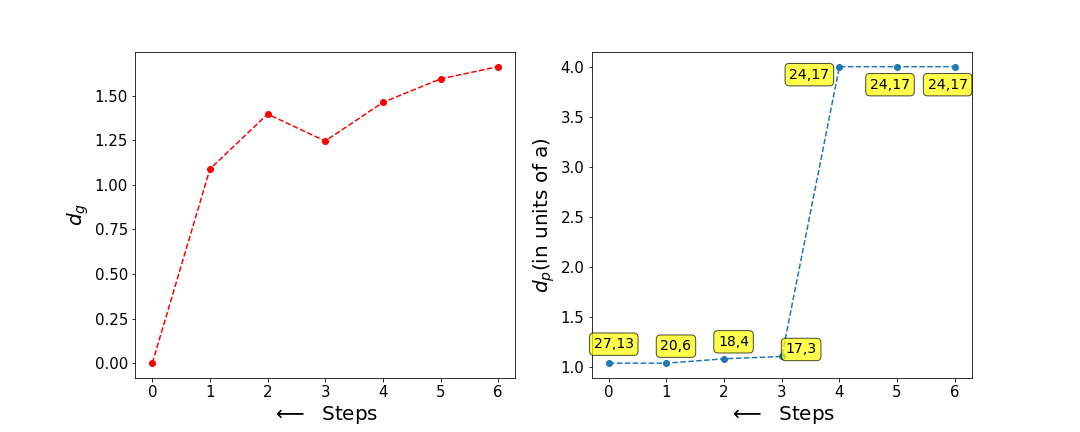} 
\caption{\label{fig:12}Left Panel: Scaling of the information geodesic $d_{g}$. Right Panel: Scaling of the corresponding physical distance $d_{p}$ between the strongest entangled pseudospin pairs.}
\end{figure}

\pin
In Fig.\ref{fig:12} (left panel), we present the decrement of the geodesic distance $d_{g}$ ($d_{g}\to 0$) under RG flow across the EHM network (Fig.\ref{EHM}). This signals the emergence of strongly entangled pairs with high $MI$ content $I_{max}(i:j)=2\ln 2$ and $d_{g}=0$) at the low energy stable fixed point of the RG. Complementary insight is obtained in the right-hand panel of Fig.\ref{fig:12}, which shows the concomitant reduction in real-space physical distance $d_{p}$ between the strongest entangled pseudospins. We have also labelled the pairs carrying the highest MI at each RG step. Initially, the pseudospin pair $(24,17)$ lying to one side of the FS (see Fig.\ref{GSrep}) and connected by tangential scattering vertex $L$, carry the highest $MI$ and comprise the geodesic $d_{g}=1.66$, while the $d_{p}$ between them is 4 lattice spacings. However, beyond the second RG step, the system undergoes a crossover into a phase comprised of high MI pseudospin pairs $(17,3)$, $(18,4)$ etc. connected via backscattering vertex ($K$ vertex). The pair $(17,3)$ is formed by states belonging to opposite side of the FS: the physical distance between these pairs has thus shrunk to $d_{p}=1$ lattice spacing. In this way, we observe the emergence of $d_{p}=1, d_{g}=0$ (short distance) strongly entangled pairs $(6,20)$, $(13,27)$ amidst competition among various entangled pairs connected via forward scattering ($V$ vertex), backscattering and tangential scattering. Additionally, we note that the emergence of such pairs is correlated with the quantization of the Wilson loop expectation value signalling the onset of T.O. (Fig.\ref{fig:11}).
 
\begin{figure}
\centering
\includegraphics[width=0.8\textwidth]{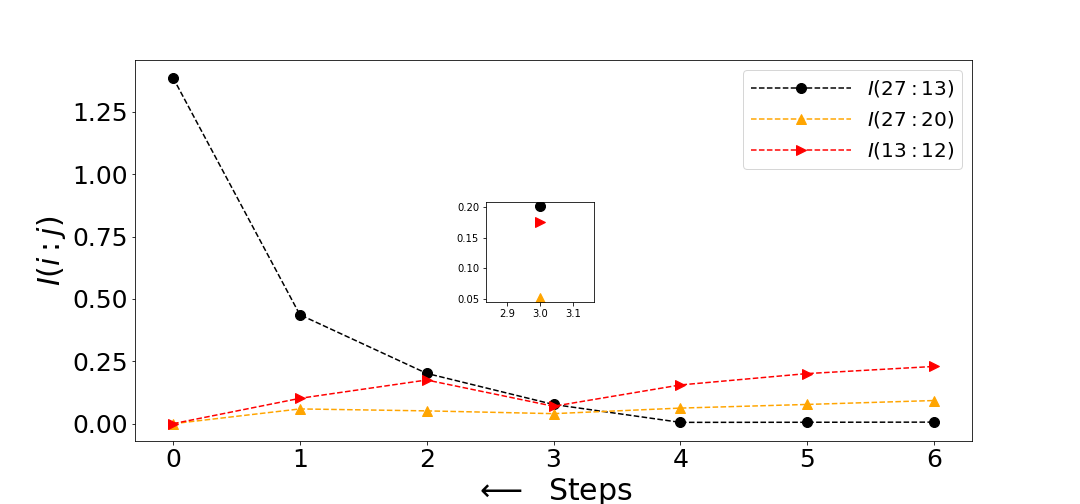}
\caption{\label{fig:13}Variation of Mutual information (MI) $I(i:j)$ for pseudospin pairs connected via tangential scattering (pair $(27,20)$), forward scattering (pair $(13,12)$) and backward scattering (pair:$27:13$) processes. Inset shows the MI values at the third RG step.} 
\end{figure}
\par\noindent
Fig.\ref{fig:13} shows the MERG analysis of the MI content of the following pairs: $(13,27)$-connected via backscattering ($K$), $(20,27)$-connected via tangential scattering ($L$) and $(13,12)$-connected via forward scattering ($V$). The analysis reveals that, starting from the third RG step, $I(13:27)$ dominates over that of the other two. This again substantiates the emergence of strongly entangled pairs $(13,27)$ and $(6,20)$ in the bulk of EHM, and is a direct outcome of the RG relevant backscattering vertex ($K$) together with irrelevant forward ($V$) and tangential ($L$) scattering vertices (see eq.\ref{RGeqns} and discussion below).  
 \begin{figure}
  \hspace*{-0.65cm}
\centering
\includegraphics[width=0.55\textwidth]{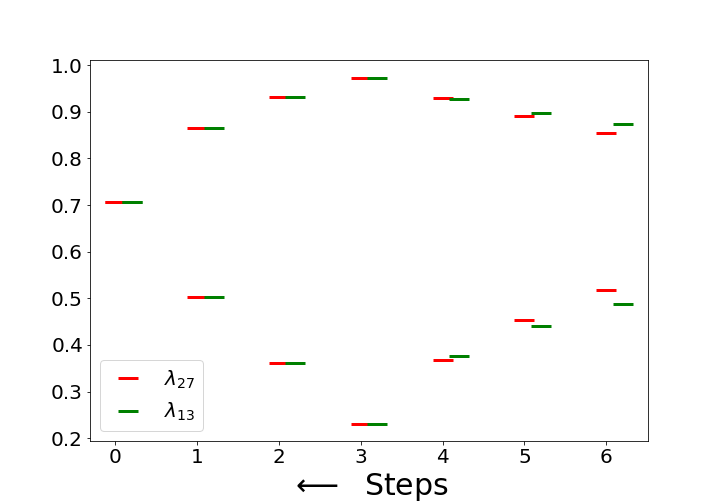}\includegraphics[width=0.55\textwidth]{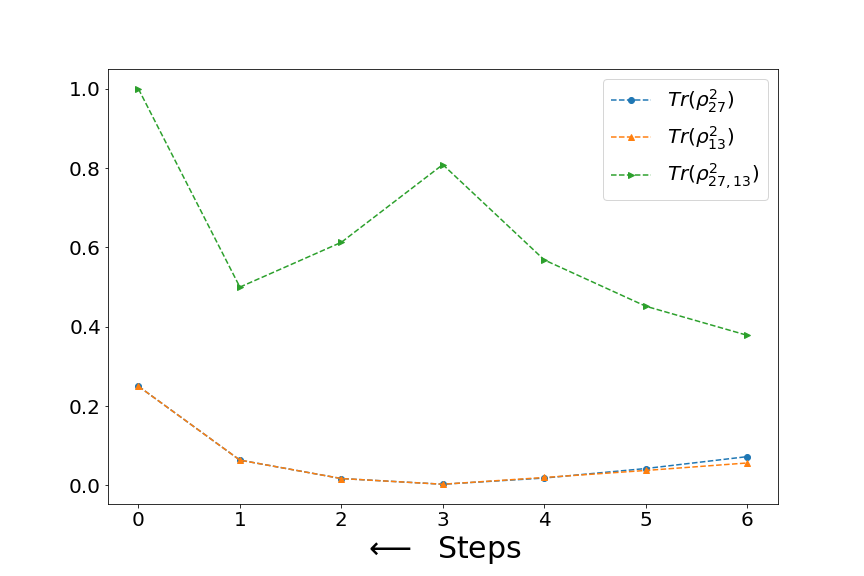}
\caption{\label{fig:14}Left Panel: Schmidt spectrum renormalization for the pseudospin pair $(27,13)$. Right Panel: Purity for the one- and two- pseudospin reduced density matrix.} 
\end{figure} 
Finally, we track the RG contribution of the 1-pseudospin and 2-pseudospin Schmidt spectra in quantifying the growth of MI for the pair $(13,27)$. The left panel of Figure~\ref{fig:14} represents the RG flow for the Schmidt spectra $\{\lambda_{13}\},\{\lambda_{27}\}$ of pseudospins 13 and 27. The plot clearly shows that both pseudospins follow an (almost) identical pattern: the distance between the two eigenvalues of, say, $\{\lambda_{13}\}$ initially grows under the RG till the third step, then showing a dramatic collapse towards a degeneracy at the fixed point. The orange and green curve in the right panel represents the purity of the respective RDMs ($Tr(\rho_{13}^{2})$ and $Tr(\rho_{27}^{2})$). Both purities display the crossover of the spread in probability eigenvalues from low to high, leading to increase in their individual entropies. Further, the purity of the pair of the pseudospins $Tr(\rho_{27,13}^{2})$ increases eventually to $1$, corresponding to a decrease in the joint entropy for these two pairs of pseudospins: $-Tr(\rho_{27,13}\ln\rho_{27,13})$. Taken together, the increase in individual entropies and the decrease in joint entropy signals the increase in the MI of the individual pseudospins. This is another way in which we witness the isolation of strongly entangled pseudospin pairs from the rest in the bulk of the EHM.    
\par\noindent
Thus, the nature of RG flow equations at a given quantum fluctuation energy scale $\omega$ can be seen to dictate the \emph{entanglement geometry} content of the EHM network leading to the T.O. ground state. In the next subsection, we discuss the RG evolution of the entanglement geometry, information entropy content and the distinct nature between the gapless and gapped states it give rise to. 
\subsection{Entanglement scaling features for the Mott liquid, normal metal and Neel antiferromagnetic insulating phases}
 \begin{figure}
 \centering
 \includegraphics[width=0.7\textwidth]{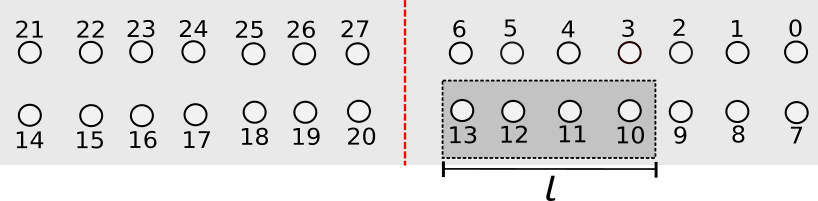}
 \caption{\label{fig:15}Momentum space partitioning (red dotted line) across blocks belonging to opposite sides of the Fermi surface. $l$ is the block size.}
 \end{figure}
\pin
We compute here the entanglement entropy across a partition in momentum-space. As shown in  Fig.\ref{fig:15}), we construct momentum-space blocks of increasing lengths ($l$) on one side of the FS, with the dark/light grey regions being the two members of the partition.
 \begin{figure}
 \hspace*{-1.35cm}
 \centering
 \includegraphics[width=0.6\textwidth]{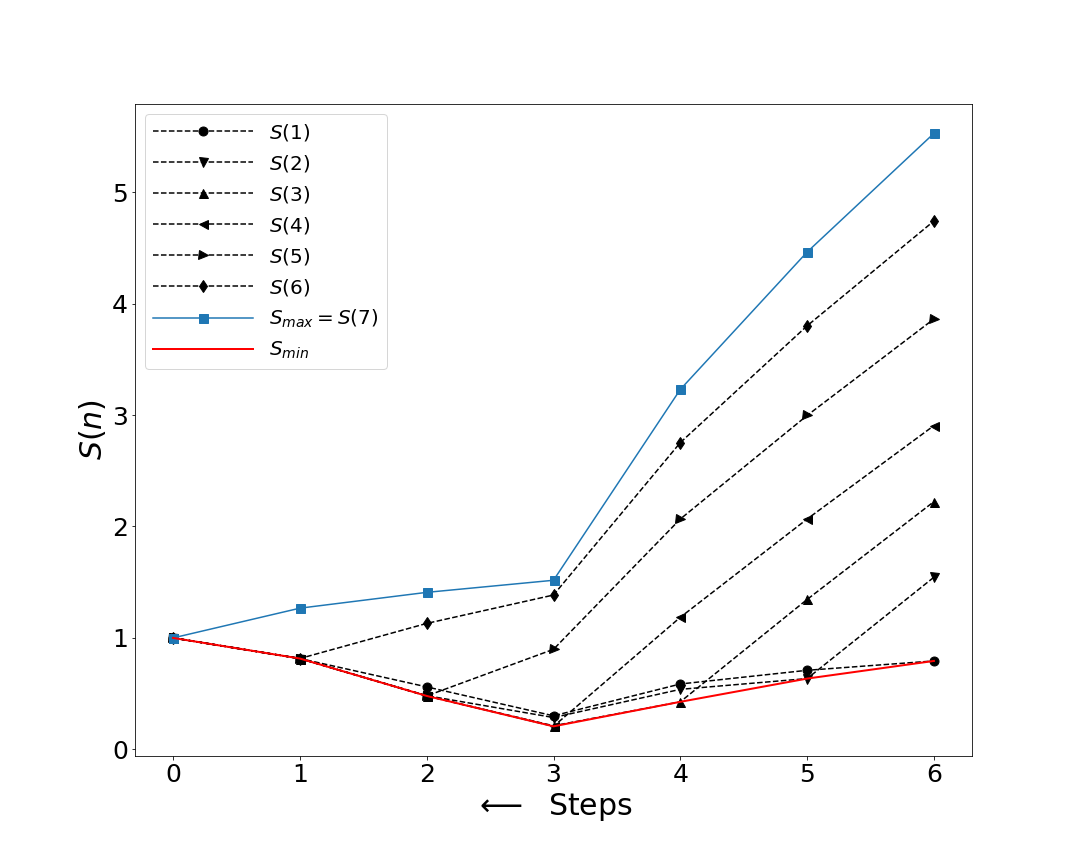}\hspace*{-0.7cm}
 \includegraphics[width=0.6\textwidth]{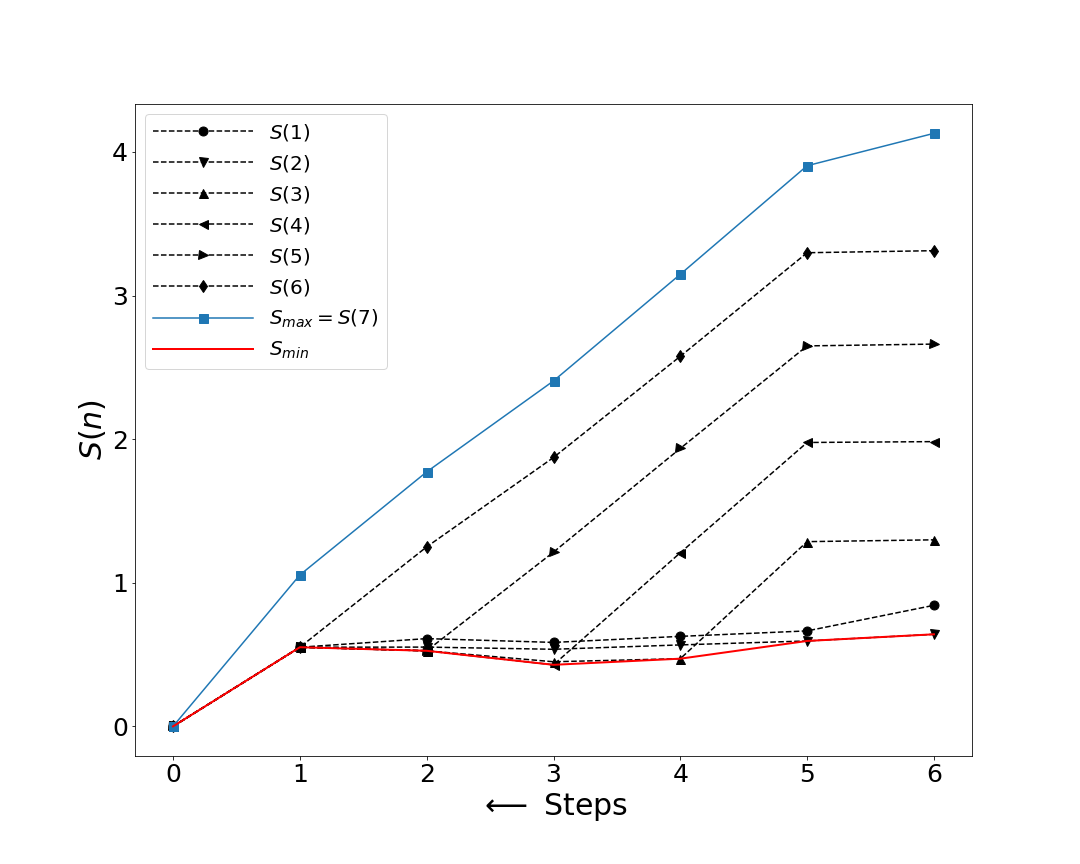}
 \caption{\label{fig:16}Entanglement entropy (EE) renormalization across the EHM for momentum space partition block sizes $1$-$7$ on one side of FS for Mott liquid (left panel) and for normal phase (right panel).}
 \end{figure}
For the Mott liquid state, the RG flow of the entanglement entropy (EE, in units of $\log 2$) in shown in the left panel of Fig.\ref{fig:16}. The figure shows an initial decrease of EE with different block sizes (ranging from $l=7$ to $1$), followed by an increase from the third step to the final value of $EE^{*}=\log 2$ at the fixed point. As presented earlier, this gradual crossover of EE is manifested by the proliferation of strongly entangled (short distance) pairs formed via pseudospin backscattering across the FS. Importantly, we note that at the fixed point, the entanglement entropy $EE^{*}=\log 2$ is independent of subsystem sizes greater than the momentum space shell width, characterizing the perfect disentanglement of the states outside the emergent space. This clearly suggests a \emph{topological} origin of this entanglement entropy. We also carry out a reverse MERG construction (Fig.\ref{URGflowScheme}) for the many-body states in the normal metallic phase for $W>W/2-\omega>W/2$ of the phase diagram Fig.\ref{phDiag}. For this, we start from the ground state eq.\ref{NP} shown in Appendix \ref{App-2}. In this case, the EE in right panel of Fig.\ref{fig:16} is seen to decrease for various block sizes along the RG flow such that at the RG fixed point, $EE=0$. This is consistent with the fact that the state eq.\ref{NP} is separable in terms of momentum space pseudospins. 
The distinct nature of EE RG flows across the EHM for the gapped T.O and gapless metallic states describes a entanglement phase transition in going from the normal metal phase at $\omega=-W/2$ to Mott insulating phase at $\omega=W/2$.
\begin{figure}
\hspace*{-1.35cm}
 \centering
 \includegraphics[width=0.6\textwidth]{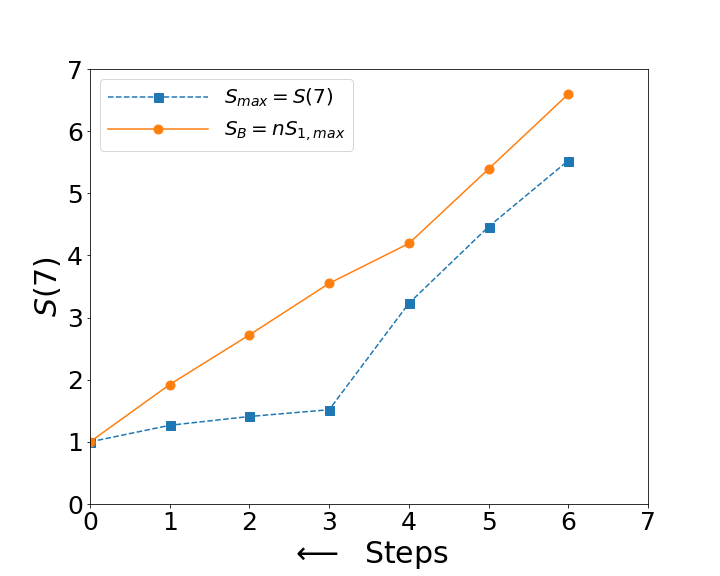}\hspace*{-0.7cm}
 \includegraphics[width=0.6\textwidth]{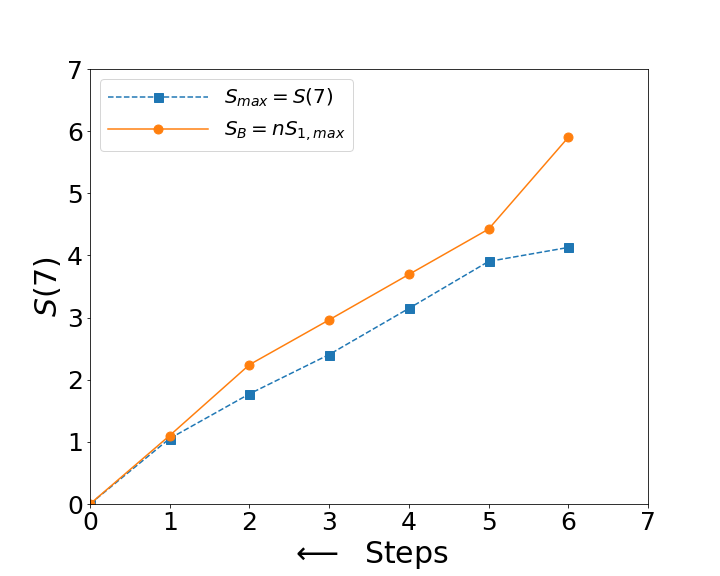}
 \caption{\label{entbound}Blue curve: Entanglement entropy (EE) renormalization across the EHM for momentum space partition block size $7$ on one side of FS for Mott liquid (left panel) and for normal phase (right panel). Orange curve: Renormalisation of the entanglement entropy bound obtained from the causal cone/minimal surface of block size $7$ for Mott liquid (left panel) and for normal phase (right panel).}
 \end{figure}
\par\noindent
In order to investigate further the holographic features of the EHMs constructed for the insulating and metallic ground states, we identify at every RG step the causal cone (dashed red line in Fig.\ref{EHM}) that extends across the bulk of the EHM for a given block of the EHM boundary (grey shaded area in Fig.\ref{EHM}). For the block of boundary pseudospins $7-13$ in Fig.\ref{EHM}, we first compute the maximum single-pseudospin entanglement entropy ($S_{1,max}$). We recall that the Ryu-Takayanagi formula~\cite{ryu2006} was shown to hold for MERA~\cite{evenbly2011,swingle2012b} and EHM~\cite{lee2016} respectively, such that at any given RG step we expect
\begin{equation}\centering
S(7) \leq n~S_{1,max}~,
\end{equation}
where $n$ corresponds to the perimeter of the causal cone/minimal surface at that RG step, i.e., the number of links that must be cut in order to isolate the boundary block of length $7$. Indeed, we find in Fig.\ref{entbound} that the above relation is satisfied by both the Mott insulating and normal metallic ground states. Importantly, we find that the quantity $S_{1,max}$ arises from the entanglement of the pseudospin $13$ with all the other pseudospins ($7-12$). Pseudospin $13$ resides deep in the IR, i.e., proximate to the Fermi surface of the normal phase, and a member of the singlet pair $(13,27)$ that is part of the topologically ordered insulating ground state. This shows that the last link that is cut by the causal cone deep within the IR corresponds to the degrees of freedom proximate to the nodal Fermi surface (of the 2D tight-binding problem on the square lattice). For the metallic ground state, this reflects the holography arising from the Fermi surface~\cite{swingle2010}. On the other hand, for the insulator ground state, the causal cone is a holographic witness to the emergence of the nonlocal Wilson loop (eq.\ref{wilsonloop}, see also Fig.\ref{fig:11})) at the IR fixed point.
 \begin{figure}
 \hspace*{-0.85cm}
\centering
\includegraphics[width=1.1\textwidth]{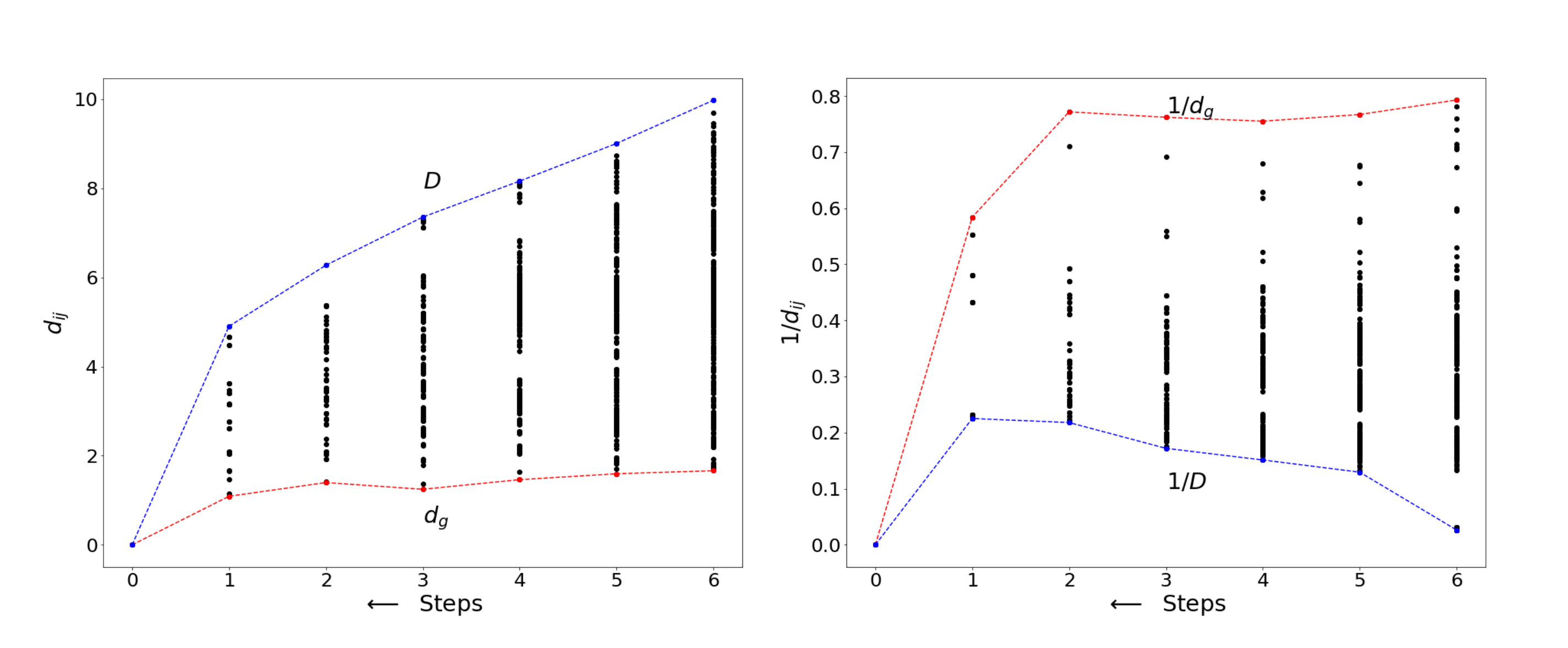}
\caption{\label{infoDistEvolve}Evolution of the information distances between various pseudospin pairs across the EHM for the normal state (right panel) and Mott insulating state (left panel). $D$ and $d_{g}$ are the largest and smallest (i.e., geodesic) information distances respectively.} 
\end{figure}
\par\noindent 
 The entanglement phase transition between the two phases can also be seen from the contrasting entanglement geometry evolution across the EHM for the two ground states. The left panel of Fig.\ref{infoDistEvolve} shows the RG evolution of information distances $d_{ij}$ in the Mott insulating regime, while the right panel of Fig.\ref{infoDistEvolve} shows the evolution of inverse information distances $1/d_{ij}$ in the marginal Fermi liquid metal. The blue curve ($D$ and $1/D$ in the two figures) tracks the RG evolution of the largest information distance (minimal MI pair) in the emergent space geometry. The red curve ($d_{g}$ and $1/d_{g}$ in the two figures) tracks the geodesic, while the black dots represents the information distances between various $(i,j)$ pairs. The increase of MI in the emergent window  of the Mott liquid is described by a shrinking space, i.e., $D-d_{g}$ reduces under RG flow towards the stable fixed point where all distances collapse to zero.  On the other hand, the decrease of MI in the normal metal is described by an expanding space as both $d_{g},D\to \infty$ under the RG flow. It is important to note that the information distance $d(a,b)$ in the IR limit is related to the negative logarithm of the two pseudospin (i.e., two electron-two hole) Green's function in momentum space~\cite{lee2016} 
 \begin{eqnarray}\centering
 d(a,b)& \sim &-\log G(a,b)=-\log(\langle \Psi_{(j)}|A^{+}_{a}A^{-}_{b}|\Psi_{(j)}\rangle)~,
 \end{eqnarray}
where $A^{+}_{a}$ and $A^{-}_{b}$ are pseudospin raising/lowering operators. Recall that in the Mott insulating phase Fig.\ref{fig:12} (also the lower boundary in Fig.\ref{infoDistEvolve} left panel), the information geodesic $d_{g}(a,b)$ is related to the nodal pair $(13,27)$ that forms the singlet at the IR fixed point in the bulk of the EHM. Following Lee and Qi~\cite{lee2016}, we note that in our case the information geodesic at the IR fixed point is given by 
\begin{eqnarray}\centering
d_{g}(13,27)=-\log \exp(-\xi|2k_{F}|) =\xi |2k_{F}|~,
\end{eqnarray}
where $\xi$ is the real-space correlation length between the constituents of the Mott pseudospin singlets eq.\ref{many-body-state}. This indicates the equivalence between the correlator deep in the EHM and the the 4-particle Green's function proximate to the Fermi surface. Similarly, in the EHM constructed for the metallic ground state, the divergence of $d_{g}=-\lim_{k\to k_{F}}\ln(|k-k_{F}|)$ is tied to the RG scaling of the entanglement towards the gapless Fermi surface~\cite{swingle2010}.
\begin{figure}
\includegraphics[width=0.48\textwidth]{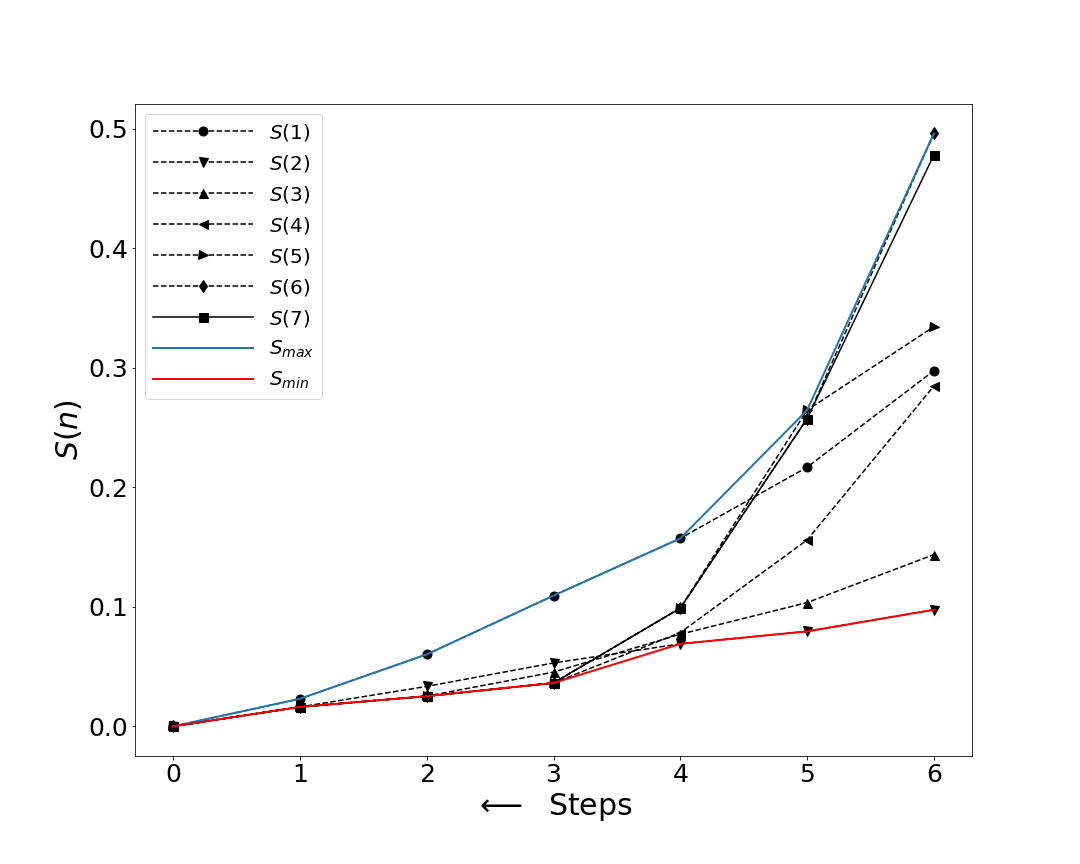}
\includegraphics[width=0.48\textwidth]{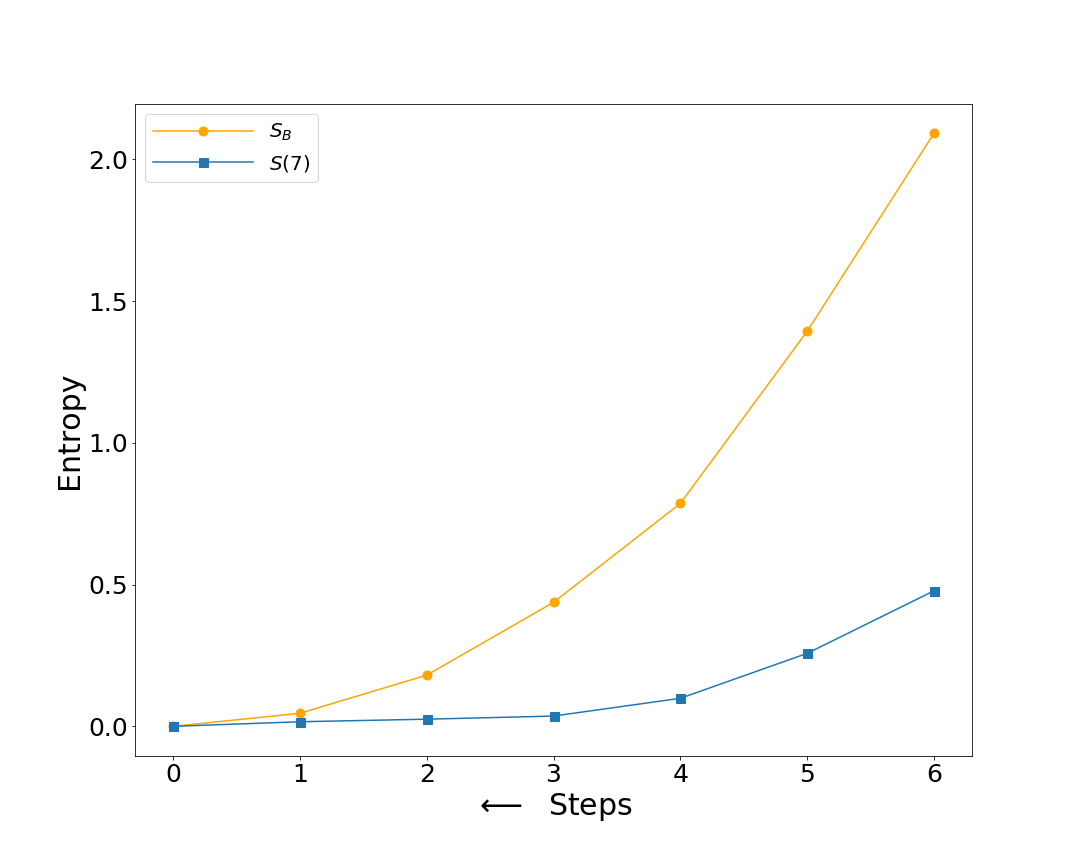}
\caption{Block EE renormalisation for the passage towards the symmetry-broken Ne\'{e}l SDW phase. Left panel: EE for momentum-space block sizes ranging from 1 to 7. Right panel: holographic EE bound (orange curve) and the momentum-space partitition block of size 7 (blue curve).}\label{symbreak}
\end{figure}
\pin
We now turn to the study of a third, and related, state of matter. In Ref.\cite{anirbanmotti}, a study of the influence of symmetry-breaking pertubations under URG revealed that the Mott liquid insulating phase of the half-filled 2D Hubbard model is unstable towards the formation of Ne\'{e}l antiferromagnetic ordering. Thus, in order to study the effects of $(\pi,\pi)$ checkerboard spin-density waves on the Mott liquid, we first apply the URG method to treat the competiton between the staggered magnetic field $\sum_{\mathbf{r}}(-1)^{i+j}hS^{z}_{\mathbf{r}}$ and Umklapp scattering processes (strength $V$). This is shown in detail in Appendix~\ref{URG_ML_SDW}. The nonperturbative RG flow equation eq.\ref{nonpertSDWRG} at weak coupling $V/h<(\log\frac{\Lambda_{0}}{\Lambda})^{-2}$ yields a exponential square root dependence on the coupling strength that is characteristic of the Ne\'{e}l SDW phase~\cite{fradkin2013field}. Further, the RG flow equation ends at a IR fixed point with the coupling strength $V^{*}=h(\log\frac{\Lambda_{0}}{\Lambda})^{-2}$. Next, we carry out the MERG construction (see Fig.\ref{URGflowScheme}) starting from the Ne\'{e}l antiferromagnet (AFM) symmetry-broken state (eq.\ref{SDWstate} of the effective Hamiltonian in eq.\ref{effH}). The block entanglement entropy computed for the Ne\'{e}l SDW from the MERG is shown in the left panel of Fig.\ref{symbreak}: it is seen to decrease for various block sizes along the RG flow such that at the RG fixed point, $EE\to 10^{-5}\log 2$. This is consistent with the fact that, in the Neel SDW phase, the entanglement is considerably lower compared to the Mott liquid phase (which we found earlier to have $EE^{*}=\log 2$). A similar flow towards a vanishing value is also observed for the holographic EE bound (right panel of Fig.\ref{symbreak}). In this way, the MERG flows of both the block entanglement entropy and the holographic EE bound of the Ne\'{e}l SDW show features that are distinct compared to those observed for the Mott liquid and the non-Fermi liquid phases (Figs.\ref{entbound}). This distinction marks the entanglement phase transition between the Mott liquid and Neel SDW phases.
\pin 
We summarise by noting that the entanglement features and geometry of the EHM for the topologically ordered insulating, metallic and symmetry-broken phases we have studied here are markedly different: the T.O. phase is marked by a nontrivial $EE=\log 2$ and vanishing $d_{g}$, the normal phase is characterised by $EE=0$ and a diverging $d_{g}$ and the symmetry-broken SDW state is characterised by a vanishing $EE$. This confirms that our EHM network carries important information with regards to the topological nature of a many-body state, and can sense a transition between the two phases through their entanglement features. In a later section, we will quantify this information flow across the EHM network using measures from information theory and deep learning (DL). We will thus show the equivalence between our MERG based EHM and a deep neural network (DNN) architecture.
\section{Probing the entanglement of the QCP in a hole-doped Mott liquid}\label{pathtodwave}
\par
\begin{figure}[!ht]
\centering
\hspace*{-2.5cm}
\includegraphics[width=1.3\textwidth]{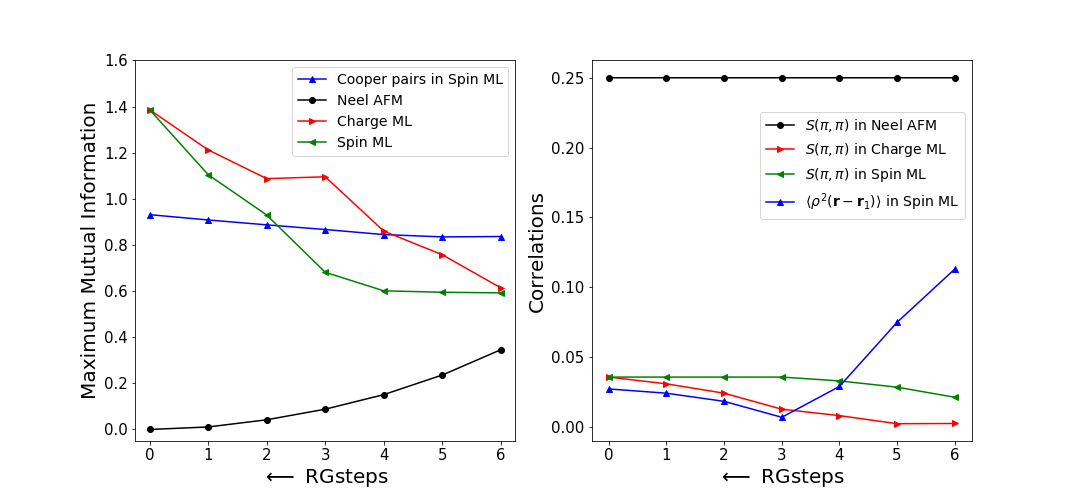}
\caption{Left panel: RG flow for the maximum MI between different pairs in the ground state of the half-filled Hubbard Model. The red/green curve represents the MMI between electron-hole pairs of opposite spin consituting the charge/spin-type Mott liquid ground states respectively. The blue curve represents the MMI between different $\mathbf{p}=0$ Cooper pairs present within the spin-type Mott liquid. The black curve represents the MMI betweeen different electron-hole pairs of the Neel AFM, where each pair has electronic states in same up/down configuration. Right panel: RG flow of correlation functions in the ground state of the half-filled Hubbard model. The red/green curve represents the longitudinal spin structure factor $S(\pi,\pi)$ of the $\mathbf{Q}=\pi,\pi$ checkerboard Ne\'{e}l AFM order within the symmetry-unbroken charge/spin type Mott liquids respectively. The blue curve represent the superconducting off-diagonal long-range order parameter $\langle\rho^{2}(\mathbf{r}-\mathbf{r}')\rangle$ (averaged over all sites) in the spin-type Mott liquid.} \label{Mottliquid correlations}
\end{figure}

\noindent
In Ref.\cite{anirbanmott2}, we observed that the emergence of d-wave superconducting off-diagonal long-ranged order (ODLRO) at $T=0$ in the 2D Hubbard model involved the divergent quantum fluctuations at a novel quantum critical point (QCP) associated with the collapse of Mottness upon doping holes into the system. This raises the question: \textit{how does the many-particle entanglement of the Mott liquid evolve with doping towards the QCP, such that its instability towards a Ne\'{e}l AFM ground state at half-filling~\cite{anirbanmotti} is replaced instead with that towards superconducting ODLRO at the QCP~\cite{anirbanmott2}?} In this section, we aim to provide an answer to this question through a coherent understanding of entanglement based features and two-point correlation functions for the Mott quantum liquid and Ne\'{e}l SDW ground states at half-filling (hole-doping fraction $f_{h}=0$), as well as the Mott quantum liquid and superconducting ground states at quantum critical doping ($f_{h}=0.25$ for $U_{0}=8t$).
\pin
At half-filling, we study the URG evolution of the longitudinal spin structure factor defined as
\begin{eqnarray}\centering
S(\mathbf{Q})=\langle s^{z}_{\mathbf{Q}}s^{z}_{-\mathbf{Q}}\rangle, s^{z}_{\mathbf{Q}}=\sum_{\mathbf{r}}(-1)^{i+j}s^{z}_{\mathbf{r}}\label{Spipi}
\end{eqnarray}
for the wavevector $\mathbf{Q}=(\pi,\pi)$ within the two-fold degenerate symmetry-preserved ground states $|\Gamma_{+}\rangle$ (eq.\ref{many-body-state}) and $|\Gamma_{-}\rangle$ (eq.\ref{many-body-state2}), as well as the Ne\'{e}l AFM ground state (eq.\ref{SDWstate} in Appendix \ref{URG_ML_SDW}). As shown in the right panel of Fig.\ref{Mottliquid correlations}, the UV to IR evolution of $S(\pi,\pi)$ within the ground states $|\Gamma_{+}\rangle$ (red curve) and $\Gamma_{-}\rangle$ (green curve) is towards a much smaller value compared to that observed in the Ne\'{e}l antiferromagnet (black curve). Importantly, the strongly suppressed values of $S(\pi,\pi)$ for the $|\Gamma_{\pm}\rangle$ quantum liquid ground states are manifestations of their high entanglement content. In the left panel of Fig.\ref{Mottliquid correlations}, the many-particle entanglement is quantified by the maximum mutual information (MMI) $\max_{i,j}I(i:j)$(eq.\ref{MI}) between charge pseudospins in $|\Gamma_{+}\rangle$ (eq.\ref{many-body-state}, red curve) and spin pseudospins in $|\Gamma_{-}\rangle$ (eq.\ref{many-body-state2}, green curve). As the RG progress from UV to IR, the entanglement content is distilled into strongly entangled pseudospin pairs for the two quantum liquid ground states $|\Gamma_{\pm}\rangle$. On the other hand, the MMI for the Ne\'{e}l AFM state (eq.\ref{SDWstate}, black curve in left panel of Fig.\ref{Mottliquid correlations}) between pseudospin pairs (eq.\ref{Dpseudospins}) decreases monotonically and eventually vanishes upon scaling towards the IR. 
\par\noindent
From the spin Mott liquid phase (eq.\ref{many-body-state2}), we also derive an effective low-energy theory for Cooper pair pseudospins (see Appendix \ref{Coop}). From this, we compute the MMI between Cooper pair pseudospins (blue curve in left panel of Fig.\ref{Mottliquid correlations}), observing that it grows mildly in passage from UV to IR. On the other hand, the value of the real-space averaged ODLRO $\langle\rho^{2}(\mathbf{r}-\mathbf{r}_{1})\rangle$    
\begin{eqnarray}\centering
\rho^{2}(\mathbf{r}-\mathbf{r}_{1})=-\frac{1}{vol}\sum_{\mathbf{k},\mathbf{p}} e^{i\mathbf{p}\cdot(\mathbf{r}-\cdot\mathbf{r}_{1})}\langle c^{\dagger}_{\mathbf{k}_{\Lambda,\hat{s}}\uparrow}c_{\mathbf{k}_{-\Lambda,T\hat{s}},\downarrow}c^{\dagger}_{\mathbf{p}-\mathbf{k}_{\Lambda\hat{s}},\downarrow}c_{\mathbf{p}-\mathbf{k}_{-\Lambda,T\hat{s}},\uparrow}\rangle\label{ODLRO}
\end{eqnarray}
is observed to reduce in magnitude under RG evolution from UV towards IR (blue curve in right panel of Fig.\ref{Mottliquid correlations}). At the UV end, the large ODLRO $\langle\rho^{2}(\mathbf{r}-\mathbf{r}_{1})\rangle=0.125$ arises from the presence of $p\neq 0$ Cooper pairs present in the spin Mott liquid that scatter via tangential, forward and back scattering processes~\cite{anirbanmott2}. As can be seen from the blue and green curves in the left panel, these scattering processes reduce the inter-Cooper pseudospin and inter-spin psuedospin entanglements in the UV. Similarly, the enhanced inter-spin pseudospins and inter-Cooper pseudospins entanglement at the IR scale (as quantum fluctuations related to various scattering processes is resolved under RG) coincides with a reduction in the ODLRO. 
\par
\begin{figure}[h!]
\centering
\hspace*{-2.5cm}
\includegraphics[width=1.3\textwidth]{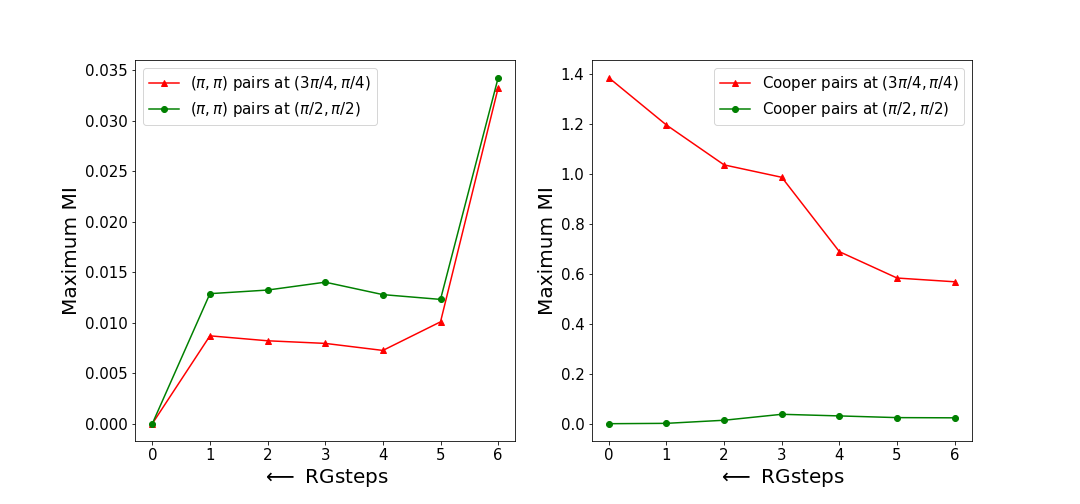}
\caption{Left panel: RG flow of MI between charge-pseudospin pairs (with $\pi,\pi$ net pair momentum) along the nodal ($\pi/2,\pi/2$) direction (green curve) and antinodal ($3\pi/4,\pi/4$) (red curve) directions. Right panel: RG flow of MI between Cooper pairs (with $0$ net-momentum) within the symmetry-preserved spin Mott liquid along the antinodal (red curve) and nodal (green curve) directions.} \label{MIatQCP1}
\end{figure}

\noindent
We now turn to a similar investigation of entanglement and many-body correlations at the QCP. As shown in Appendix \ref{AppQCP}, at the QCP,
excitations normal to the Fermi surface in the nodal direction (i.e., along the $(\pi/2,\pi/2)$ Fermi point) are gapless, and described by a marginal Fermi liquid state. Instead, the antinodal direction (along the $(3\pi/4,\pi/4)$ Fermi point) is gapped via strong Cooper pair backscattering. Starting from the ground states at the fixed point eq.\ref{QCP_State} and employing the MERG program Fig.\ref{URGflowScheme}, we include the backscattering, forward scattering and tangential scattering processes for constructing the eigenstates in passage towards the UV. As before, we compute from these eigenstates the RG evolution of entanglement and many-body correlations. In the left panel of Fig.\ref{MIatQCP1}, the MMI value between $(\pi,\pi)$ net-momentum electronic pairs is seen to decrease under RG along both the antinodal (red curve) and nodal directions (green curve), vanishing altogether at the IR fixed point. This was referred to as the \textit{collapse of Mottness} in Ref.\cite{anirbanmott2}. On the other hand, the MMI between $(0,0)$ net-momentum Cooper pairs (red curve) in Fig.\ref{MIatQCP1} (right panel) is seen to increase along the antinodal direction, stopping at the maximum value $2\log 2$; this corresponds to the singlet pairing between pseudospins $6$ and $20$. Along the nodal direction, the Cooper pseudospins $13$ and $27$  within the gapless marginal Fermi liquid comprise a separable state. Therefore, as seen in the green curve in Fig.\ref{MIatQCP1} (right panel), the small MMI between Cooper pairs along the nodal direction in the UV reduces further under RG flow, and vanishes in the IR. 
\begin{figure}[h!]
\centering
\hspace*{-2.5cm}
\includegraphics[width=1.3\textwidth]{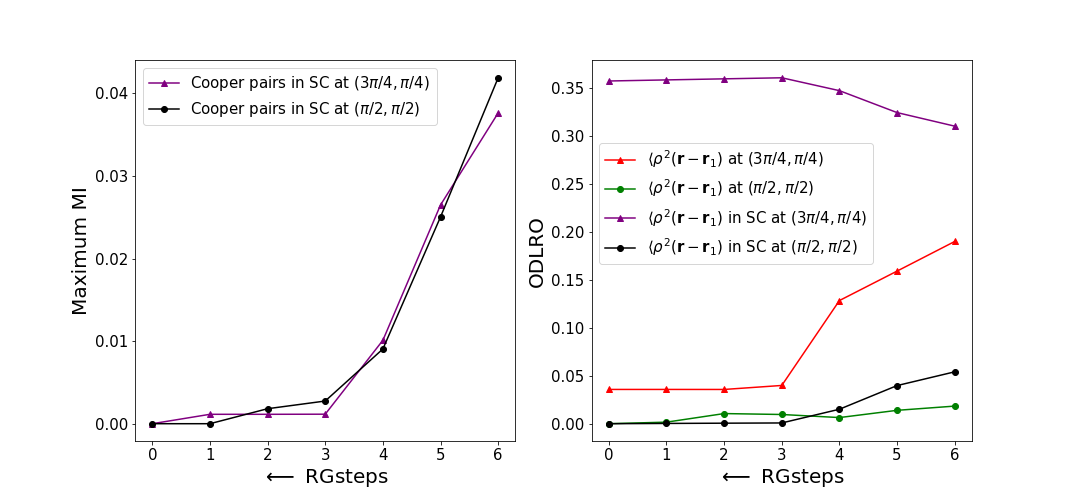}
\caption{Left panel: RG flow of MI between Cooper pairs within the symmetry-broken d-wave superconductor at the QCP along the nodal (black curve) and antinodal (purple curve) directions. Right panel: RG flow of the ODLRO within the symmetry-preserved spin Mott liquid at the QCP along the antinodal (red) and nodal (green) directions. The black and purple curves represent the ODLRO along the nodal and antinodal directions within the symmetry-broken d-wave superconductor respectively.}\label{MottatQCP2}
\end{figure}
\pin
By following Ref.\cite{anirbanmott2}, we include a $U(1)$ phase-rotation symmetry-breaking field $\Delta(c^{\dagger}_{\mathbf{k}_{\Lambda\hat{s}},\uparrow}c^{\dagger}_{-\mathbf{k}_{\Lambda\hat{s}},\downarrow}+h.c.)$ and perform the URG once again, obtaining the d-wave symmetry broken superconductivity phase. As shown in the right panel Fig.\ref{MottatQCP2}, the Cooper pairs do not condense along the nodal direction  both in the presence (black curve) as well as absence (green curve) of symmetry breaking. This ensures that ODLRO is absent in the IR in both ground states, displaying that the nodal marginal Fermi liquid is protected against symmmetry-breaking. However, along the antinodal direction in the $U(1)$ symmetry-broken state, the ODLRO increases under RG flow from UV to IR (purple curve in the right panel of Fig.\ref{MottatQCP2}) even as the MMI of Cooper pairs reduces and eventually vanishes at the IR fixed point (purple curve in left panel of Fig.\ref{MottatQCP2}). As a consistency check, we note that within the $U(1)$ phase-rotation symmetry-preserved Cooper pair quantum liquid ground state (eq.\ref{QCP_State}), the ODLRO reduces substantially from UV to IR (red curve in right panel of Fig.\ref{MottatQCP2}), and finally terminates at a small value of $\langle\rho^{2}(\mathbf{r}-\mathbf{r}_{1})\rangle=0.05$.
\pin
We summarise by noting that our investigations reveal a clear anticorrelation between the RG evolution of the MMI and many-body correlations (such as the spin structure factor and the ODLRO) for the various symmetry-preserved as well as symmetry-broken ground states we have encountered in the 2D Hubbard model at half-filling and the QCP. These results display how the mutual-information based entanglement of the quantum liquid ground states evolves with hole-doping: the spin and charge pseudospins that condensed from the dominant backscattering processes at half-filling to form the Mott liquid ground states are replaced by their Cooper-pairing pseudospin counterparts at the QCP via a collapse of Mottness. Concomitantly, the associated dominant symmetry-broken ground states changes hands from the Ne\'{e}l AFM to d-wave superconductivity (whose nodes are gapless marginal Fermi liquids).  
\section{A Deep Neural Network based on EHM}\label{EHMtoDNN}
\pin
Motivated by the equivalences between tensor network RG and deep learning \cite{beny2013,mehta2014exact,hu2019machine}, we propose a deep neural network based on the EHM tensor network for classifying the correlated quantum liquid ground states of the 2D Hubbard model~\cite{anirbanmotti,anirbanmott2}. We begin, however, with a brief discussion of the concepts of the deep neural network (DNN)~\cite{salakhutdinov2009} pertinent to our presentation. A DNN is a sequence of layers of mathematical operations that transform an input feature vector $X$ into $\hat{X}$ while keeping only the essential parts relevant to the target output vector $Y$. An optimal $\hat{X}$ can be obtained via Lagrangian minimization~\cite{tishby2015deep}
\begin{equation}\centering\centering
\mathcal{L}[p(\hat{x}|x)]=\mathcal{I}(X:\hat{X})-\beta\mathcal{I}(\hat{X}:Y)~,
\end{equation}
where the \textit{mutual information like quantity} $R=\mathcal{I}(\hat{X}:X)$ quantifies the complexity of the representation, and $\mathcal{I}_{Y}=\mathcal{I}(\hat{X}:Y)$ is the amount of information relevant to $Y$ preserved in $\hat{X}$. The quantity $\beta$ is the tradeoff parameter, while the function $p(\hat{x}|x)$ is the conditional probability. An optimal representation $\hat{X}$ is one for which the MI ($R$) is reduced and $\mathcal{I}_{Y}$ is preserved. This is known as the information bottleneck (IB) principle \cite{tishby2015deep}. A DNN following the IB principle can be considered optimal. In the section below, we will demonstrate that EHM based MERG fulfills the IB principle, and is therefore on par with a optimally functioning supervised DNN.
\subsection{Demonstrating the IB principle for EHM based on MERG}
\pin
We have seen earlier that the MERG is a family of unitary disentanglement transformations with an input many-body eigenstate (Fig.\ref{MERG}), leading to simpler representations with a lower number of entangled qubits. It was also demonstrated that the EHM of Fig.\ref{EHM} is a tensor network representation of MERG. An important question to ask in this regard is:~\emph{does the EHM network architecture naturally follow the IB principle?} 
\par\noindent
In order to perform the information bottleneck analysis, we first prepare the entanglement RG results as a data model. Every pair of qubits in the many-body state can be labelled as $(a,b)$. With every pair $(a,b)$ is associated one \emph{feature}: $F(a,b)=I(a:b)$ (the MI defined in eq.\ref{MI}). The larger the MI, the stronger is the entanglement within the pair. Next, we perform a classification of all pairs into two classes, strongly and weakly entangled, by using a \emph{classifier} $C(a,b)$
\begin{eqnarray}\centering
C(a,b)=  \begin{array}{cc} 
      0 & 0\leq F(a,b)<\log 2 \\
      1 & \log 2\leq F(a,b)\leq 2\log 2~.
   \end{array}
\end{eqnarray} 
In order to compute the set ($R,\mathcal{I}_{y}$) at each step of the RG, we first 
prepare the input vector ($X$), RG transformed vectors $\hat{X}$ and target vector $Y$. 
The input feature vector $X=\lbrace F(a,b)\rbrace$ is built from the state $|\Psi_{(6)}
\rangle$, constituted of MI values for ${4 \choose 2}=6$  pairs $(a,b)$ made of $4$ 
pseudospin qubits ($6$, $13$, $20$, $27$). $S$ is a list of these pairs, where $S_{i}$ 
is the ith element in the list
\begin{eqnarray}\centering
S=\lbrace (13,6), (20,6), (27,6), (20,13),(27,13),(27,20)\rbrace\label{pair-set}
\end{eqnarray}
Note that the above four qubits (Fig.\ref{GSrep}) have been chosen so as to eventually compose the emergent subspace at the RG fixed point, and therefore constitute the entanglement features of the T.O. ground state.
We similarly construct the transformed feature vector $\hat{X}_{j}$ from $|\Psi_{(j)}\rangle$ at each RG step. The target vector $Y=\lbrace C(a,b)\rbrace$ is constructed for pairs eq.\ref{pair-set} in $S$ using the state $|\Gamma_{+}\rangle$ at the fixed point eq.\ref{many-body-state}. From these datasets, we compute the information plane coordinates \cite{tishby2015deep,shwartz2017opening} $R=\mathcal{I}(\hat{X}_{j}:X)$ and $\mathcal{I}_{Y}=\mathcal{I}(Y:\hat{X}_{j})$. 
\begin{figure}
\centering
\includegraphics[width=\textwidth]{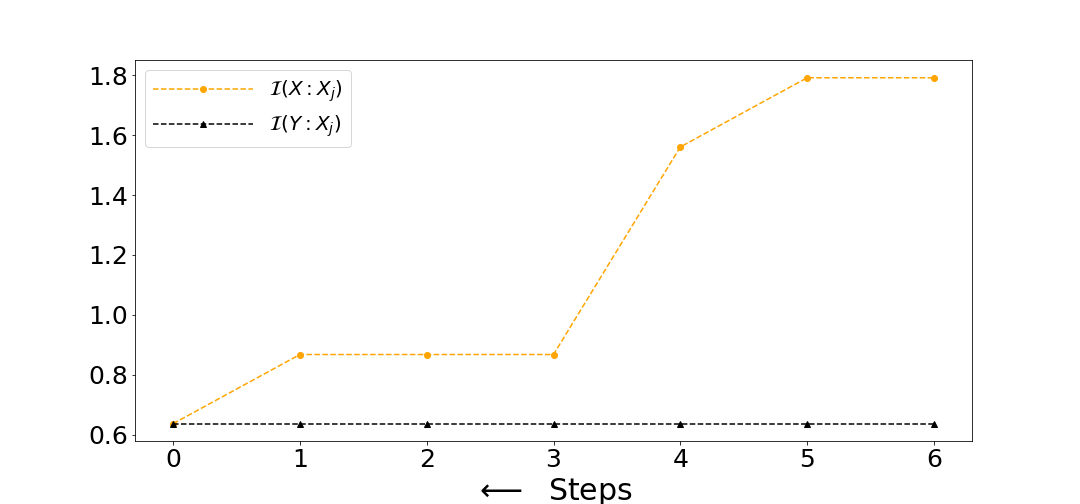} 
\caption{\label{EHM_IB} Testing the information bottleneck principle for EHM network. Orange curve represents the representation complexity $R=\mathcal{I}(\hat{X}_{j}:X)$, while black curve represents $\mathcal{I}_{Y}=\mathcal{I}(Y:\hat{X}_{j})$, the information about class $Y$ present in the representation.}
\end{figure}
 \par\noindent
The representational complexity $R$ (orange curve) displayed in Fig.\ref{EHM_IB} is seen to decrease across the EHM, while the mutual information $I_{Y}$ (eq.\ref{MI}, black curve) is constant throughout at a value of 0.636. This demonstrates clearly that the IB principle is met by the MERG based EHM. This allows us to make the following mapping between the MERG based EHM and a DNN: the disentanglement transformations (yellow blocks in Fig.\ref{EHM}) are equivalent to the \emph{hidden layers} of a DNN that outputs simpler representations. A constant value of $I_{Y}$ across the EHM suggests a strong dependence between $Y$ and representation $\hat{X}_{j}$ obtained at various RG transformation steps.
\begin{figure}
\hspace*{-1.45cm}
\centering
\includegraphics[width=1.2\textwidth]{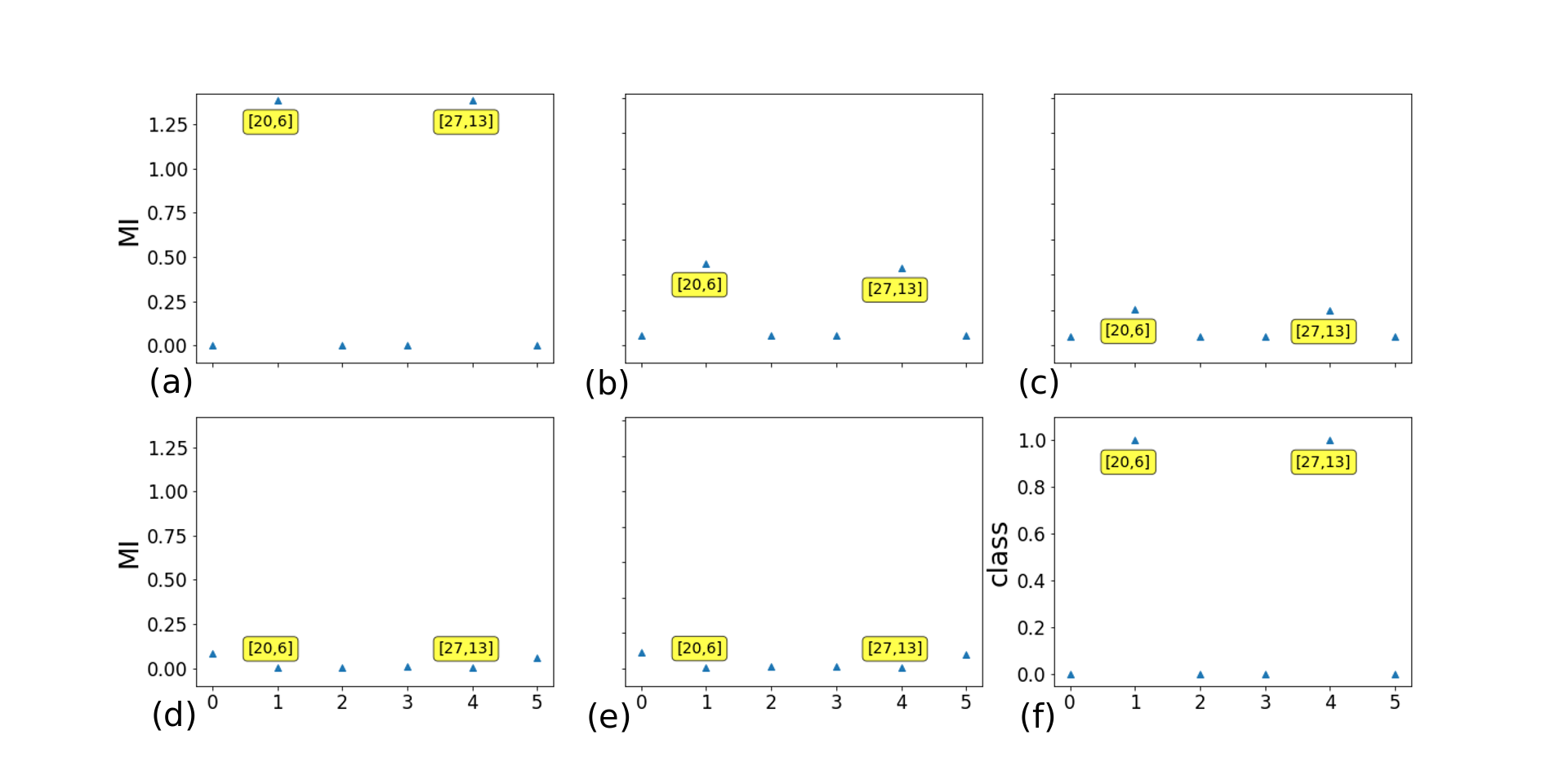}
\caption{\label{MI_class}(a-e) MI values $\hat{X}_{j}$ of the list of pairs $S$ for states $|\Gamma_{+}\rangle$, $|\Psi_{1}\rangle$, $|\Psi_{3}\rangle$, $|\Psi_{4}\rangle$, $|\Psi_{6}\rangle$. (f) Class $Y$ for the pairs in $|\Gamma_{+}\rangle$.}
\end{figure}
Fig.\ref{MI_class} verifies the strong dependence between the representation of MI values $\hat{X}_{j}$ (Figs. \ref{MI_class} (a-e)) at each RG step and the class $Y$ (Fig.\ref{MI_class}(f)). We observe that the MI values for the pairs $(6,13)$ and $(20,27)$ either belong simultaneously to the strongly entangled class~$C(6,20)=C(13,27)=1$, or to the weakly entangled class ~$C(6,20)=C(13,27)=0$. This results in the preservation of mutual information $\mathcal{I}(Y:\hat{X}_{j})$ at every RG step (black line in Fig.\ref{EHM_IB}).
\subsection{Constructing a DNN model for classifying \emph{entangledness} of a pair}
\pin
Given the demonstration of the IB principle for the EHM based on MERG, we construct a conditional probability DNN model for predicting the eventual fate of a given pair $(a,b)$ to be either strongly ($C(a,b)=1$) or weakly entangled ($C(a,b)=0$). This is done in the following way: given any pair $(a,b)$ and its feature $F(a,b)$ at each RG step $j$, we compute the conditional probability $P(C=1|F(a,b))$ by performing a statistical analysis of the MI dataset. Below we list the specifics of the MERG based DNN model:
\begin{enumerate}
\item[1.] The bare representation $X_{6}$ is constructed from initial state $|\Psi_{(6)}\rangle$ by collecting the MI values of all ${28 \choose 2}=378$ $(a,b)$ pairs.
\item[2.] Unitary transformations equivalent to \emph{hidden layers} are linear maps that act on the state $|\Psi_{(6)}\rangle$, followed by state $|\Psi_{(5)}\rangle=U_{6}|\Psi_{(6)}\rangle$ and so on. This is in contrast with standard deep learning, where nonlinear filters and weight matrices are chosen variationally in constructing the transformation layers~\cite{parr2018matrix}. In our formalism, the unitary operators are constructed exactly (see Sec.\ref{URG-Formal}). Furthermore, because of the exact nature of the construction, there is no requirement of a training dataset for the  construction of the DNN.
\item[3.] From each of the many-body states, we compute the MI values (eq.\ref{MI}) using one- and two-pseudospin Schmidt spectra (eq.\ref{SchmidtSpec}). The collection of $378$ MI values at each RG step $j$ is referred to as the bottlenecked representation ($X_{j}$) of $X_{6}$.
\item[4.] At each RG step $j$, we compute the joint probability $P(F(a,b),C(a,b))$ between the bottlenecked representation $X_{j}$ and the target classifier $\hat{Y}$. From this, we compute the conditional probability for the eventual fate at $j^{*}=0$ of the \emph{entangledness} of a pair given its MI value at RG step $j$.   
\end{enumerate}
\par\noindent
We now set the criterion for a DNN model ideal from our perspective: such a DNN should predict that in the bulk of the EHM, the pairs $(13,27)$ and $(6,20)$ are strongly entangled($C=1$) while all others are weakly entangled ($C=0$), as observed from the state $|\Gamma_{+}\rangle$ (eq.\ref{many-body-state}). Note that we have already shown via entanglement RG that the formation of strongly entangled short distance pairs demonstrates the onset of T.O.~(Sec.\ref{EmStrongShort}). Therefore, if the DNN model correctly predicts the eventual fate of the pairs, it is then equivalent to predicting the onset of T.O.
\par
\begin{figure}
\centering
\includegraphics[width=1\textwidth]{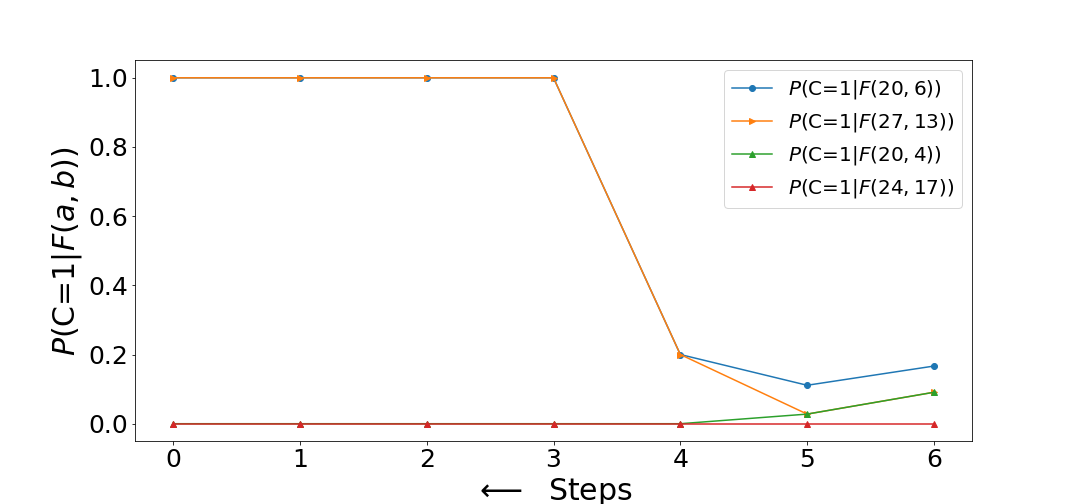}
\caption{\label{cond}Predicting the eventual entanglement strength (either strong or weak) of a subset of pairs after each RG step using conditional probability.}  
\end{figure}

\begin{figure}
\centering
(a)\includegraphics[width=0.4\textwidth]{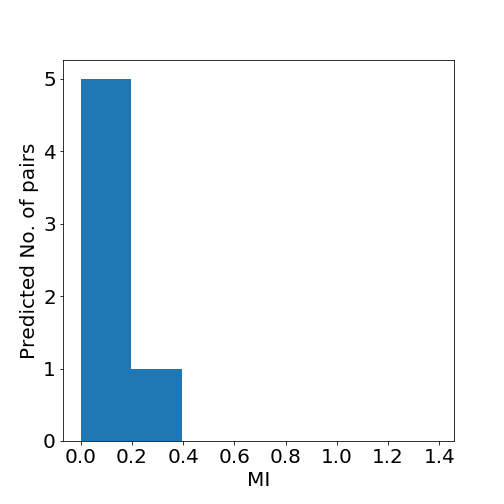} (b)\includegraphics[width=0.4\textwidth]{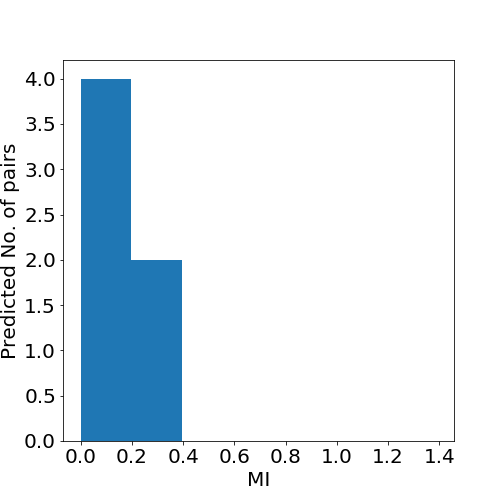}\\ 
(c)\includegraphics[width=0.4\textwidth]{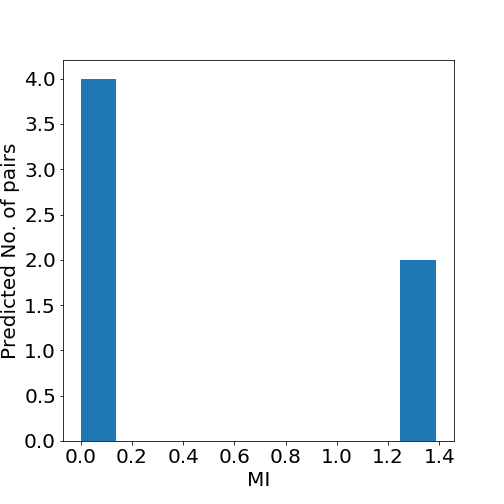} 
(d)\includegraphics[width=0.4\textwidth]{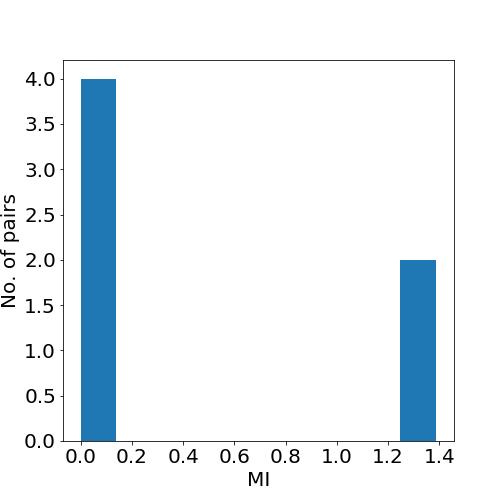}  
\caption{\label{HistPredict}(a-c) Prediction of MI histogram distribution among pairs in list $S$ from conditional probability model after (a) 0, (b) 2 and (c) 3 layers of MERG transformation. (d) MI histogram obtained after 6 layers of MERG transformation.}
\end{figure}

\begin{figure}
\centering
\includegraphics[width=0.9\textwidth]{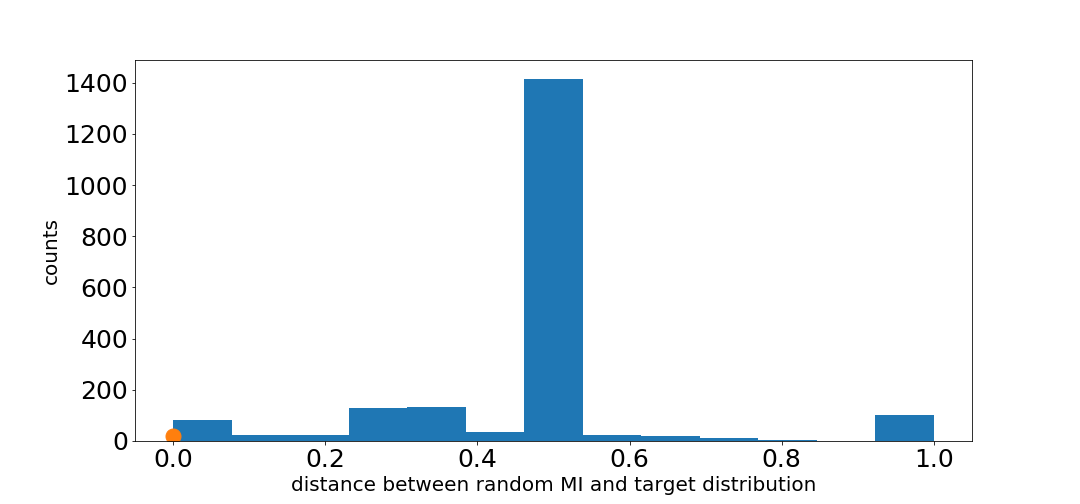}  
\caption{\label{nullHypothesis}Histogram plot for the statistical distance between MI distribution of a random tensor network and the DNN's predicted MI distribution at the RG fixed point.}
\end{figure}

\noindent
Fig.\ref{cond} shows that the conditional probability $P(C=1|F(a,b))$ for the pairs $(6,20)$ and $(13,27)$ to be strong ($C=1$) are small to start with during the RG flow; however, after the third unitary RG step, it rises to value 1. On the other hand, the conditional probability vanishes for two other pairs $(20,4)$ and $(24,17)$. In this way, we demonstrate that layers near the UV scale of the EHM network are \emph{holographic witnesses} to the emergence of T.O. at the IR scale. The existence of such holographic witnesses is further demonstrated in Fig.\ref{HistPredict} by the prediction of the eventual histogram distribution of MI from the initial layers of the EHM. Specifically, Fig.\ref{HistPredict}(a-c) is the MI histogram distribution among pairs in $S$ (formed from the pseudospins $6$,$13$,$20$,$27$) and predicted from the zeroth, second and third unitary transformation layers of the EHM network. We find that the MI distribution predicted after only $3$ layers already matches with that obtained after $6$ layers Fig.\ref{HistPredict}(d). This prediction process is further validated via a null hypothesis test: can the MI distribution of a random tensor network simulate the predicted MI distribution of the target? The statistical distance histogram displayed in Fig.\ref{nullHypothesis} shows that there is only $4\%$ probability for the predicted MI distribution to be obtained from a random tensor network. This proves the uniqueness of the target tensor network obtained from the DNN.
\par
\begin{figure}
\centering
\includegraphics[width=0.9\textwidth]{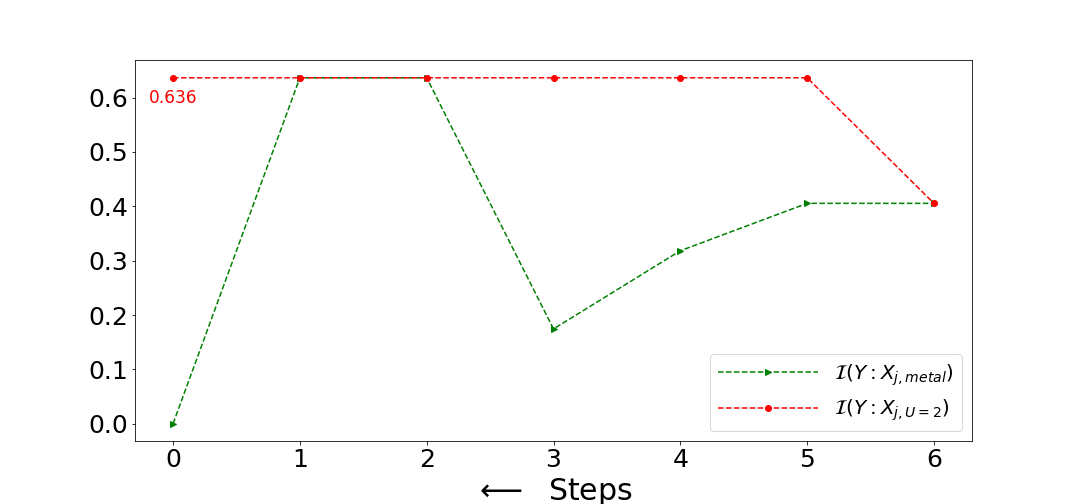}
\caption{\label{mutualInformationMetIns} The red curve shows the mutual information content between the MI feature vector of a Mott insulator $X_{j,U=2}$ for $U_{0}=2$, $\omega=4$ and target vector $Y$ (Mott insulator at $U_{0}=8$, $\omega=4$. The green curve show the mutual information content between $Y$ and the feature vector of a metal at $U_{0}=8$, $\omega=-4$.}
\end{figure}

\noindent 
 Finally, we demonstrate two example cases to check the DNN's ability to distinguishing a metal from a insulator. For this, we compute the mutual information content $\mathcal{I}(X_{j},X_{j,U=2})$ between the target vector $Y$, obtained from the final RG step for the case of strong repulsion $U_{0}=8$, and feature vectors $X_{j,U=2}$ obtained at each step for a case of weak repulsion $U_{0}=2$. The red curve in Fig.\ref{mutualInformationMetIns} represents $\mathcal{I}(Y:X_{j,U=2})$, showing saturation at a finite value $\mathcal{I}(X_{j}:X_{j,U=2})=0.636$. We note that this is equal to the saturation value between the mutual information of the representation and class $\mathcal{I}(X_{j}:Y)$ obtained for the case of strong repulsion $U_{0}=8$ (seen as the value of the MI (black curve) in Fig.\ref{EHM_IB}). This precise match of results implies that the constructed DNN is able to predict that, even at weak repulsion ($U_{0}=2$) and low quantum fluctuation scales ($4-\omega=0$), the phase is a Mott insulator. On the other hand, the green curve in Fig.\ref{mutualInformationMetIns} is obtained from the mutual information between the MI distribution of a metal ($U_{0}=8, \omega=8$) and the target distribution of a insulator. It shows a final sharp dip to the value $\mathcal{I}(X_{j,metal}:Y)=0$, implying that the gapless phase is not an insulator. This allows us to conclude that the DNN (constructed from the MERG based EHM) can successfully distinguish between RG flows that lead to a gapless metal and a gapped topological insulator. Such a DNN can, therefore, be employed for characterising the metal-insulator transition between the two phases.  
\section{Conclusions and discussions}\label{conclusions}
\pin
In summary, we have demonstrated that MERG flow can be used to track the emergence of a topologically ordered (T.O.), gapped insulating state at low energies in the 2D Hubbard model~\cite{anirbanmotti}. Upon obtaining the T.O. ground state wavefunction at the Mott insulator fixed point of Ref.\cite{anirbanmotti}, we perform the inverse unitary transformations to re-entangle the emergent T.O. state with the decoupled degrees of freedom. The fixed point ground state, and the intermediate states reconstructed thereform, are shown to have quantum circuit representations. In this way, we obtain the tensor network RG equivalent to MERG, called the entanglement holographic mapping (EHM). Each layer of the EHM is shown to be composed of two-local unitary disentanglers, and has a finite depth quantified by the circuit complexity. We show that MERG functions as a topological quantum error correcting code, leading to quantization of a nonlocal Wilson loop characterising the T.O. Mott insulating fixed point deep in the bulk of the EHM. It does so by resolving exactly the quantum fluctuations intrinsic to the system. The robustness of MERG quantum error correcting code against external quantum fluctuations is left for future study. 
\par\noindent 
We have also computed the mutual information for all pseudospin pairs at each RG step, thereby extracting the entanglement geometry content from the EHM. Both the geodesic of the emergent geometry and the physical distance between strongest entangled pairs are shown to shrink along the RG direction, finally vanishing at the IR fixed point. In this way, we show that the emergence of short ranged strongly entangled pairs is concurrent with the emergence of topological order. On the other hand, for the normal metal phase of the 2D Hubbard model, the entanglement geometry is seen to expand and finally approach a separable state in the IR, demonstrating thereby the entanglement content of the EHM in this case is markedly different from that of the T.O. Mott insulating state. We also observe the scaling of momentum space block entanglement entropy (EE) for the metal and T.O. phases. In the T.O. phase the EE displays a nonmonotonic RG flow, i.e., an initial decreases followed by a gradual upsurge to a finite value. Additionally, at the fixed point, the EE is found to be constant for all block sizes. This is in contrast with the monotonic decay of EE in the normal metal phase. Thus, we have realized a concrete example of an EHM network that is witness to T.O. Further, the entanglement content and the $(\pi,\pi)$ structure factor of the Ne\'{e}l antiferromagnet at half-filling are observed to vanish and grow respectively under RG. A similar study of the entanglement features and d-wave superconducting ODLRO of the Cooper pair liquid at the critical hole-doping related to the QCP reveal that the collapse of Mottness is tied to the growth of ODLRO. Importantly, we observe a clear anticorrelation between the total quantum correlation content (i.e., the maximum quantum mutual information) in a correlated state of quantum matter, and the tendency for symmetry-broken orders to arise from it. In a recent cold-atom experimental realisation of the fermionic 2D Hubbard model~\cite{cocchi2017}, it was observed that electrons are delocalised over neighbouring lattice sites even in the strongly correlated half-filled Mott insulating state at very low temperatures. This is consistent with our finding of short-distance strongly entangled pairs in the Mott liquid (see Fig.\ref{fig:12} (right panel)). It would be interesting to test some of our other findings in a similar experiment.
\par\noindent 
Finally, we show that the EHM network functions as an optimal deep neural network (DNN) satisfying the information bottleneck (IB) principle. The unitary transformations lead to successive disentanglement of qubits, naturally reducing the entanglement spread and leading eventually to simpler representations of the many-body state. However, the essential topological content of the IR fixed point is retained across the EHM. The predictive power of the DNN is tested via formulation of the conditional probability model that can classify a given pair to be either strongly or weakly entangled. The model is found to predict the entanglement features in the bulk of the EHM from the layers near its boundary. This is realised by a successful prediction of the onset of T.O. during the RG flow well before the stable fixed point is reached. Further, the DNN is successfully able to distinguish the gapless normal metal ground state from the T.O. gapped insulating state, as well as classify insulating ground states reached at two different values of the Hubbard coupling as belonging to the same class. In this way, the DNN is able to classify different phases of the 2D Hubbard model, as well as distinguish between the RG flows that lead to them. Our work sets the stage for yet deeper investigations of the many-particle entanglement of novel gapped as well as gapless quantum liquids that arise from systems of strongly interacting electrons. It also heralds the development of applications of the MERG formalism for the creation of novel quantum error correcting codes and deep neural network architectures based on the quantum entanglement of many-particle wavefunctions.
\vspace*{0.5cm}
\par \noindent {\bf Acknowledgments}
The authors thank R. K. Singh, A. Dasgupta, S. Patra, A. Bhattacharya, A. Taraphder, N. S. Vidhyadhiraja, P. Majumdar, A.-M. S. Tremblay, S. R. Hassan and M. Patra for several discussions and feedback. A. M. thanks the CSIR, Govt. of India for funding through a junior and senior research fellowship. S. L. thanks the DST, Govt. of India for funding through a Ramanujan Fellowship during which a part of this work was carried out.
\appendix
\section{Derivation of the unitary operator}\label{App-1}
\pin
We present briefly the formalism developed in Ref.\cite{anirbanmotti}. $H$ is a Hamiltonian belonging to the space $\mathcal{A}(\mathcal{H}_{2}^{\otimes N})$, where $\mathcal{H}_{2}$ is a SU(2) Hilbert space spanned by the generators  
\begin{eqnarray}\centering
c^{\dagger}_{N},c_{N},\hat{n}_{N}-1/2=c^{\dagger}_{N}c_{N}-1/2~,
\end{eqnarray}
and $\mathcal{A}$ is the antisymmetrizer. We want to find a unitary operation $U_{N}$ that block diagonalizes the Hamiltonian $H$ in the occupation number space of $N:\lbrace|1_{N}\rangle, |0_{N}\rangle\rbrace$
\begin{eqnarray}\centering
&&(1-\hat{n}_{N})U_{N}HU_{N}^{\dagger}\hat{n}_{N}=0\Rightarrow P_{N}H(1-P_{N})=0~,
\end{eqnarray}
where $P_{N}=U^{\dagger}_{N}\hat{n}_{N}U_{N}$. Using the definition of $P_{N}$, we construct the Hamiltonian $H'=P_{N}HP_{N}$, leading to the block equation
\begin{eqnarray}\centering
P_{N}HP_{N}=H'&&\Rightarrow (P_{N}HP_{N}+(1-P_{N})HP_{N})P_{N}=H'P_{N}\nonumber\\
&&\Rightarrow HP_{N}=H'P_{N}\label{Hrot}~.
\end{eqnarray}
It is important to note that
\begin{eqnarray}\centering
P_{N}H^{'}(1-P_{N})=0, [H',P_{N}]=0~,\label{propH'}
\end{eqnarray}
implying that the Hamiltonian $H'$ is block diagonal in the rotated basis. A state $|\Psi\rangle$, belonging to the projected space generated by $P_{N}$, can be written in the occupation number basis of $N$ 
\begin{equation}\centering\centering
|\Psi\rangle = a_{1}(1+\eta_{N})|1_{N}\Psi_{1_{N}}\rangle = a_{1}|1_{N}\Psi_{1_{N}}\rangle+a_{0}|0_{N}\Psi_{0_{N}}\rangle~,
\end{equation}
where $\eta_{N}$ is an (as of now) undetermined operator that connects a many-body state $|1_{N}\Psi_{1_{N}}\rangle$ (with $N$ occupied, $|1_{N}\rangle$) to another state $|0_{N}\Psi_{0_{N}}\rangle$ (where state $N$ is unoccupied, $|0_{N}\rangle$), and the subsystem configuration of the remaining $(1,...,N-1)$ electrons ($|\Psi_{1_{N}}\rangle$) is rotated into the configuration $|\Psi_{0_{N}}\rangle$. 
Taking cue from this decomposition, we can define the projection operator $P_{N}$ using the $\eta_{N}$ operator and its Hermitian conjugate
 \begin{eqnarray}\centering
 P_{N}=\mathcal{N}(1+\eta_{N}+\eta^{\dagger}_{N})~.\label{nonlocalProj}
\end{eqnarray}  
Next, we solve for $\eta_{N}$ by putting $P_{N}$ (eq.\ref{nonlocalProj}) into eq.\ref{Hrot}, and projecting onto the state $|1_{N},\Psi_{1_{N}}\rangle$ 
\begin{eqnarray}\centering
\hat{n}_{N}H\hat{n}_{N}|1_{N},\Psi_{1_{N}}\rangle +\hat{n}_{N}H(1-\hat{n}_{N})\eta_
{N}|1_{N},\Psi_{1_{N}}\rangle &=&H'|1_{N},\Psi_{1_{N}}\rangle~,\label{SimEq1}\\
(1-\hat{n}_{N})H(1-\hat{n}_{N})\eta_{N}|1_{N},\Psi_{1_{N}}\rangle +(1-\hat{n}_{N})H\hat{n}_{N}|1_{N},\Psi_{1_{N}}\rangle &=&H'\eta_{N}|1_{N},\Psi_{1_{N}}\rangle~,\label{SimEq2}\\
\Rightarrow\eta_{N}=\frac{1}{H'-(1-\hat{n}_{N})H(1-\hat{n}_{N})}(1-\hat{n}_{N})H\hat{n}_{N}~.\label{eta_N}
\end{eqnarray}
Given the symmetry properties of $H'$ (eq.\ref{propH'}), it supports the decomposition into a diagonal piece $H^{'D}$, and an off-diagonal piece $H^{'X}_{N}$ constituting all electronic states apart from $N$
\begin{eqnarray}\centering
H'=H^{'X}_{N}+H^{'D}~.\label{decomp2}
\end{eqnarray}
Using eq.\ref{decomp2}, we can obtain a simple description of $\eta_{N}$
\begin{eqnarray}\centering
\eta_{N}=\frac{1}{\hat{\omega}-Tr_{N}(H^{D}(1-\hat{n}_{N}))(1-\hat{n}_{N})}Tr_{N}(c^{\dagger}_{N}H)c_{N}~,
\end{eqnarray}
where $\hat{\omega}=H^{'D}+H^{'X}_{N}-H^{X}_{N}$, and $H^{X}_{N}$ has a definition similar to $H^{'X}_{N}$. $Tr_{N}(c^{\dagger}_{N}H)c_{N}$ is the collection of selected off-diagonal pieces of $H$ involving the state $N$. Additionally, we note that a form for $\eta^{\dagger}_{N}$ can similarly be obtained by projecting onto state $|0_{N},\Psi_{0_{N}}\rangle$
\begin{eqnarray}\centering
\eta^{\dagger}_{N}=\frac{1}{\hat{\omega}-Tr_{N}(H^{D}\hat{n}_{N})\hat{n}_{N}}c^{\dagger}_{N}Tr_{N}(Hc_{N})~.\label{etaDag_N}
\end{eqnarray}
In the above equation $Tr_{N}(.)$ represents the partial tracing of the electronic state $N$, which is carried out in the electronic fock space. Putting $\eta_{N}$eq.\ref{eta_N} into eq.\ref{SimEq1}, and using the form for $\eta^{\dagger}_{N}$, we obtain the algebra for these operators
\begin{eqnarray}\centering
\lbrace\eta^{\dagger}_{N},\eta_{N}\rbrace=1~,~ [\eta^{\dagger}_{N},\eta_{N}]=2\hat{n}_{N}-1~.  \label{eta-algebra}
\end{eqnarray} 
An additional set of relations for $\eta_{N}$ and $\eta^{\dagger}_{N}$ are: $\eta^{2}_{N}=\eta^{\dagger 2}=0$. The state $|\Psi\rangle$ can now be connected via a similarity transformation to the state $|1_{N}\Psi_{1_{N}}\rangle$
\begin{eqnarray}\centering
|\Psi\rangle = a_{1}\exp(\eta_{N})|1_{N}\Psi_{1_{N}}\rangle~.
\end{eqnarray}
Note that in $|1_{N}\Psi_{1_{N}}\rangle$, the state $N$ is now disentangled. Finally, we can construct the form for the unitary transformation $U_{N}$ from the similarity transformation~\cite{suzuki1982construction,shavitt1980quasidegenerate}
\begin{eqnarray}\centering
U_{N}=\frac{1}{\sqrt{2}}(1+\eta_{N}-\eta^{\dagger}_{N})
\end{eqnarray}
Using the algebra eq\ref{eta-algebra} one can verify the identity $UU^{\dagger}=U^{\dagger}U=1$.
\section{Derivation of the normal phase Hamiltonian in pseudospin subspace}\label{App-2}
\pin
A projection of the gapless marginal Fermi liquid normal phase Hamiltonian (eq.\ref{normalState}) of the 2D Hubbard model at $1/2$-filling~\cite{anirbanmotti} in the charge pseudospin subspace gives 
\begin{eqnarray}\centering
H^{*}=\sum_{l=1,i=(1,2)}^{j^{*}}\epsilon_{\Lambda_{l}\hat{s}_{i}}A^{z}_{\Lambda_{l}\hat{s}_{i}}-\sum_{i,j,k=1,2}R^{*}_{k}A^{z}_{\Lambda_{i},\hat{s}_{k}}A^{z}_{\Lambda_{i},\hat{s}_{k}}~.\label{ChargePS}
\end{eqnarray}
The ground state of the gapless normal phase with 14 electrons in 28 states can then be written as
\begin{eqnarray}\centering
|\Psi\rangle_{NP} = \prod_{i=0,n=(0,1)}^{6}|\downarrow_{i+21n}\rangle\prod_{i=7,n=(0,1)}^{13}|\uparrow_{i+7n}\rangle~.\label{NP}
\end{eqnarray}
Upon projecting the Hamiltonian in the basis of Cooper pair pseudospins $B^{+}_{\Lambda,\hat{s}}=c^{\dagger}_{\mathbf{k}_{\Lambda,-\hat{s}},-\sigma}c^{\dagger}_{\mathbf{k}_{\Lambda\hat{s}},\sigma}$, $B^{z}_{\Lambda,\hat{s}}=\frac{1}{2}(\hat{n}_{\mathbf{k}_{\Lambda,-\hat{s}},-\sigma}+\hat{n}_{\mathbf{k}_{\Lambda,\hat{s}},\sigma}-1)$, we attain the form
\begin{eqnarray}\centering
\hspace*{-2cm}
H^{*}=\sum_{\Lambda<\Lambda^{*},i=1,2}\epsilon_{\Lambda\hat{s}_{i}}B^{z}_{\Lambda\hat{s}_{i}}+\sum_{\Lambda,\hat{s}}R^{*}_{\hat{s}}\left(B^{z}_{\Lambda,\hat{s}}+\frac{1}{2}\right)\left(B^{z}_{-\Lambda,T\hat{s}}+\frac{1}{2}\right)\left(B^{z}_{\Lambda',\hat{s}}+\frac{1}{2}\right)~.\label{ChargePS1}
\end{eqnarray}
\section{URG study for the symmetry broken Neel SDW state}\label{URG_ML_SDW}
\pin
In an earlier work~\cite{anirbanmotti}, we showed that upon introducing a staggered magnetic field $(h)$ (by adding a term $h\sum_{i,j}e^{i\pi(i+j)}S^{z}_{\mathbf{r}}$ to the Hamiltonian) in the Mott liquid (ML) leads to the symmetry-broken Ne\'{e}l antiferromagnetic phase. In order to study the competition between the tendency towards symmetry-breaking and the effects of nested Umklapp scattering, we apply the unitary RG formalism (eq.\ref{URG-Formal}) to a simplified model Hamiltonian $H$ which includes only the Umklapp scattering processes (with coupling $V$, whose bare value is the Hubbard $U_{0}$) between electrons in the neighbourhood of the antinodes of the underlying tight-binding Fermi surface, and checkerboard spin density waves (SDW) involving momentum transfer $\mathbf{Q}=(\pi,\pi)$  
\begin{eqnarray}\centering
\hspace*{-2cm}
H&=&h\sum_{\Lambda,\hat{s},\sigma=\pm 1}\sigma (c^{\dagger}_{\Lambda,\hat{s},\sigma}c_{\mathbf{k}_{\Lambda,\hat{s}}-\mathbf{Q},\sigma}+h.c.)+V\sum_{\Lambda,\Lambda',\hat{s}}
c^{\dagger}_{\mathbf{k}_{\Lambda,\hat{s}},\sigma}c^{\dagger}_{\mathbf{k}_{-\Lambda,T\hat{s}},-\sigma}c_{\mathbf{k}_{\Lambda,-\hat{s}},-\sigma}c_{\mathbf{k}_{-\Lambda,-T\hat{s}},\sigma}\nonumber\\
\hspace*{-2cm}
&=&h\sum_{\Lambda,\hat{s},\sigma=\pm 1}\sigma (c^{\dagger}_{\Lambda,\hat{s},\sigma}c_{\mathbf{k}_{-\Lambda,-T\hat{s}},\sigma}+h.c.)+V\sum_{\Lambda,\Lambda',\hat{s}}
c^{\dagger}_{\mathbf{k}_{\Lambda,\hat{s}},\sigma}c_{\mathbf{k}_{-\Lambda,-T\hat{s}},\sigma}c^{\dagger}_{\mathbf{k}_{-\Lambda,T\hat{s}},-\sigma}c_{\mathbf{k}_{\Lambda,-\hat{s}},-\sigma}~~~,~
\end{eqnarray}
where the Umklapp scattering processes are restricted to the $\mathbf{Q}=(\pi,\pi)$ net-momentum pairs, as they carry the dominant spectral weight and lead to the condensation of pseudospins in the IR (eq.\ref{pseudospins})~\cite{anirbanmotti}. The symmetry-breaking field $h$ breaks explicitly the $SU(2)$ spin-rotation symmetry of the $H$, and fixes the quantization axis along the eigen-direction of $\sigma(c^{\dagger}_{\Lambda,\hat{s},\sigma}c_{\mathbf{k}_{-\Lambda,-T\hat{s}},\sigma}+h.c.)$
\begin{eqnarray}\centering
|\psi^{\sigma,\pm}_{\Lambda,\hat{s}}\rangle = \frac{1}{\sqrt{2}}\left[|1_{\mathbf{k}_{\Lambda\hat{s}},\sigma}0_{\mathbf{k}_{-\Lambda,-T\hat{s}},\sigma}\rangle \pm |0_{\mathbf{k}_{\Lambda\hat{s}},\sigma}1_{\mathbf{k}_{-\Lambda,-T\hat{s}},\sigma}\rangle\right]~.\label{eigenDirection}
\end{eqnarray} 
The unitary disentanglement RG operation block-diagonalizes the Hamiltonian iteratively, and the disentangled electronic states are oriented along the eigen-directions (eq.\ref{eigenDirection}) governed by the projection operators $\rho_{j,\hat{s},\sigma}=|\psi^{\sigma,\pm}_{\Lambda,\hat{s}}\rangle\langle\psi^{\sigma,\pm}_{\Lambda,\hat{s}}|$. This leads to the Hamiltonian RG equation
\begin{eqnarray}\centering
\hspace*{-2cm}
U_{(j)}H_{(j)}U^{\dagger}_{(j)}-H_{(j)}=\Delta H_{(j)}=\frac{2l_{j}V^{2}_{(j)}}{h-l^{2}_{j}V_{(j)}}\sum_{\Lambda<\Lambda_{j}}c^{\dagger}_{\mathbf{k}_{\Lambda,\hat{s}},\sigma}c^{\dagger}_{\mathbf{k}_{-\Lambda,T\hat{s}},-\sigma}c_{\mathbf{k}_{\Lambda,-\hat{s}},-\sigma}c_{\mathbf{k}_{-\Lambda,-T\hat{s}},\sigma}~,~~~~~\label{HamRG}
\end{eqnarray}
here $l_{j}$ is the number-density of electronic states on the momentum-shell located at distance $\Lambda_{j}$ from the Fermi surface. The van Hove singularities at the antinodes ($(\pi,0)$ and $(0,\pi)$) induces a logarithmic dependence on the density of states $l_{j}=\log\frac{\Lambda_{0}}{\Lambda_{j}}$. The dependence on $l_{j}$ in the RG  eq.\ref{HamRG} arises by summing the RG contribution of tangential scattering processes between electronic states in the transverse direction to the Fermi surface~\cite{anirbanmotti}. From here, we extract the Umklapp scattering vertex RG flow
\begin{eqnarray}\centering
\frac{\Delta V_{(j)}}{\Delta\log\Lambda/\Lambda_{0}}=\frac{2l_{j}V^{2}_{(j)}}{h-l^{2}_{j}V_{(j)}}~,\label{nonpertSDWRG}
\end{eqnarray}
where $\Delta\log\Lambda/\Lambda_{0}=1$. In the continuum, this RG equation attains the form ($\bar{V}=V/h$)
\begin{eqnarray}\centering
\frac{d\bar{V}}{d\log\Lambda}=\frac{2l(\Lambda)\bar{V}}{1-l^{2}(\Lambda)\bar{V}}~.
\end{eqnarray}
For weak coupling $V<<1/l^{2}(\Lambda)$, the above RG equation has the perturbative one-loop RG equation form $\frac{d\bar{V}}{d\log\Lambda}=2\log\frac{\Lambda_{0}}{\Lambda}\bar{V}^{2}$. This leads to the gap function $\Lambda^{*}/\Lambda_{0}=\exp\left(-\sqrt{\frac{2}{\bar{V}_{0}}}\right)$, where $\bar{V}_{0}=V_{0}/h$ and corresponds to the Ne\'{e}l SDW phase~\cite{fradkin2013field}. From the non-perturbative RG formulation of eq.\ref{nonpertSDWRG} for the passage between the Mott liquid phase and SDW phase, we can construct the effective Hamiltonian at the RG fixed point
\begin{eqnarray}\centering
\centering
H^{*}=\frac{h}{2}(D^{x}_{*,\uparrow}-D^{x}_{*,\downarrow})+V^{*}\sum_{\Lambda<\Lambda^{*},\hat{s}}\mathbf{D}_{\Lambda,\hat{s},\uparrow}\cdot\mathbf{D}_{\Lambda,-\hat{s},\downarrow}~,~V^{*}=\frac{h}{2\log\frac{\Lambda_{0}}{\Lambda^{*}}}~.\label{effH}
\end{eqnarray}
The form of the paired electron-hole pseudospin operators are as follows
\begin{eqnarray}\centering
\centering
D^{x}_{\Lambda,\hat{s},\sigma}&=&\frac{1}{2}(c^{\dagger}_{\mathbf{k}_{\Lambda\hat{s}},\sigma}c_{\mathbf{k}_{-\Lambda,-T\hat{s}},\sigma}+h.c.)~,\nonumber\\
D^{y}_{\Lambda,\hat{s},\sigma}&=&\frac{i}{2}(c^{\dagger}_{\mathbf{k}_{\Lambda\hat{s}},\sigma}c_{\mathbf{k}_{-\Lambda,-T\hat{s}},\sigma}-h.c.)~,\nonumber\\
D^{z}_{\Lambda,\sigma}&=&\frac{1}{2}(\hat{n}_{\mathbf{k}_{\Lambda\hat{s}},\uparrow}-\hat{n}_{\mathbf{k}_{-\Lambda,-T\hat{s}},\uparrow})~,\label{Dpseudospins}
\end{eqnarray}
and $\mathbf{D}_{*,\sigma}=\sum_{\Lambda\leq \Lambda^{*}}\mathbf{D}_{\Lambda,\sigma}$~. For the case $D_{*,\sigma}=1/2$, the ground state of the $H^{*}$ has the form
\begin{eqnarray}\centering
\hspace*{-2cm}
&&|\Psi\rangle=\cos\frac{\theta}{2}|\Uparrow\Downarrow\rangle-\sin\frac{\theta}{2}|\Downarrow\Uparrow\rangle~, \nonumber\\
\hspace*{-2cm}
&&|\Uparrow\rangle=\frac{1}{\sqrt{2}}\left[|1_{\mathbf{k}_{\Lambda^{*}\hat{s}},\uparrow}0_{\mathbf{k}_{-\Lambda^{*},-T\hat{s}},\uparrow}0_{\mathbf{k}_{\Lambda^{*}\hat{s}},\downarrow}0_{\mathbf{k}_{-\Lambda^{*},-T\hat{s}},\downarrow}\rangle-|0_{\mathbf{k}_{\Lambda^{*}\hat{s}},\uparrow}1_{\mathbf{k}_{-\Lambda^{*},-T\hat{s}},\uparrow}0_{\mathbf{k}_{\Lambda^{*}\hat{s}},\downarrow}0_{\mathbf{k}_{-\Lambda^{*},-T\hat{s}},\downarrow}\rangle\right]~,\nonumber\\
\hspace*{-2cm}
&&|\Downarrow\rangle=\frac{1}{\sqrt{2}}\left[|1_{\mathbf{k}_{\Lambda^{*}\hat{s}},\uparrow}0_{\mathbf{k}_{-\Lambda^{*},-T\hat{s}},\uparrow}0_{\mathbf{k}_{\Lambda^{*}\hat{s}},\downarrow}0_{\mathbf{k}_{-\Lambda^{*},-T\hat{s}},\downarrow}\rangle+|0_{\mathbf{k}_{\Lambda^{*}\hat{s}},\uparrow}1_{\mathbf{k}_{-\Lambda^{*},-T\hat{s}},\uparrow}0_{\mathbf{k}_{\Lambda^{*}\hat{s}},\downarrow}0_{\mathbf{k}_{-\Lambda^{*},-T\hat{s}},\downarrow}\rangle\right]~,\nonumber\\~~~~\label{SDWstate}
\end{eqnarray}
where
\begin{eqnarray}\centering
\cos\theta=\frac{h}{\sqrt{h^{2}+\frac{(V^{*})^{2}}{4}}}=\frac{1}{\sqrt{1+\frac{1}{16\log^{2}\frac{\Lambda_{0}}{\Lambda^{*}}}}}~.
\end{eqnarray} 
For an IR fixed-point with a momentum-space width around the Fermi surface of $\Lambda^{*}<\Lambda_{0}$, the superposition coefficient $\cos\frac{\theta}{2}\to 1$ results in a highly polarized Ne\'{e}l antiferromagnetic configuration~\cite{anirbanmotti}. Upon simulating the RG equation for bare $V\equiv U_{0}=8$ on a $512\times 512$ momentum-space grid, we found that the magnitude of the coupling at RG fixed point is $V^{*}=8\times 10^{-5}t$, the fixed point width $\Lambda^{*}/\Lambda_{0}=0.988$ and polarization coefficient given by $\cos\theta/2=0.999$. From here, we can perform the MERG scheme using the inverse unitary transformations $U^{\dagger}_{j+1}|\Psi_{(j)}\rangle = |\Psi_{(j+1)}\rangle$ (see Fig.\ref{URGflowScheme}), and obtain the many-body states across RG scales from UV to IR. 
\section{Effective Cooper pair Hamiltonian obtained from the Mott liquid fixed point at half filling}\label{Coop}
\pin
We begin by writing the effective Hamiltonian for the spin-type Mott liquid (eq.\ref{fixed_point_Ham_mott}) 
\begin{eqnarray}\centering
\hat{H}^{*}_{2}&=&\sum_{\Lambda,\hat{s}}\epsilon_{\Lambda,\hat{s}}\hat{n}_{\Lambda,\hat{s},\sigma}-\sum_{\hat{s}}K^{*}_{\hat{s}}\mathbf{S}_{*, \hat{s}}\cdot\mathbf{S}_{*, -\hat{s}}~,
\end{eqnarray}
where the spin-type pseudospins ($\mathbf{S}_{*, \hat{s}}$) within the emergent fixed-point window are defined as
\begin{eqnarray}\centering
\mathbf{S}_{\Lambda,\hat{s}}=f^{s;\dagger}_{\Lambda,\hat{s}}\frac{\boldsymbol{\sigma}}{2}f^{s}_{\Lambda,\hat{s}}~,f^{s;\dagger}_{\Lambda,\hat{s}} = \left[c^{\dagger}_{\Lambda,\hat{s},\sigma}~ c^{\dagger}_{\Lambda - 2\Lambda^{*}_{\hat{s}},T\hat{s},-\sigma}\right]~,
\end{eqnarray}
and demonstrate the presence of Cooper pairs within the spin-type Mott liquid. For this, we rewrite the above effective Hamiltonian as $\hat{H}^{*}_{2}=H_{0}+H_{1}$, where
\begin{eqnarray}\centering
H_{0}&=&\sum_{\Lambda,\hat{s}}\epsilon_{\Lambda,\hat{s}}\hat{n}_{\Lambda,\hat{s},\sigma}+\sum_{\hat{s},\Lambda}K^{*}_{\hat{s},\mathbf{p}=0}c^{\dagger}_{\Lambda,\hat{s},\sigma}c^{\dagger}_{\Lambda,-\hat{s},-\sigma}c_{\Lambda-2\Lambda^{*},T\hat{s},-\sigma}c_{\Lambda-2\Lambda^{*},-T\hat{s},\sigma}~,\nonumber\\
H_{1}&=&\sum_{\Lambda\neq\Lambda'}K^{*}_{\hat{s},\mathbf{p}}c^{\dagger}_{\Lambda,\hat{s},\sigma}c^{\dagger}_{\Lambda',-\hat{s},-\sigma}c_{\Lambda-2\Lambda^{*},T\hat{s},-\sigma}c_{\Lambda'-2\Lambda^{*},-T\hat{s},\sigma}~,
\end{eqnarray}
and where $H_{0}$ denotes the physics of zero-momentum electron pairs ($|\mathbf{k}_{\Lambda,\hat{s}},\sigma\rangle$, $|\mathbf{k}_{\Lambda,-\hat{s}},-\sigma\rangle$), $H_{1}$ the physics of non-zero net-momentum electron pairs ($|\mathbf{k}_{\Lambda,\hat{s}},\sigma\rangle$, $|\mathbf{k}_{\Lambda',-\hat{s}},-\sigma\rangle$) and $\mathbf{p}=\mathbf{k}_{\Lambda_{j},\hat{s}}+\mathbf{k}_{\Lambda_{j}-\delta,-\hat{s}}$. Here $\delta>0$, such that we account for only the states within the  window $\Lambda'=\Lambda-\delta<\Lambda_{j}$, where $\Lambda_{N}=\Lambda^{*}>\Lambda_{N-1}>\ldots >0$. Also, $K^{*}_{\hat{s},\mathbf{p}}=K^{*}_{\hat{s},\mathbf{p}=0}\equiv K^{*}_{\hat{s}}$ at the bare level. We now perform a second renormalization group analysis of the Hamiltonian $H^{*}_{2}$ in order to study the effect of the non-zero momentum pairs on their zero-momentum counterparts. For this, we write down the unitary disentangling operator
\begin{eqnarray}\centering
U_{\mathbf{k}_{\Lambda\hat{s}},\sigma}=\frac{1}{\sqrt{2}}\left[1+\eta_{\mathbf{k}_{\Lambda\hat{s}},\sigma}-\eta^{\dagger}_{\mathbf{k}_{\Lambda\hat{s}},\sigma}\right]~,
\end{eqnarray}
where
\begin{eqnarray}\centering
\eta_{\mathbf{k}_{\Lambda\hat{s}},\sigma}&=&\frac{1}{\hat{\omega}-Tr_{\mathbf{k}_{\Lambda\hat{s}},\sigma}(H^{D}_{\mathbf{k}_{\Lambda\hat{s}},\sigma}\hat{n}_{\mathbf{k}_{\Lambda\hat{s}},\sigma)}}c^{\dagger}_{\mathbf{k}_{\Lambda\hat{s}},\sigma}Tr_{\mathbf{k}_{\Lambda\hat{s}},\sigma}(H_{\mathbf{k}_{\Lambda\hat{s}},\sigma}c_{\mathbf{k}_{\Lambda\hat{s}},\sigma})~.
\end{eqnarray} 
From the Hamiltonian RG flow equation $H_{(j-1)}=U_{(j)}H_{(j)}U^{\dagger}_{(j)}$, we obtain the coupling RG equation
\begin{eqnarray}\centering
K_{\hat{s},\mathbf{p}}^{*,(j-1)}-K_{\hat{s},\mathbf{p}}^{*,(j)}=\frac{(K^{*,(j)}_{\hat{s},\mathbf{p}})^{2}}{\omega-\frac{1}{2}(\epsilon_{\Lambda_{j},\hat{s}}+\epsilon_{\Lambda_{j}-\delta,-\hat{s}})-\frac{K_{\hat{s},\mathbf{p}}^{*,(j)}}{4}}~.
\end{eqnarray}
For $\omega>0$, the denominator $\omega-\frac{1}{2}(\epsilon_{\Lambda_{j},\hat{s}}+\epsilon_{\Lambda_{j}-\delta,-\hat{s}})$ attains its highest value for $\mathbf{p}=0$, thereby indicating the domination of the $\mathbf{p}=0$ pairs under RG. By accounting solely for the dominant scattering processes, the effective Hamiltonian at the IR fixed point ($V^{*}=4(\omega -\epsilon_{\Lambda^{**}\hat{s}})$) is then given by 
\begin{eqnarray}\centering
H^{*}_{0}&=&\sum_{\Lambda<\Lambda^{**},\hat{s}}\epsilon_{\Lambda,\hat{s}}B^{z}_{\Lambda,\hat{s}}+\epsilon_{\Lambda-2\Lambda^{*},T\hat{s}}B^{z}_{\Lambda-2\Lambda^{*},T\hat{s}}+\sum_{\hat{s},\Lambda<\Lambda^{**}}K^{**}_{\hat{s}}B^{+}_{\Lambda,\hat{s}}B^{-}_{\Lambda-2\Lambda^{*},T\hat{s}}~,\label{CooperPairML}
\end{eqnarray} 
where $B^{+}_{\Lambda,\hat{s}}=c^{\dagger}_{\mathbf{k}_{\Lambda,-\hat{s}},-\sigma}c^{\dagger}_{\mathbf{k}_{\Lambda\hat{s}},\sigma}$,~ $B^{z}_{\Lambda,\hat{s}}=\frac{1}{2}(\hat{n}_{\mathbf{k}_{\Lambda,-\hat{s}},-\sigma}+\hat{n}_{\mathbf{k}_{\Lambda,\hat{s}},\sigma}-1)$ etc.
\section{Effective Hamiltonian and eigenstates at critical doping}\label{AppQCP}
\pin
At the QCP of the doped 2D Hubbard model ($\omega=4$ and $\Delta\mu_{eff}=-4$)~\cite{anirbanmott2}, the nodal direction $(\pi/2,\pi/2)$ is gapless while the antinodal regions centered around $(0,\pi)$ and $(\pi,0)$ are gapped. The effective Hamiltonian for the QCP was derived in Ref.\cite{anirbanmott2}  
and is given by $H_{QCP}$
\begin{eqnarray}\centering
\hspace*{-2cm}
H_{QCP}&=&\sum_{\hat{s}=\hat{s}_{AN},\Lambda,\Lambda'}^{\hat{s}'}K^{*}_{\hat{s}}(B^{+}_{\Lambda,\hat{s}}B^{-}_{\Lambda',-\hat{s}}+h.c.)\nonumber\\
\hspace*{-2cm}
&+&\sum_{\Lambda<\Lambda^{*}}\epsilon_{\Lambda\hat{s}_{N}}B^{z}_{\Lambda\hat{s}_{N}}+\sum_{\Lambda,\hat{s}}R^{*}_{\hat{s}}\left(B^{z}_{\Lambda,\hat{s}_{N}}+\frac{1}{2}\right)\left(B^{z}_{-\Lambda,\hat{s}_{N}}+\frac{1}{2}\right)\left(B^{z}_{\Lambda',\hat{s}_{N}}+\frac{1}{2}\right)~.\label{QCP}
\end{eqnarray}
Note that we have written the two electron-one hole vertices $R^{*}_{ll'}\hat{n}_{j, l}\hat{n}_{j, l'}(1-\hat{n}_{j', l})$ describing the marginal Fermi liquid (eq.\ref{normalState}) in terms of Anderson pseudospins defined below eq.\ref{CooperPairML} with $l=\hat{s}_{N}$ (the nodal direction). The first term in eq.\ref{QCP} reflects the physics of the off-diagonal backscattering vertices between opposite directions normal to the Fermi surface, and generates a gap in the antinodal region. The ground state wavefunction can be written in terms of the two Fermi points, one near the antinodes $\hat{s}=(\pi/4,3\pi/4)$ and another along the nodes $\pi/2,\pi/2$, as follows
\begin{eqnarray}\centering
|\Psi\rangle= \frac{1}{2}\prod_{n=[0,3]}\prod_{i=[7n,7n+2],j=[7n+3,7n+5]}|0_{i}1_{j}\rangle(|0_{6}1_{20}\rangle-|1_{6}0_{20}\rangle)|1_{13}1_{27}\rangle~.\label{QCP_State}
\end{eqnarray}
The Cooper pair (Anderson) pseudospins qubits $6$ and $20$ within the antinodal patch along $(3\pi/4,\pi/4)$ form a singlet ground state $|0_{6}1_{20}\rangle-|1_{6}0_{20}\rangle$ via the backscattering vertex in eq.\ref{QCP}. Importantly, we note that at the QCP, the Umklapp charge backscattering vertices are RG irrelevant. This is due to the presence of a term for the doublon-holon disparity caused by hole doping in eq.\ref{fixed_point_Ham_mott}~\cite{anirbanmott2}: $-\Delta\mu_{eff}(A^{z}_{*,\hat{s}}+A^{z}_{*,-\hat{s}})$. This can be simply seen from the modified RG equations (eq.\ref{RGeqns}) for Umklapp backscattering $K^{(j-1)}_{l}$ at the QCP $\Delta\mu_{eff}=-\omega=4$
\begin{eqnarray}\centering
\Delta K_{l}^{(j-1)}&=&\frac{(K_{l}^{(j)})^{2}}{\omega - \frac{1}{2}(\epsilon_{\Lambda_{j}\hat{s}}+\epsilon_{\Lambda_{j} -\hat{s}})+\Delta\mu_{eff}-\frac{1}{4}K_{l}^{(j)}}~.
\end{eqnarray}
The negative signature in the denominator $\omega - \frac{1}{2}(\epsilon_{\Lambda_{j}\hat{s}}+\epsilon_{\Lambda_{j} -\hat{s}})+\Delta\mu_{eff}<0$ renders the coupling $K_{l}$ RG irrelevant.
\bibliographystyle{plain} 
\bibliography{netbib}

\end{document}